\documentclass[12pt]{article}

\usepackage[normalem]{ulem}
\usepackage{epsfig}
\usepackage{natbib}
\usepackage{amsfonts}
\usepackage{amsmath}
\usepackage{amssymb}
\usepackage{hyperref}
\usepackage{colortbl}
\usepackage{upgreek}

\newtheorem{lemma}{Lemma}

\newcommand{\RomanNumeralCaps}[1]

\usepackage{tikz}
\usetikzlibrary{decorations.markings}
\usepackage{mathrsfs}
\usepackage[all,cmtip]{xy}
\usepackage{mleftright}
\usepackage{booktabs}
\usepackage{enumitem}
\usepackage{subfigure}
\usepackage{placeins}
\usepackage{comment}
\usepackage{pdflscape}

\usepackage[ruled]{algorithm}%
\usepackage{algorithmic}

\setlist[enumerate]{labelsep=*, leftmargin=1.5pc}
\setlist[enumerate]{label=\normalfont(\roman*), ref=\roman*}

\DeclareFontFamily{U}{mathx}{\hyphenchar\font45}
\DeclareFontShape{U}{mathx}{m}{n}{
      <5> <6> <7> <8> <9> <10>
      <10.95> <12> <14.4> <17.28> <20.74> <24.88>
      mathx10
      }{}
\DeclareSymbolFont{mathx}{U}{mathx}{m}{n}
\DeclareFontSubstitution{U}{mathx}{m}{n}
\DeclareMathAccent{\widecheck}{0}{mathx}{"71}

\newtheorem{thm}{Theorem}[section]

\newtheorem{theorem}[thm]{Theorem}

\newtheorem{remark}[thm]{Remark}

\newtheorem{question}[thm]{Question}
\newtheorem{problem}[thm]{Problem}

\newcommand{\IGNORE}[1]{}
\newcommand{\ignore}[1]{}

\newcommand{\re}{\operatorname{Re}}

\newcommand{\mbb}[1]{\mathbb{#1}}

\newcommand{\jt}{\textstyle}

\newcommand{\la}{\langle}
\newcommand{\ra}{\rangle}

\newcommand{\n}{\mathbf{\tilde{n}}}

\newcommand{\bds}{\boldsymbol}

\newcommand{\sech}{\operatorname{sech}}

\def\bnabla{\boldsymbol{\nabla}}

\def\e{\boldsymbol{e}}
\def\f{\boldsymbol{f}}
\def\g{\boldsymbol{g}}
\def\k{\boldsymbol{k}}
\def\u{\boldsymbol{u}}
\def\v{\boldsymbol{v}}

\def\n{\boldsymbol{n}}
\def\x{\boldsymbol{x}}
\def\y{\boldsymbol{y}}
\def\z{\boldsymbol{z}}

\def\bF{{\mathbf F}}
\def\bG{{\mathbf G}}
\def\bU{{\mathbf U}}
\def\bW{{\mathbf W}}

\def\0{{\mathbf 0}}

\def\bomega{\boldsymbol{\omega}}

\def\RR{\mathbb{R}}
\def\TT{\mathbb{T}}
\def\CC{\mathbb{C}}
\def\ZZ{\mathbb{Z}}
\def\NN{\mathbb{N}}

\def\hu{\widehat{\u}}

\def\teta{\widetilde{\boldsymbol{\eta}}}

\newcommand{\argmin}{\operatorname{argmin}}
\newcommand{\argmax}{\operatorname{argmax}}

\newcommand{\esssup}{\operatorname{ess\,sup}}

\renewcommand{\L}{\mathcal L}

\def\B{\mathcal{B}}

\def\E{\mathcal{E}}
\def\H{\mathcal{H}}
\def\K{\mathcal{K}}
\def\M{\mathcal{M}}
\def\N{\mathcal{N}}
\def\O{\mathcal{O}}
\def\P{\mathcal{P}}
\def\Q{\mathcal{Q}}
\def\R{\mathcal{R}}
\def\T{\mathcal{T}}
\def\X{\mathcal{X}}

\newcommand{\tu}{\widetilde{\mathbf{u}}}
\newcommand{\tuE}{\widetilde{\mathbf{u}}_{\E_0}}
\newcommand{\tuB}{\widetilde{\mathbf{u}}_{B}}

\newcommand{\tuET}{\widetilde{\u}_{0;\E_0,T}}

\newcommand{\tuBT}{\widetilde{\u}_{0;B,T}}

\newcommand{\tuEtT}{\widetilde{\u}_{0;\E_0,\tTE}}
\newcommand{\tuBtB}{\widetilde{\u}_{0;B,\tTB}}
\newcommand{\tuT}{\widetilde{\u}_{0;T}}

\newcommand{\uST}{\u_{0;S,T}}
\newcommand{\uBT}{\u_{0;B,T}}

\newcommand{\tuuE}{\widetilde{u}_{\E_0}}
\newcommand{\tuuET}{\widetilde{u}_{0;\E_0,T}}
\newcommand{\Tmax}{T_{\text{max}}}
\newcommand{\tuuETm}{\widetilde{u}_{0;\E_0,\Tmax^{\E_{0}}}}

\def\tTE{\widetilde{T}_{\E_0}}
\def\tTB{\widetilde{T}_{B}}

\newcommand{\wn}{\omega_{\nu}}
\newcommand{\psin}{\psi_{\nu}}
\newcommand{\chin}{\chi_{\nu}}
\newcommand{\chinck}{\widecheck{\chi}^T_{\nu}}
\newcommand{\phichk}{\widecheck{\varphi}_{\nu}^T}
\def\talpha{\widetilde{\alpha}}
\def\tC{\widetilde{C}}

\newcommand{\intO}{\int_{\Omega}}

\definecolor{Gray}{gray}{0.9}

\def\Bmp#1{ \begin{minipage}{#1} }
\def\Emp{ \end{minipage} }
\def\Bmpc#1{ \begin{minipage}[c]{#1} }
\def\Bmpt#1{ \begin{minipage}[t]{#1} }
\def\Bmpb#1{ \begin{minipage}[b]{#1} }

\begin{document}
\title{Extreme flows: where physics meets mathematically rigorous bounds}

\author{Bartosz Protas\thanks{Email address for correspondence: bprotas@mcmaster.ca} 
\\ \\ 
Department of Mathematics and Statistics, McMaster University \\
Hamilton, Ontario, L8S 4K1, Canada
}

\date{\today}

\maketitle

\begin{abstract} 
Extreme flows realize the largest possible growth, either 
instantaneously or in finite time, of certain quantities of interest 
which is achieved by a suitable choice of the initial condition or the 
applied forcing. The quantities of interest usually measure some 
small-scale properties and therefore provide information about the 
regularity of the flow. Extreme behavior is at the heart of several 
open problems in fluid mechanics including the dissipation anomaly in 
turbulence and formation of singularities in various models of fluid 
flow. In this essay we describe a framework making it possible to study 
such extreme behavior systematically by combining mathematical 
analysis, scientific computation and physics. As a first step, one aims 
to deduce rigorous upper bounds on the growth of the quantities of 
interest in the solutions of a given model. These inequalities express 
fundamental limitations on the most extreme behavior possible among 
{\em all} admissible solutions. However, given how they are obtained, 
these bounds may be conservative and overestimate the growth actually 
realizable in the system. In order to probe this possibility, as the 
next step, we set up variational optimization problems where the growth 
of the quantity of interest is maximized under suitable constraints. 
Solution of such problems is enabled by modern methods of numerical 
optimization. When properties of the thus obtained maximizers match the 
bounds, the bounds are declared sharp and therefore cannot be 
fundamentally improved. Finally, properties of the solutions saturating 
the bounds reveal insights about the physical mechanisms realizing the 
extreme behavior in a given problem. We survey problems where this 
research program has produced sharp bounds together with extreme flows 
saturating these bounds, such that these problems can be considered 
closed. A collection of open problems, either still under investigation 
or amenable to the application of the proposed framework with the 
promise of new insights, is then presented. We close the essay with a 
discussion of a handful of methodological improvements. 
\end{abstract}

\begin{flushleft}
  Mathematical Foundations: Navier-Stokes equations, Variational
  methods; 
\end{flushleft}  


\section{Introduction}
\label{sec:intro}

The standard mathematical model describing the motion of a viscous 
incompressible fluid is the Navier-Stokes system \citep{dg95}
\begin{subequations}\label{eq:NS}
\begin{alignat}{2}
\partial_t\u + \left(\u\cdot\bnabla\right)\u + \bnabla p - \nu\Delta\u & = \f & &\qquad\mbox{in} \,\,(0,T]\times\Omega, \label{eq:NSa} \\
\bnabla\cdot\u & = 0 & & \qquad\mbox{in} \,\,(0,T]\times\Omega, \label{eq:NSb} \\
\u(0) & = \u_0 &   & \qquad\mbox{in} \,\,\Omega,  \label{eq:NSc}
\end{alignat}
\end{subequations}
where $T > 0$ is the length of the time window of interest and $\Omega 
\subseteq \RR^{d}$, $d = 2,3$, is the flow domain. With $t \in [0,T]$ 
and $\x = [x_{1}, x_{2}, x_{3}]^{T} \in \Omega$ denoting, respectively, the 
time and the position vector, $\u = [u_{1}, u_{2}, u_{3}]^{T} = \u(t,\x)$ 
represents the velocity field, $p = p(t,\x)$ is the scalar pressure 
field and $\nu > 0$ the coefficient of kinematic viscosity, a constant. 
The source term $\f = \f(t,\x)$ models the bulk force applied to the 
fluid and $\u_{0}$ is the initial condition assumed to be 
divergence-free ($\bnabla \cdot \u_{0} = 0$). Equation \eqref{eq:NSa} 
represents Newton's second law of mechanics (conservation of momentum), 
whereas the divergence-free condition \eqref{eq:NSb} expresses the 
conservation of mass. Without the loss of generality, formulation 
\eqref{eq:NS} implicitly assumes the fluid density is constant and 
equal to unity ($\rho \equiv 1$, with "$\equiv$" meaning "identically 
equal"). System \eqref{eq:NS} is equipped with suitable boundary 
conditions whose form may depend on the domain $\Omega$. In the 
unforced case with $\f \equiv \0$, there are three physical parameters 
determining solutions of system \eqref{eq:NS}: the characteristic 
length scale $L$ (which can often be taken as the "size" of the domain, 
i.e., $L = |\Omega|^{-1/d}$), the characteristic velocity $U$ 
(typically determined by the initial data $\u_{0}$) and the kinematic 
viscosity $\nu$, which can be combined into a single nondimensional 
similarity parameter, the Reynolds number $Re := U L / \nu$ ("$:=$" 
means "equal to by definition"). Thus, for a fixed flow domain $\Omega$, 
solutions of the unforced system \eqref{eq:NS} depend on this one 
parameter only. In the presence of forcing, when $\f \not\equiv \0$, 
another relevant parameter is the Grashof number $Gr := L^{3 - d/2} \left[ 
\| \f \|_{L^{2}} \right]_{T}/ (\rho \nu^{2})$, where $[\cdot]_{T} := 
(1/T) \int_{0}^{T} \cdot \, dt$ is the time-averaging operator, which 
also controls the Reynolds number. In general, we are interested in the 
large $Re$ and/or large $Gr$ regimes where nonlinear effects dominate 
the linear viscous damping. The pressure $p$ can be regarded as a 
Lagrange multiplier providing an additional degree of freedom at each 
point necessary to accommodate the incompressibility condition 
\eqref{eq:NSb}. It can be recovered from the velocity field by solving 
the problem
\begin{equation}
-\Delta p = \bnabla\cdot\left[ \left(\u\cdot\bnabla\right)\u \right] = \bnabla \u : \left( \bnabla \u \right)^{T} 
\qquad \text{in} \ \Omega
\label{eq:p}	
\end{equation}
obtained by applying the divergence operator $(\bnabla\cdot)$ to 
\eqref{eq:NSa} and using \eqref{eq:NSb}. Problem \eqref{eq:p} is 
subject  to pressure boundary conditions deduced from the velocity 
boundary conditions imposed on $\partial \Omega$.

Since being introduced by Navier in 1822 and by Stokes in 1842, system 
\eqref{eq:NS} has found a wide range of applications in different areas 
of science, engineering and, more recently, in medicine, where one 
needs to study flows of fluids. However, despite decades of concerted 
research efforts, we are still quite far from understanding some of the 
basic mathematical and physical properties of solutions of the 
Navier-Stokes system \eqref{eq:NS}, especially in three dimensions (3D, 
$d = 3$). This is very unsatisfactory given the significance of this 
model for both fundamental and applied research. In this essay, we 
consider in detail two such open problems where the common theme is the 
question about the most "extreme" behavior possible in solutions of 
system \eqref{eq:NS} and of related problems. 

Since many questions concerning solutions of the Navier-Stokes system 
\eqref{eq:NS}, especially in 3D, are currently considered intractable, 
research has also focused on its various simplified models. One way to 
obtain such models is to simplify the "physics" the model describes, 
e.g., by neglecting one or more physical effect. Eliminating the 
viscous dissipation from the Navier-Stokes system \eqref{eq:NS} by 
setting $\nu = 0$ and adjusting the boundary conditions, one obtains 
the Euler system 
\begin{subequations}\label{eq:Eu}
\begin{alignat}{2}
\partial_t\u + \left(\u\cdot\bnabla\right)\u + \bnabla p & = \f & &\qquad\mbox{in} \,\,(0,T]\times\Omega, \label{eq:Eua} \\
\bnabla\cdot\u & = 0 & & \qquad\mbox{in} \,\,(0,T]\times\Omega, \label{eq:Eub} \\
\u(0) & = \u_0 &   & \qquad\mbox{in} \,\,\Omega.  \label{eq:Eucc}
\end{alignat}
\end{subequations}
While this system too has received a lot of attention 
\citep{gbk08,DrivasElgindi2023,Elgindi2025}, our general understanding 
of the properties of its solutions is arguably not much better than for 
the original Navier-Stokes system \eqref{eq:NS}. Another possibility is 
to consider models of system \eqref{eq:NS} in a smaller spatial 
dimension. Setting $d = 1$ and relaxing the incompressibility condition 
\eqref{eq:NSb}, we obtain the one-dimensional (1D) Burgers equation 
\begin{subequations}\label{eq:B}
\begin{alignat}{2}
\partial_t u  + u \partial_{x} u - \nu\partial_{xx} u & = f & &\qquad\mbox{in} \,\,(0,T]\times\Omega, \label{eq:Ba} \\
u(0) & = u_0, &   & \qquad\mbox{in} \,\,\Omega,  \label{eq:Bc}
\end{alignat}
\end{subequations}
where $x$, $f$ and $u$ are the scalar counterparts of $\x$, $\f$ and 
$\u$. We observe that for all nontrivial (non-constant) solutions we have 
$\partial_{x} u \not\equiv  0$. In the unforced setting with $f \equiv 
0$ system \eqref{eq:B} admits closed-form solutions via the Cole-Hopf 
transform. Since we have a good understanding of its solutions 
\citep{kl04}, system \eqref{eq:B} has served as a useful testbed for 
exploring different ideas about turbulence and extreme behavior. The 
inviscid Burgers system
\begin{subequations}\label{eq:B0}
\begin{alignat}{2}
\partial_t u  + u \partial_{x} u  & = f & &\qquad\mbox{in} \,\,(0,T]\times\Omega, \label{eq:B0a} \\
u(0) & = u_0 &   & \qquad\mbox{in} \,\,\Omega,  \label{eq:B0c}
\end{alignat}
\end{subequations}
is also of interest. 

When there is no risk of confusion we will use the simplified notation 
$\u(t) = \u(t,\cdot)$ and $u(t) = u(t,\cdot)$ for the solutions of 
systems \eqref{eq:NS}--\eqref{eq:Eu} and \eqref{eq:B}--\eqref{eq:B0}, 
respectively. In some circumstances it will also be useful to 
explicitly indicate the dependence of the solution on the initial 
condition, i.e., $\u = \u(t;\u_{0})$ and $u = u(t;\u_{0})$. Hereafter 
and unless stated otherwise we assume the flow domain is a 
$d$-dimensional torus $\Omega = \TT^{d} := [0,1]^{d}$, $d = 1,2,3$, 
i.e., solutions of systems \eqref{eq:NS}--\eqref{eq:Eu} ($d = 2,3$) and 
\eqref{eq:B}--\eqref{eq:B0} ($d = 1$) are periodic in all $d$ Cartesian 
directions; in the former case, it is additionally assumed that the 
pressure $p$, and hence also its gradient $\bnabla p$, are periodic as 
well.

\subsection{Two Questions of Interest}
\label{sec:questions}

Here we formulate two general questions of both mathematical and 
physical interest that remain open as regards solutions of the 
Navier-Stokes system \eqref{eq:NS} together with a suitable motivation. 
The first question is whether for any smooth initial condition $\u_{0}$ 
systems \eqref{eq:NS} and \eqref{eq:Eu} in 3D always admit smooth 
solutions $\u(t)$ for arbitrarily long times $t$; in other words, the 
question is about the possibility of a finite-time "blow-up". The 
second question concerns the so-called "dissipation anomaly", a 
scenario where the suitably normalized rate of energy dissipation may 
not vanish in the inviscid limit (i.e., when $Re \rightarrow 
\infty$).

\subsubsection{Global Existence of Solutions versus Finite-time Blow-up}
\label{eq:blowup}

\begin{figure}
\centering
   \mbox{
     \subfigure[]{\includegraphics[width=0.35\textwidth]{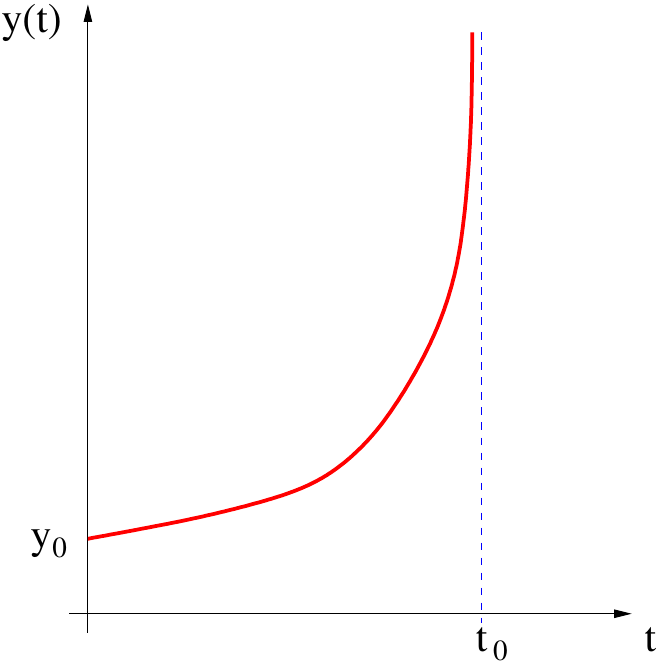}}\qquad
     \subfigure[]{\includegraphics[width=0.5\textwidth]{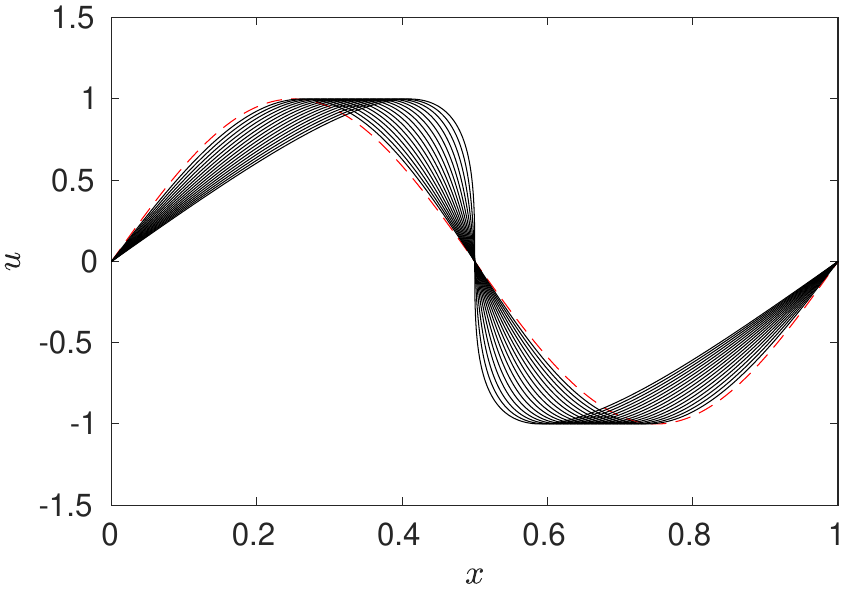}}}
        \caption{(a) Solution $y(t)$ of model problem \eqref{eq:y2} 
   exhibiting blow-up as $t \rightarrow t_0 = 1/y_{0}$; (b) Solutions 
   $u(t_{i},x)$ of system \eqref{eq:B0} with the initial condition 
   $u_{0}(x) = \sin(x)$ at different times $t_{i} \in [0,1]$; a 
   singularity occurs at $x_{s} = 1/2$ where $\lim_{t \rightarrow 1} 
   \big| \partial_{x} u(t,x_{s}) \big| = \infty$.}
     \label{fig:blowup}
\end{figure}

By "blow-up" we mean a situation where a solution of an evolutionary 
differential equation spontaneously develops a singularity such that 
the equation can no longer be satisfied in the classical pointwise 
sense. In this section we present some very elementary arguments  
illustrating why, in principle, this scenario could occur in nonlinear 
systems such as \eqref{eq:NS} and \eqref{eq:Eu}. For simplicity, we 
will focus here on the latter and consider its vorticity form obtained 
applying the curl operator ($\bnabla\times\cdot$) to \eqref{eq:Eua}
\begin{subequations}\label{eq:EuVort}
\begin{alignat}{2}
\partial_t\bomega + \left(\u\cdot\bnabla\right)\bomega & = (\bnabla \u) \bomega & &\qquad\mbox{in} \,\,(0,T]\times\Omega, \label{eq:EuVorta} \\
\bomega & = \bnabla \times \u & & \qquad\mbox{in} \,\,(0,T]\times\Omega, \label{eq:vort} \\
\bomega(0) & = \bnabla \times \u_0 &   & \qquad\mbox{in} \,\,\Omega,  \label{eq:EuVortc}
\end{alignat}
\end{subequations}
where \eqref{eq:vort} defines the vorticity vector $\bomega$ and the 
term on the right-hand side (RHS) of \eqref{eq:EuVorta} represents 
vortex stretching. Velocity can be expressed in terms of vorticity by 
inverting relation \eqref{eq:vort} which gives $\u = - \bnabla 
\Delta^{-1} \bomega = \int_{\Omega} \bG(\cdot,\x')\bomega(\x')\, d\x' 
=: \bG * \bomega$ where $\Delta^{-1}$ is the inverse Laplacian 
associated with suitable boundary conditions whereas $\bG(\x,\x')$ is 
the Biot-Savart kernel obtained applying the gradient operator 
$\bnabla$ to Green's function of the Laplacian on the domain $\Omega$. 
We can thus rewrite system \eqref{eq:EuVorta}--\eqref{eq:vort} as
\begin{equation}
\frac{d \bomega}{dt} = \left[ \bnabla (\bG * \bomega) \right] \,\bomega,
\label{eq:Gw}
\end{equation}
where $d/dt$ represents the Lagrangian derivative, which clearly 
reveals its quadratic nature. To see 
what this could imply, we will consider what is arguably the simplest 
model problem for \eqref{eq:Gw}, namely, the ordinary differential 
equation (ODE) 
\begin{equation}
\frac{d y}{dt}  = y^2, \qquad y(0) = y_0 > 0, 
\label{eq:y2}
\end{equation}
where  $y(t) \in \RR$ with the closed-form solution $y(t) = y_0 /(1 - 
y_0 \, t)$. We see that this solution becomes unbounded, "blows up", as 
the time approaches the blow-up time $t_0$, i.e., $y(t) \rightarrow 
\infty$ as $t \rightarrow t_0 = 1/y_0$, cf.~figure \ref{fig:blowup}a. 
Clearly, the equation is not satisfied at the blow-up time $t_0$ and 
the solution $y(t)$ is not defined for $t \ge t_0$. Although this 
simple example shows that in the ODE setting a quadratic nonlinearity 
leads to singularity formation, the original Euler equation 
\eqref{eq:Gw} is clearly more nuanced: first, while there is no space 
dependence in \eqref{eq:y2}, the presence of the convolution operator 
$\bG * \bomega$ implies the nonlinearity in the Euler 
system is in fact nonlocal (in the sense that the value of the 
vortex-stretching term at a point depends on the vorticity $\bomega$ 
{\em everywhere} in the flow domain); second, the Euler equation 
\eqref{eq:EuVorta} involves vector quantities and the actual rate of 
vorticity amplification $d\bomega/dt$ at a point $\x$ depends on how 
the vorticity vector $\bomega$ is aligned with respect to the 
eigendirections of the tensor $\bnabla \u$ at $\x$ and on the 
corresponding eigenvalues.  We add that in two dimensions (2D, $d = 2$) 
the vorticity vector is always perpendicular to the plane of motion, 
$\bomega = \omega \e_{3}$, where $\e_{j}$, $j=1,2,3$, are the unit 
vectors of the Cartesian coordinate system, such that the vortex 
stretching term in \eqref{eq:EuVorta} vanishes identically.

Our second example concerns the inviscid Burgers system \eqref{eq:B0} 
whose solutions can be expressed in the implicit form as $u(t,x) = 
u_{0}(x - u(t,x)\, t)$ \citep{kl04}. If the initial condition $u_{0}$ 
is smooth and such that there is a point $x \in \Omega$ where $d 
u_{0}(x) / dx < 0$, then this solution develops a shock singularity at 
the time $t_0 = -1 / (\min_x d u_{0}(x)/dx)$ where it ceases to be 
differentiable, in the sense that $\lim_{t \rightarrow t_0} \big| 
\partial_{x} u(t,x_{s}) \big| = \infty$ at the singularity location 
$x_{s}$, cf.~figure \ref{fig:blowup}b. Thus, equation \eqref{eq:B0a} is 
not satisfied in the pointwise sense at $t = t_0$ and  the solution 
is again not defined for $t \ge t_0$. Here, we mean "classical" 
solutions satisfying equation \eqref{eq:B0a} in the pointwise sense for 
{\em all} $(t,x) \in (0,t_0) \times \Omega$. On the other hand, it is 
possible to construct "weak" solutions of system \eqref{eq:B0} defined 
in a certain integral sense using suitable test functions \citep{kl04}. 
They exist for arbitrarily long times $t \ge t_0$  and may involve 
shock-type singularities of the type discussed above. In general, weak 
solutions are not unique.

Above we have seen examples of singularity formation in two very simple 
models which nevertheless share some properties with the Navier-Stokes 
and Euler systems \eqref{eq:NS}--\eqref{eq:Eu}. The Big Question is 
therefore whether such a behavior could also occur in these systems. We 
thus have the following open question
\begin{question}
Given a sufficiently smooth initial condition $\u_0$, do the 
Navier-Stokes and Euler systems \eqref{eq:NS}--\eqref{eq:Eu} always 
admit smooth solutions $\u(t)$ satisfying these systems in the 
classical pointwise sense for arbitrarily long times $t>0$? 
\label{Q1}
\end{question}

Needless to say, solutions which are not smooth are not physically 
meaningful. This is because the emergence of a singularity would imply 
the breakdown of the equation at the blow-up time, invalidating it as a 
model of natural phenomena. Given the significance of this issue, both 
theoretically and in practice, resolution of the first part of Question 
\ref{Q1}, concerning Navier-Stokes flows, has been recognized by the 
Clay Mathematics Institute as one of its seven "Millennium Problems" 
posed as challenges to the mathematics community at the beginning of 
the 20th century \citep{f00}.

\subsubsection{Dissipation Anomaly}
\label{sec:anomaly}
Our second open problem is motivated by what is commonly referred to as 
the "zeroth law of turbulence" \citep{frisch1995turbulence}, which at 
present is an empirical result. We are interested in the time-averaged 
rate of energy dissipation \citep{BuariaPumir2026}
\begin{equation}
\epsilon_{\nu}^{T}(\f) := \left[ \nu \int_0^T \int_\Omega | \bnabla \u(\x,t) |^2 \, d\x\,dt \right]_{T}
\label{eq:eps}
\end{equation}
and the question is then about the long-time limit 
$\epsilon_{\nu}^{\infty} := \lim_{T \rightarrow \infty} 
\epsilon_{\nu}^{T}$ of the normalized rate of energy dissipation $D_{\nu}(\f) 
:= \epsilon_{\nu}^{\infty} / (U^{3} L^{2})$ as the viscosity vanishes $\nu 
\rightarrow 0$ ($\re \rightarrow \infty$), namely, whether this 
quantity approaches a nonzero value. Thus, we have the following
\begin{question}[dissipation anomaly]
Assume the 3D Navier-Stokes system \eqref{eq:NS} admits 
smooth classical solutions for an arbitrarily long time $T$ and for 
different viscosities $\nu > 0$. Are there forcings $\f = \f(t,\x)$ that 
may depend on the viscosity $\nu$ with a uniformly bounded magnitude 
$\left[ \| \f \|_{L^{2}} \right]_{T} \le K$ for some $K > 0$, such that
\begin{equation}
D_{\nu}(\f) \rightarrow C > 0 \qquad \text{as} \qquad \nu \rightarrow 0?
\label{eq:D}	
\end{equation}
\label{Q2}
\end{question}

The question about the dissipation anomaly is also pertinent in the 
context of unforced flows (with $\f \equiv \0$). In such cases, one 
considers a finite-time horizon with $T < \infty$ and the energy 
dissipation rate is a function of the initial condition 
$\epsilon_{\nu}^{T}(\u_{0})$, the latter subject to suitable 
constraints. In addition to its practical consequences, the zeroth law 
of turbulence, if it indeed holds, will have far-reaching theoretical 
implications. More specifically, if the answer to Question \ref{Q2} is 
affirmative, then this implies an unbounded growth of the velocity 
gradients $\bnabla\u$ as the viscosity $\nu$ vanishes and therefore a 
nonzero energy dissipation rate in the limiting inviscid Euler flows, a 
phenomenon referred to as "anomalous dissipation". This eventuality is 
closely related to Onsager's conjecture concerning the possibility of 
energy dissipation in Euler flows that are sufficiently irregular (in 
the H\"older class $C^{\alpha}(\Omega)$ with $\alpha < 1/3$, to be 
precise). There has been a lot of progress as regards Onsager's 
conjecture lately that led to the construction of several families of 
weak solutions to the Euler system exhibiting anomalous dissipation as 
well as nonuniqueness \citep{Eyink2024}.

Proposition \eqref{eq:D} in Question \ref{Q2} is at the heart of 
Kolmogorov's statistical theory of homogeneous isotropic turbulence 
\citep{frisch1995turbulence}, referred to as "K41". As such, it has 
received a lot of attention in the turbulence community  with many 
efforts to test it both experimentally and with numerical computations 
(in experiments, the role of forcing is played by various stirring 
mechanism). The state-of-the-art rigorous bounds on the rate of energy 
dissipation were obtained by \citet{DoeringFoias2002} and allow for 
dissipative anomaly. As discussed by \citet{IyerPRL2025}, most of the 
empirical evidence in support of this possibility comes from external 
flows (such as wakes, jets and grid turbulence). On the other hand, 
\citet{IyerPRL2025} carefully reexamined recent numerical computations 
of forced Navier-Stokes flows on a 3D periodic box and conjectured that 
the normalized energy dissipation rate $D_{\nu}(\f)$ is in fact a 
decreasing function of the Reynolds number $Re$, albeit decaying  
slower than $\O\left(Re^{-1}\right)$, a behavior referred to as a {\em 
weak} dissipation anomaly.

The problem of the dissipative anomaly is well understood in the 
context of unforced 1D Burgers flows governed by systems 
\eqref{eq:B}--\eqref{eq:B0} with $f \equiv 0$, where the counterpart of 
Question \ref{Q2} does have an affirmative answer \citep{Eyink2024}; 
for consistency with the original reference in the remainder of this 
subsection we assume the domain is $\Omega = [-\pi,\pi]$. This can be 
demonstrated by performing explicit calculations based on the exact 
"sawtooth" solution $u_{\nu}(t,x) = (1/t)\left[ x - \pi \tanh(\pi x/(2 
\nu t) )\right]$ of the viscous Burgers equation \eqref{eq:Ba} which in 
the inviscid limit $\nu \rightarrow 0$ becomes
\begin{equation}
u(t,x) = \begin{cases}
\frac{x}{t} + \frac{1}{2} \delta u, \qquad -\pi \le x < 0 \\
\frac{x}{t} - \frac{1}{2} \delta u, \qquad 0 < x \le \pi \\
\end{cases},
\label{eq:u0}
\end{equation}
where $\delta u := 2 \pi / t$ is the jump discontinuity at the origin. 
Noting that $\partial_{x} u_{\nu}(t,x) = 1/t - \pi^{2} /(2 \nu t^{2} ) 
\sech^{2}(\pi x/(2 \nu t) )$, the energy dissipation rate can for 
$\nu \rightarrow 0$ and $t$ fixed be approximated as follows 
\begin{equation}
\epsilon_{\nu}(t) = \frac{\nu}{2} \int_{-\pi}^{\pi} \left[ \partial_{x} u_{\nu}(t,x) \right]^{2}\,dx 
\approx \frac{1}{2\pi} \frac{\pi^{4}}{4 \nu t^{4}} \frac{2 \nu t}{\pi} \frac{4}{3} 
\quad \xrightarrow[\nu \rightarrow 0]{} \quad \frac{(\delta u)^{2}}{12 t},
\label{eq:e1D}
\end{equation}
where we noted that the energy dissipation rate is dominated by large 
values $\O(1/(\nu t^{4}))$ attained by the integrand expression in a 
small region of size $\O(\nu t / L)$ around the shock discontinuity. 
Clearly, the energy dissipation rate converges to a well defined finite 
quantity as $\nu \rightarrow 0$. In fact, the same expression 
is also obtained computing the rate of change of the kinetic energy of 
the inviscid solution given in \eqref{eq:u0}: $(d/dt) \int_{-\pi}^{\pi} 
u^{2}(t,x) \, dx = - (\delta u)^{2} / (12 t)$. In addition to Question 
\ref{Q2} motivated by the zeroth law of turbulence in the 
real-life 3D setting, it is also interesting to know if some form of 
dissipation anomaly can happen in the more idealized setting of 2D Navier-Stokes 
flows.

\subsection{Why Bounds?}
\label{sec:bounds}

Our approach to studying Questions \ref{Q1} and \ref{Q2} relies on a 
priori bounds on the behavior of the relevant quantities. These are 
inequalities typically obtained from the governing equations using 
techniques of harmonic and functional analysis collectively referred to 
as "energy methods". Since blow-up in the solution of a PDE is usually 
signaled by an unbounded growth of the solution or its derivatives, in 
the study of extreme behavior we are usually interested in obtaining 
bounds on the temporal growth of different norms of the solution 
characterizing its regularity in terms of some norms of the initial 
data $\u_{0}$; in the study of the dissipation anomaly one is 
interested in bounding the time-averaged energy dissipation rate 
$\epsilon_{\nu}^{T}(\f)$ in terms of the viscosity $\nu$ or, 
equivalently, the Reynolds number $Re$. Such bounds are rigorously 
derived inequalities expressing fundamental limitations on the behavior 
of {\em all} possible solutions of a given problem. They therefore 
offer a more general way to study the problems of interest here than 
equalities that can normally describe the behavior of individual 
solutions only \citep[footnote in the Preface]{dg95}. It should be 
emphasized that the PDE problems introduced above admit only very few 
exact solutions that can be expressed in a closed form (and those that 
can are usually uninteresting from the point of view of extreme 
behavior).

In this essay we describe a research program that aims to provide 
insights about the open questions stated above by identifying forms of 
the most extreme behavior possible in fluid flows. In fact, it has 
already led to resolution to certain simpler versions of Questions 
\ref{Q1} and \ref{Q2}. This approach consists of the following three 
main steps:
\begin{itemize}
\item[(S1)] first, one needs to obtain a priori bounds for a given 
problem which is typically done using energy methods; however, there 
can be many such bounds and we are interested in identifying the "best" 
ones which offer the "sharpest", i.e., the least conservative, estimate 
of the quantity of interest,

\item[(S2)] second, once the best bounds are identified, one is 
interested in verifying their sharpness; a bound is deemed "sharp" if 
there are solutions with behavior saturating this bound; such "extreme" 
solutions can be sought systematically by formulating and solving 
variational PDE optimization problems where the quantities the bounds 
apply to are maximized subject to certain constraints; if solutions 
saturating the bound are found, this means the bound is sharp and hence 
cannot be fundamentally improved  (i.e., except for, perhaps, numerical 
prefactors, etc.); otherwise, this signals the bound may not be sharp 
leaving room for its improvement,

\item[(S3)] finally, if flows are found that saturate the relevant bounds, 
they can be analyzed to reveal the key physical mechanisms 
realizing this extreme behavior.
\end{itemize}
\noindent We remark that optimization techniques have been used to 
solve a variety of mostly applied problems in fluid mechanics 
involving, e.g., drag reduction and lift enhancement, data 
assimilation, etc. Here we leverage these techniques for an 
entirely different purpose, namely, to give us a glimpse into the inner 
workings of the equations describing fluid flows.

The structure of this essay is as follows. Following a brief 
introduction to the notation, our first objective in 
\S\,\ref{sec:twoproblems} is to survey two model problems, one related 
to the extreme growth and the other to the dissipation anomaly, where 
the research program outlined above has been brought to fruition 
effectively providing resolutions of these problems. Then, in 
\S\,\ref{sec:search}, we discuss the recent progress with the related 
problem of systematic search for extreme, possibly singular, behavior 
in 3D Navier-Stokes and Euler flows. Next, in \S\,\ref{sec:misc}, we 
present a forward-looking miscellany of open problems related to 
extreme growth and different manifestations of dissipation anomaly 
amenable to investigation within the proposed framework. This is 
followed by a discussion of possible technical improvements in 
\S\,\ref{sec:improve}, whereas some final comments are deferred to 
\S\,\ref{sec:final}. Finally, to make his essay self-contained, some 
more technical material is collected in a number of appendices.

\section{Notation}
\label{sec:notation}

In this section we introduce the notation we will use in this essay. 
The key concept is that of a function space as it allows one to treat 
infinite-dimensional objects such a functions of space and/or time in a 
manner analogous to vectors in  $\RR^{d}$, although there are also 
important differences. A function is said to belong to a certain space 
if its "size" measured using the norm in this space is finite. To fix 
attention, we will consider here space-dependent vector fields with 
time dependence amenable to an analogous treatment. 

The first family of function spaces we will need are the 
Lebesgue spaces $L^q(\Omega)$, $q \in [1, \infty]$. A vector field 
$\u \; : \; \Omega \; \rightarrow \RR^{d}$ belongs to 
$L^{q}(\Omega)$ if and only if $\| \u \|_{L^{q}} < \infty$, where
\begin{equation}
\| \u \|_{L^{q}} := \begin{cases}
\left( \int_{\Omega} |\u^{q}(\x)|\,d\x \right)^{1/q}, \qquad & 1 \le q < \infty \\
{\esssup_{\x \in \Omega}} |\u(\x)|, \qquad & q = \infty 
\end{cases}
\label{eq:Lq}
\end{equation}
with $|\cdot|$ representing the Euclidean norm. For time-dependent 
vector fields $\u = \u(t,x)$, $L^{p}([0,T];L^{q}(\Omega))$ will denote 
the space of functions such that $\int_{0}^{T} \| \u(t) \|_{L^{q}}^{p} \, 
dt < \infty$. In the special case when $q = 2$, the space 
$L^{2}(\Omega)$ is endowed with the Hilbert structure encoded in the 
inner product $\langle \z_{1}, \z_{2} \rangle_{L^{2}} := 
\int_{{\Omega}} \z_{1} \cdot \z_{2}\, d\x$, with $"\cdot"$ denoting the 
dot product of vectors in $\RR^{d}$, such that we have $\| \u 
\|_{L^{2}} = \sqrt{\langle \u,\u \rangle_{L^{2}}}$. This allows us to 
define the kinetic energy as 
\begin{equation}
\K(\u) := \frac{1}{2} \| \u \|_{L^{2}}.
\label{eq:K}
\end{equation}
The Lebesgue norms characterize the magnitude of functions, namely, 
how "big" they can locally grow (or how slowly they decay at large 
distances from the origin on unbounded domains); as such, they are 
insensitive to the regularity of the function describing how rapidly it 
varies. The latter property is quantified by their norms in Sobolev 
spaces $H^{s}(\Omega)$, $s \in \RR$, which can be defined in terms of 
the Parceval identity as \citep{af05} 
\begin{equation} \| \u \|_{H^{s}} := 
\sqrt{\sum_{\k \in \ZZ^{d}} \left[ 1 + (2 \pi k)^{s} \right]^{2} \left| \hu_{\k} \right|^{2}}, 
\label{eq:Hs} 
\end{equation} 
where $\hu_{\k} := \int_{\Omega} \u(\x) e^{-2 \pi i \k\cdot \x} \, 
d\x$, with $i := \sqrt{-1}$ the imaginary  unit, are the Fourier 
coefficients corresponding to  the wavenumber $\k = [k_{1}, \dots, 
k_{d}]^{T} \in \ZZ^{d}$  and $k := |\k|$. Thus, a vector field $\u$ belongs 
to the Sobolev space $H^{s}(\Omega)$ if and only if $\| \u \|_{H^{s}} < 
\infty$, where $s$ is the number of its weak (distributional) 
derivatives that are square integrable. Here, we focus on the $L^{2}$-based 
Sobolev spaces endowed with the Hilbert structure with the inner 
product $\langle\cdot,\cdot\rangle_{H^{s}}$ such that $\forall \u \in 
H^{s}(\Omega)$ $\| \u \|_{H^{s}} = \sqrt{\langle\u,\u\rangle_{H^{s}}}$; 
clearly,  $H^{0}(\Omega) = L^{2}(\Omega)$. We will also refer to seminorms 
involving derivatives of the highest degree only and defined as 
$\| \u \|_{\dot{H}^{s}} :=  \sqrt{\sum_{\k \in \ZZ^{d}} (2 \pi k)^{2s}  
\left| \hu_{\k} \right|^{2}}$. We note that as long as the infinite
series on the RHS in \eqref{eq:Hs} converges, this definition is also 
valid for both noninteger and negative values of $s$. Definition 
\eqref{eq:Hs} reveals a key relationship between the level of Sobolev 
regularity of the vector field $\u$ and the rate of decay of its 
Fourier coefficients as $k \rightarrow \infty$. More specifically, when 
$\u \in H^{s}(\Omega)$, then the Fourier coefficients of $\u$ must 
decay no slower than $| \hu_{\k} | = \O(k^{-s-1/2-\varepsilon})$ for 
some $\varepsilon > 0$ as $k \rightarrow \infty$. Moreover, if $\u$ is 
real-analytic, then $| \hu_{\k} | = \O(e^{-\delta k})$ for some 
$\delta> 0$ referred to as the width of the analyticity strip. It 
characterizes the distance from the real axis to the nearest 
singularity in the extension of the vector field $\u$ to the 
complex domain $\CC^{d}$. The spaces $H^{s}(\Omega)$ are 
a special case of a more general family of Sobolev spaces defined for 
$k \in \NN$ and $q \in [1,\infty]$ as \citep{af05} 
\begin{equation}
W^{k,q}(\Omega) := \left\{ \u \in L^{q}(\Omega) \ : \ \bnabla^{\alpha} \u \in L^{q}(\Omega), \ \alpha \le k \right\}
\label{eq:Wkq}
\end{equation}
such that we have $H^{k}(\Omega) = W^{k,2}(\Omega)$ and $L^{q}(\Omega) 
= W^{0,q}(\Omega)$. For $q \neq 2$, the spaces $W^{k,q}(\Omega)$ are 
not endowed with the Hilbert structure.

Different Sobolev spaces are related by embedding theorems asserting 
that if $\u \in W^{k,q}(\Omega)$, then also $\u \in W^{l,p}(\Omega)$, 
which is denoted $W^{k,q}(\Omega) \hookrightarrow W^{l,p}(\Omega)$, 
provided $d, k, l, q, p$ satisfy certain conditions \citep{af05}. In 
this essay, we will refer to a few particular results of this type.

In addition to kinetic energy \eqref{eq:K}, another important quantity 
is the enstrophy\footnote{{We note that
	unlike energy, cf.~\eqref{eq:K}, enstrophy is often defined without 
	the factor of 1/2. However, for consistency with earlier studies 
	belonging to this research program 
	\citep{ap11a,ap16,KangYunProtas2020,KangProtas2021,RamirezProtas2026}, 
	we choose to retain this factor here.}}
\begin{equation}
\E(\u) := \frac{1}{2}\| \bomega \|_{L^{2}} = \frac{1}{2}\| \bnabla \u \|_{L^{2}}
\label{eq:E}
\end{equation}
where the second equality holds for divergence-free vector fields on 
periodic and unbounded domains \citep{dg95}. When there is no risk of 
confusion, we will use the simplified notation $\K(t) := \K(\u(t))$, 
$\E(t) := \E(\u(t))$ and $\K_{0} = \K(0)$, $\E_{0} = \E(0)$. For smooth 
Navier-Stokes flows, the energy and enstrophy satisfy the energy 
equation
\begin{equation}
\frac{d\K(t)}{dt} = - 2 \nu \E(t)
\label{eq:dKdt}
\end{equation}
obtained by dotting \eqref{eq:NSa} with $\u(t)$, integrating over the 
domain $\Omega$ and then performing integration by parts where we note 
that the cubic term disappears, $\int_{\Omega} \u \cdot 
(\u\cdot\bnabla)\u \, d\x = 0$, indicating that advection does not 
contribute to the global energy balance. When studying Burgers flows 
governed by \eqref{eq:B}--\eqref{eq:B0} we will consider the 1D 
counterparts of the energy \eqref{eq:K} and enstrophy \eqref{eq:E} 
given in terms of obvious definitions.

\section{Two Success Stories: Model Problems with Sharp Bounds}
\label{sec:twoproblems}

In this section we survey two problems where the research program 
described above has been brought to fruition, in the sense that 
rigorous bounds characterizing the behavior of certain relevant 
quantities have been demonstrated to be sharp. By analyzing the 
behavior of the solutions saturating these estimates, it is possible to 
identify the key physical mechanisms underlying the most extreme 
behavior allowed by these models. The first problem, discussed in 
\S\,\ref{sec:Burgers}, concerns the largest possible growth of 
enstrophy in solutions of the 1D unforced ($f \equiv 0$) Burgers system 
\eqref{eq:B} with initial data $u_{0}$ with a fixed enstrophy $\E_{0} = 
\E(u_{0}) > 0$. Even though the Burgers system  is known to be globally 
well posed and hence the enstrophy $\E(u(t))$ of its solutions is 
bounded for all times $0 \le t \le \infty$ \citep{kl04}, obtaining 
sharp a priori bounds on this quantity is quite relevant since some of 
the techniques used to derive these estimates are also employed to 
obtain similar estimates in the study of Question \ref{Q1}. The second 
problem, discussed in \S\,\ref{sec:noanomaly}, is related to Question 
\ref{Q2}, but formulated in the context of unforced 2D Navier-Stokes 
flows governed by system \eqref{eq:NS} with $d = 2$ and $\f \equiv \0$. 
Such flows feature a direct enstrophy cascade combined with an inverse 
energy cascade and the main quantity of interest is the rate of 
enstrophy dissipation (rather than the rate of energy dissipation which 
is the relevant quantity in 1D and 3D flows). While it is known that in 
such a setting the dissipation anomaly is not possible, we derive a 
bound on how this quantity must vanish in the inviscid limit $\nu 
\rightarrow 0$. We then construct families of flows saturating this 
bound, demonstrating that it is in fact sharp. The extreme flows found 
in this way make it possible to identify the dominating physical 
mechanisms for enstrophy dissipation in 2D unforced flows.

\subsection{On the Maximum Growth of Enstrophy in 1D Viscous Burgers Flows}
\label{sec:Burgers}

As will be discussed in detail in \S\,\ref{sec:search} below, the 
question about the possibility of singularity formation in 3D 
Navier-Stokes flows can be recast in terms of whether starting from a 
smooth initial condition $\u_{0}$ with a finite enstrophy $\E(\u_{0}) > 
0$, the enstrophy of the corresponding flow $\E(\u(t))$ can become 
unbounded in finite time \citep{dg95}. It turns out that this problem 
has an interesting counterpart in 1D in the context of viscous Burgers 
flows governed by \eqref{eq:B}. We know that this system admits 
globally smooth solutions with enstrophy bounded for all times, 
$\E(u(t)) < \infty$, $t \ge 0$ \citep{kl04}. However, it is a pertinent 
question how much the enstrophy can grow at most in a viscous Burgers 
flow if the initial data $u_{0}$ has enstrophy $\E_{0}$, where we are 
interested in the limit $\E_{0} \rightarrow \infty$. In other words, we 
seek an upper bound on $\max_{t > 0} \E(t)$ in terms of $\E_{0}$. 

A natural way to study this problem is to first consider bounds on the 
rate of growth $d\E/dt$ of the enstrophy and this question was first 
taken up by \citet{ld08}. Multiplying equation \eqref{eq:Ba} by 
$\partial_{xx} u$, integrating over $\Omega = \TT$, performing 
integration by parts with respect to $x$ and using periodicity, one 
obtains
\begin{align}
r(u) := \frac{d\E(u(t))}{dt} & = -\nu \| \partial_{xx} u \|_{L^2}^2 
- \int_0^1 u \, \partial_x u \, \partial_{xx} u \, dx \label{eq:r} \\
& = -\nu \| \partial_{xx} u \|_{L^2}^2 
+ \frac{1}{2} \int_0^1 \left(\partial_x u \right)^3 \, dx. \nonumber
\end{align}
The goal is now to bound this quantity in terms of the enstrophy $\E(u)$ 
itself and since similar estimates will appear later in this essay, we 
present this derivation here in some detail. To estimate the cubic term we 
use the 1D Gagliardo-Nirenberg, Poincar{\'e} and Cauchy-Schwarz 
inequalities \eqref{eq:CS}--\eqref{eq:GN}
\begin{equation}
\left| \int_0^1 u \, \partial_x u \, \partial_{xx} u \, dx \right| 
 \le \| u \|_{L^{\infty}} \left| \int_0^1 \partial_x u \, \partial_{xx} u \, dx \right| 
 \le \frac{2}{\pi} \| \partial_{xx} u \|_{L^2}^{1/2} \, \| \partial_{x} u \|_{L^2}^{5/2}.
\label{eq:cubic1D}
\end{equation}
Using this in \eqref{eq:r}, we obtain
\begin{align}
\frac{d\E(u(t))}{dt} & \le -\nu \| \partial_{xx} u \|_{L^2}^2  + \frac{C}{2} \| \partial_{xx} u \|_{L^2}^{1/2} \, \| \partial_{x} u \|_{L^2}^{5/2} \nonumber \\
& \le -\nu \| \partial_{xx} u \|_{L^2}^2  + \frac{C}{2} \left[ \frac{\beta^{4}}{4} \| \partial_{xx} u \|_{L^2}^{2} 
+  \frac{3}{4 \beta^{4/3}} \| \partial_{x} u \|_{L^2}^{10/3}\right],
\label{eq:dEdt1Da}
\end{align}
where Young's inequality \eqref{eq:Y} with $m = 4$ and $n = 4/3$ was 
used to split the product of $\| \partial_{xx} u \|_{L^2}^{1/2}$ and 
$\| \partial_{x} u \|_{L^2}^{5/2}$ in such a way that the former term 
appears squared. Finally, making the judicious choice $\beta = (8 \nu / 
C)^{1/4}$ allows us to cancel the two terms involving $\| \partial_{xx} 
u \|_{L^2}^{2}$, leading to
\begin{equation}
\frac{d\E}{dt} \leq 3 \left(\frac{1}{2 \pi^2\nu}\right)^{1/3}\E^{5/3}.
\label{eq:dEdt1D}
\end{equation}

Considering the borderline case and replacing the differential 
inequality \eqref{eq:dEdt1D} with the corresponding equation, we obtain 
an ODE of the  form $dy/dt = y^{\alpha}$, $y(0) > 0$, with $\alpha = 
5/3$. By analogy with \eqref{eq:y2}, since the exponent $\alpha > 1$, 
one may expect that its solution can become unbounded in a finite time, 
which would contradict the global well-posedness of the 
Burgers system \eqref{eq:B}. To see that this cannot, in fact, happen, 
we consider the energy equation \eqref{eq:dKdt} which has the same form
for 1D Burgers flows. Upon integration in time, it yields the bound
\begin{equation}
\K(t) - \K_{0} = - \nu \int_{0}^{t} \E(\tau)\, d\tau \qquad \Longrightarrow \qquad \int_{0}^{t} \E(\tau)\, d\tau \le \frac{1}{\nu} \K_{0}
\label{eq:Kt}
\end{equation}
which in combination with Gr{\"o}nwall's lemma \eqref{eq:Gro} applied to 
the ODE factorized as  $dy/dt = y^{\alpha - 1} y$ allows us to bound the 
enstrophy as 
\begin{equation}
\max_{t \ge 0} \E(t) \le \E_{0} \exp\left( \K_{0} / \nu \right).
\label{eq:maxEtG}
\end{equation}
In fact, an estimate of this type can be obtained as long as the 
exponent in the model problem is $\alpha \le 2$.

We now move on to consider the question of the sharpness of estimate 
\eqref{eq:dEdt1D}. An estimate such as this is said to be "sharp" if 
there exists a family of periodic functions $\tuuE \in H^{2}(\TT)$ 
parametrized by $\E_{0} = \E(\tuuE)$ which saturate the upper bound, 
i.e., for which $d\E(\tuuE)/dt = \O\left(\E_{0}^{5/3}\right)$. When a 
polynomial bound is found to be sharp with respect to the exponent, as 
defined here, then one can inquire whether it is also sharp with 
respect to the prefactor. To obtain estimate \eqref{eq:dEdt1D} we had 
to employ a number of inequalities, cf.~\eqref{eq:CS}--\eqref{eq:GN}, 
and while each of these inequalities is known to be sharp, they may be 
saturated by different functions. Therefore, when these inequalities 
are chained together as when obtaining \eqref{eq:dEdt1D}, the resulting 
bound may or may not be sharp. In what to the best of our knowledge was 
a first application of such an approach, \citet{ld08} probed this 
possibility by formulating the following optimization problem
\begin{problem}
\label{pb:maxdEdt1D}
  Given $\E_0\in\RR_+$ and the objective functional \eqref{eq:r}, find
\begin{equation*}
\tuuE  =  \mathop{\arg\max}_{u \in \Sigma_{\E_0}} \, r(u), \qquad \text{where} \qquad
\Sigma_{\E_0}  :=  \left\{u\in H^2(\Omega)\,\colon \; \int_0^1 u \, dx = 0, \ \E(u) = \E_0 \right\}.
\end{equation*} 
\end{problem}
\noindent 
Remarkably, \citet{ld08} were able to solve this problem in a closed 
form using the method of Lagrange multipliers. By analyzing the asymptotic 
behavior of these solutions for large enstrophies, they concluded that
\begin{equation}
r(\tuuE) \sim \frac{0.393}{\nu^{1/3}} \E_0^{5/3} \quad \text{as} \quad \E_0 \rightarrow \infty,
\label{eq:r1d}
\end{equation}
thus demonstrating that estimate \eqref{eq:dEdt1D} is sharp (up
to a numerical prefactor which is larger than in \eqref{eq:r} by
about 2.83). In other words, for each value of $\E_0$, the optimal
fields $\tuuE$, which have the form of steep waves with fronts
becoming sharper as $\E_0$ increases, instantaneously produce as much
enstrophy $r(\tuuE)$ as is only allowed by upper bound \eqref{eq:dEdt1D}. 

\begin{figure}[t] 
\begin{center} 
\mbox{
\subfigure[]{\includegraphics[width=0.45\textwidth]{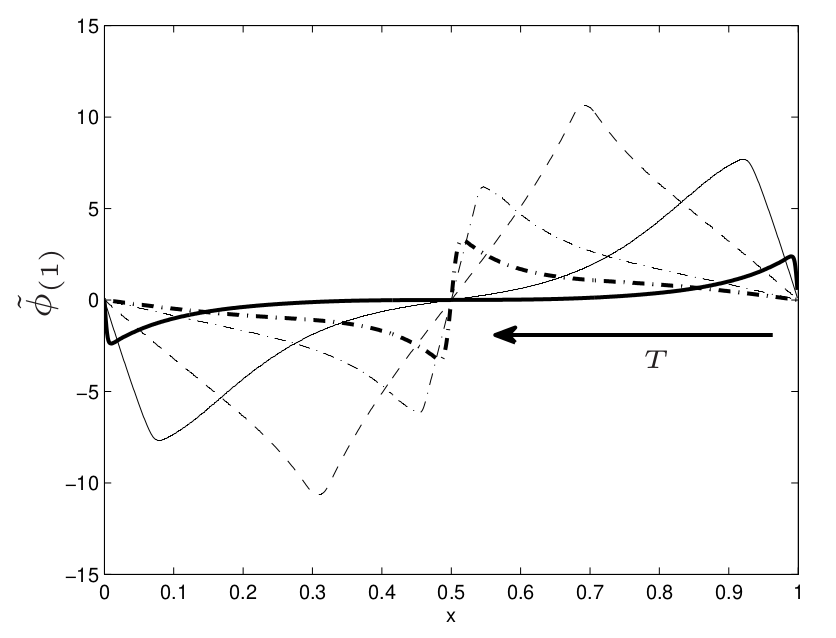}}\qquad 
\subfigure[]{\includegraphics[width=0.55\textwidth]{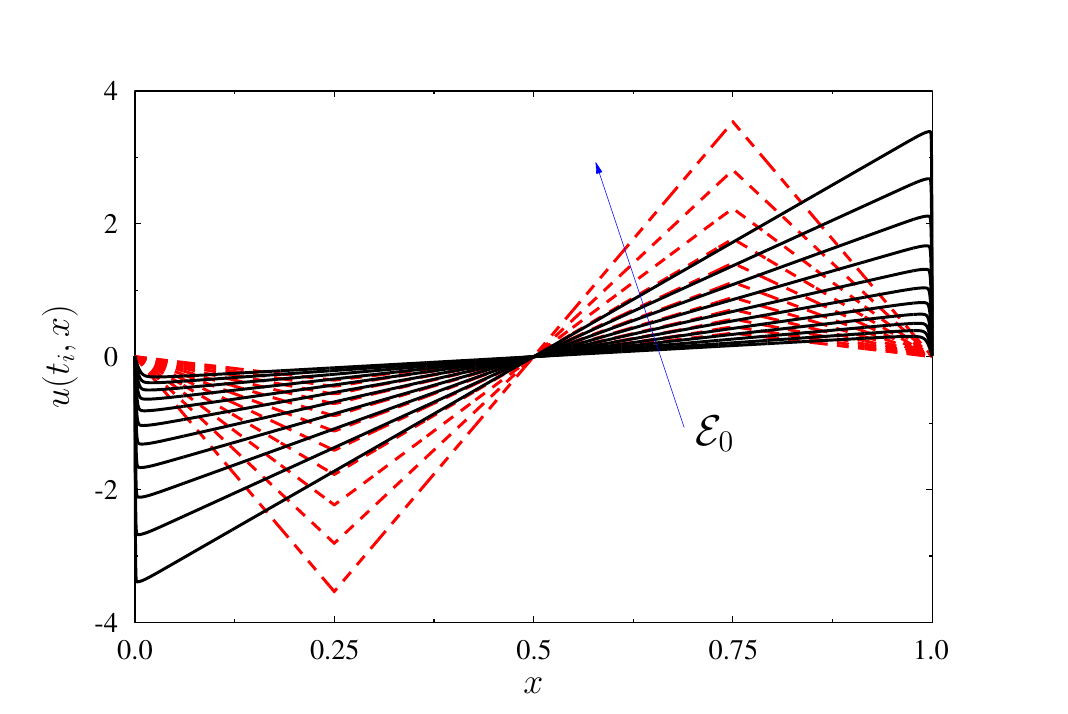}}} 
\mbox{ 
\subfigure[]{\includegraphics[width=0.55\textwidth]{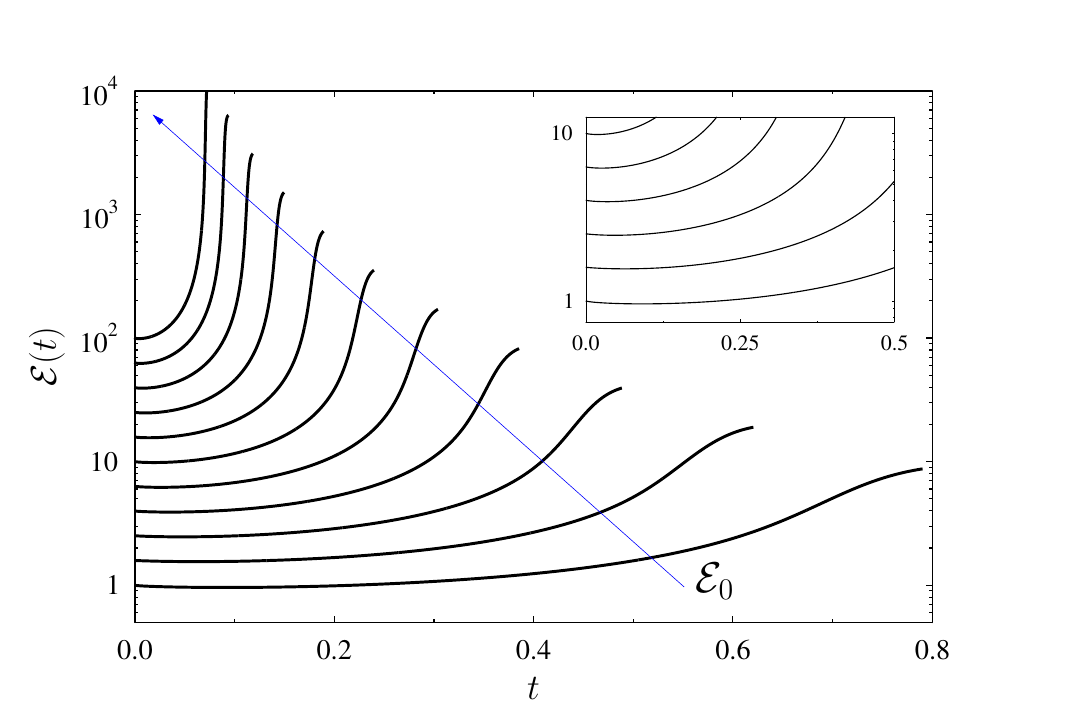}}\quad 
\subfigure[]{\includegraphics[width=0.45\textwidth]{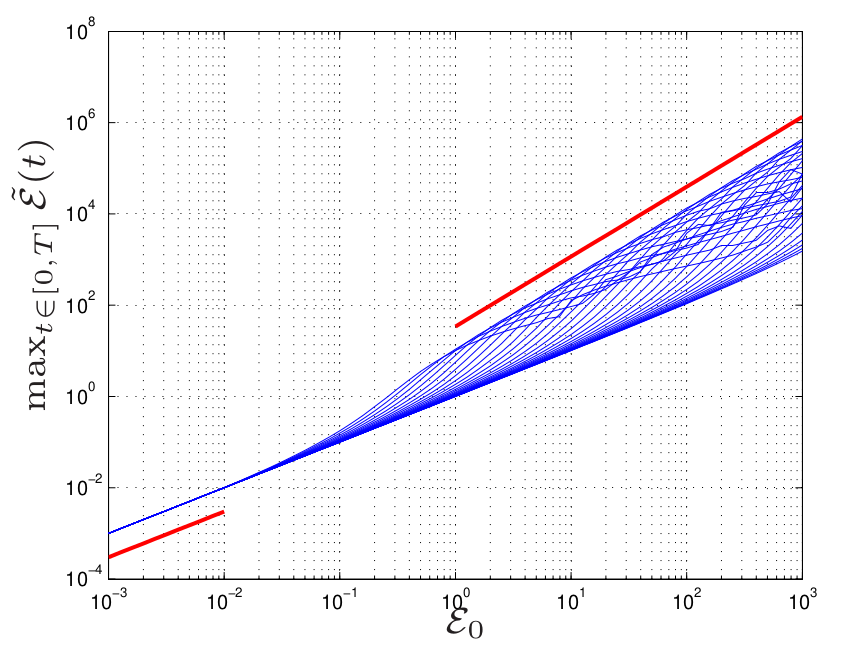}}} 
\caption{(a) Optimal initial conditions $\tuuET$ obtained by solving
  Problem \ref{pb:maxET1D} with fixed enstrophy $\E_0 = 10^3$ and 
  different time intervals: (thick solid line) $T=10^{-3}$, (thin solid 
  line) $T=10^{-2}$, (thin dashed line) $T=10^{-1.5}$, (thin dotted 
  line) $T=10^{-1}$ and (thick dotted line) $T=10^{0}$; the arrow 
  indicates the trend with increasing $T$.  (b) (red dashed lines) 
  optimal initial conditions $\tuuETm$ and (black solid lines) the 
  corresponding flow states $u\left(\Tmax^{\E_{0}};\tuuETm \right)$ at the 
  maximum time $\Tmax^{\E_{0}}$ for increasing values of $\E_{0}$ and 
  (c) the time evolutions of the enstrophy 
  $\E\left(u\left(t;\tuuETm\right)\right)$ in the corresponding flows 
  (the inset is a magnification of the region with small $\E_{0}$); the 
  arrows indicate the trends with the increase of $\E_{0}$. (d) Maximum 
  enstrophy $\max_{t\in[0,T]} \E_T\left(\tuuET\right)$ as a function of initial 
  enstrophy $\E_0$ for different $T$; two distinct power laws can be 
  observed with exponents 1 for small $\E_0$ and 3/2 for large $\E_0$, 
  cf.~\eqref{eq:maxETap}.}
\label{fig:maxET1D}
\end{center}
\end{figure}

Now that the instantaneous bound on the rate of growth of enstrophy 
has been shown to be sharp (up to a prefactor), the natural question is 
about obtaining bounds on the maximum growth of enstrophy in finite 
time (since system \eqref{eq:B} is globally well posed, we know these 
bounds must be finite). The simplest way to do this, which assumes that 
the flow evolution $u(t)$ saturates the instantaneous bound 
\eqref{eq:dEdt1D} at every instant of time $t$, is to directly integrate 
this inequality with respect to time which gives
\begin{equation}
\max_{t \in [0,T]} \E(u(t)) \leq \left[\E_0^{1/3} + \frac{1}{4}\left(\frac{1}{2\pi^2 \nu}\right)^{4/3}\E_0\right]^{3} \quad \underset {\E_0 \rightarrow \infty} {\longrightarrow} \quad
\frac{1}{64} \left(\frac{1}{2 \pi^2 \nu}\right)^{4} \E_0^{3}.
\label{eq:maxEt1D}
\end{equation}
Clearly, this bound predicts that $\max_{t \ge 0} \E(t) = 
\O\left(\E_{0}^{3}\right)$. On the other hand, solving the Burgers 
system with the maximizers of Problem \ref{pb:maxdEdt1D} used as the 
initial data $u_{0} = \tuuE$ produces maximum enstrophy which scales 
only as $\O(\E_0)$ for large $\E_0$, far below what is allowed by 
estimate \eqref{eq:maxEt1D}. There is thus a significant gap between 
the growth of enstrophy allowed by this estimate  and what is obtained 
in this case. In order to close this gap, \citet{ap11a} considered the 
following family of variational optimization problem involving flow 
evolutions over finite times 
\begin{problem}
\label{pb:maxET1D}
  Given $\E_0, T\in\RR_+$ and the objective functional $\E_T(u_0) := \E(u(T;u_{0}))$,
find 
\begin{equation*} 
\tuuET  =  \mathop{\arg\max}_{u_0 \in \Xi_{\E_0}} \, \E_T(u_0), \qquad \text{where} \qquad \Xi_{\E_0}  :=  
\left\{u_0\in H^1(\Omega)\,\colon \; \int_0^1 u_0 \, dx = 0, \ \E(u_0) 
= \E_0 \right\}. 
\end{equation*} 
\end{problem} 
\noindent 
The idea behind this problem is to find optimal initial data $\tuuET$ 
with prescribed enstrophy $\E_0$ that at the given time $T$ produces 
the largest enstrophy $\E_T(\tuuET)$. We emphasize that in involving 
the flow evolution on $[0,T]$, Problem \ref{pb:maxET1D} is 
fundamentally different, and arguably harder to solve, than Problem 
\ref{pb:maxdEdt1D} where the instantaneous only amplification of 
enstrophy is considered. Hence, unlike Problem \ref{pb:maxdEdt1D}, 
Problem \ref{pb:maxET1D} needs to be solved numerically and a 
state-of-the-art adjoint-based Riemannian gradient approach developed 
to solve this class of problems is presented in Appendix 
\ref{sec:solution}. Solutions of this problem obtained by \citet{ap11a} 
for $\nu = 10^{-3}$ and a broad range of values of $\E_0$ and $T$  are 
summarized in figure \ref{fig:maxET1D}. As is evident from figure 
\ref{fig:maxET1D}a, the optimal initial data $\tuuET$ obtained for a 
fixed enstrophy $\E_0$ and a short time window $T$ features a steep 
front and hence resembles the instantaneous maximizers $\tuuE$ found by 
\cite{ld08} by solving Problem \ref{pb:maxdEdt1D}; however, as $T$ 
increases, it gradually turns into a rarefaction wave. For each value of 
$\E_{0}$ one can define the maximal time
\begin{equation}
\Tmax^{\E_{0}} := \mathop{\arg\max}_{T>0} \E_{T}(\tuuET)
\label{eq:Tmax}
\end{equation}
which gives the time scale over which solutions of Problem 
\ref{pb:maxET1D} with fixed values of $\E_{0}$ produce the largest 
value of the enstrophy. The initial conditions $\tuuETm$ and the 
corresponding final states $u\left(\Tmax^{\E_{0}};\tuuETm\right)$ 
obtained by solving Problem \ref{pb:maxET1D} for different values of 
$\E_{0}$ and $T = \Tmax^{\E_{0}}$ are shown in figure 
\ref{fig:maxET1D}b, with the resulting time evolutions of the enstrophy 
presented in figure \ref{fig:maxET1D}c. As is evident from these plots, 
the extreme Burgers flows resulting from the optimal initial conditions 
$\tuuETm$ obtained by solving Problem \ref{pb:maxET1D} for different 
increasing values of $\E_{0}$ over the corresponding maximal times 
$\Tmax^{\E_{0}}$ appear to exhibit a self-similar structure which 
merits further investigation. In figure \ref{fig:maxET1D}d we plot 
$\max_{t \in [0,T]} \E\left(u(t;\tuuET)\right)$ as a function of 
$\E_{0}$ for different $T$ and by maximizing these quantities (i.e., 
computing an upper envelope) with respect to $T$ at fixed values of 
$\E_0$ we obtain the relation 
\begin{equation} 
\max_T \E_T\left(\tuuET\right)  \sim 11.488 \, \E_0^{1.531} \quad \text{as} \quad \E_0 \rightarrow \infty
\label{eq:maxETap} 
\end{equation} 
with the exponent of $\E_0$ lower, roughly by a factor of 2, than the 
exponent 3 in the finite-time estimate \eqref{eq:maxEt1D}.  This 
indicates that this estimate may not be sharp and could possibly be 
improved by lowering the exponent of $\E_0$. Moreover, by analyzing the 
data shown in figure \ref{fig:maxET1D}c we also discover that 
$\Tmax^{\E_{0}} = \O\left(\E_{0}^{-1/2}\right)$ \citep{ap11a,p12}

However, since Problem \ref{pb:maxET1D} is nonconvex and the numerical 
approach employed to solve it relies on local optimality conditions, 
cf.~Appendix \ref{sec:solution}, we of course cannot guarantee that the 
solutions found for any $\E_0$ and $T$, cf.~figure \ref{fig:maxET1D}, 
are global maximizers. However, the results reported by \cite{ap11a} 
were obtained following a thorough search involving the use of many 
different, mutually orthogonal (in the function space $H^1(\Omega)$), 
and random initial guesses $u^{(0)}$. The optimal initial conditions 
shown in figure \ref{fig:maxET1D}a,b are in fact nonunique maximizers 
as their rescaled copies $(1/m) \tuuET(m x)$, $x\in [0,1]$, 
$m=2,3,\dots$, were also found to be local maximizers, but 
characterized by smaller values of $\E_T$. Further support for the 
conjecture that the maximizers shown in figures \ref{fig:maxET1D}a,b 
are in fact global was provided by \citet{FantuzziGoluskin2020} who 
obtained upper bounds on $\max_{t \ge 0} \E(t)$ revealing behavior 
consistent with \eqref{eq:maxETap} using an entirely different 
approach, cf.~\S\,\ref{sec:SoS}.

The results reported by \citet{ap11a}, in particular relation 
\eqref{eq:maxETap}, inspired research aiming to improve, or sharpen, 
bound \eqref{eq:maxEt1D} by lowering its exponent. In this context we 
also mention earlier work by \citet{Biryuk2001} which implies a 
finite-time bound with a smaller exponent, namely, $\max_{t \ge 0} 
\E(u(t)) \le C_B \E_0^{3/2}$, i.e., as observed in \eqref{eq:maxETap}. 
While this approach did not rely on time integration of an 
instantaneous bound such as \eqref{eq:dEdt1D}, the prefactor in this 
estimate $C_B = C_B(\| u_0 \|_{H^2})$ requires the $H^2$ norm of the 
initial data $u_0$ to be bounded. Consequently, owing to Poincar\'e's 
inequality \eqref{eq:Poincare}, this prefactor will not remain bounded 
in the limit we are interested in, i.e., as $\E_0 \rightarrow \infty$. 
Properties of extreme Burgers flows  were also analyzed by \citet{p12}. 
In particular, subject to the additional assumption that the initial 
condition be given in terms of an odd $C^3$ function, an 
$\O(\E_0^{3/2})$ estimate was established on the maximum growth of 
enstrophy $\max_t \E(t)$, cf.~\eqref{eq:maxETap}. These results provide 
a rigorous and quantitative justification for the behavior of Burgers 
flows with initial data $\tuuET$ obtained as local maximizers of 
Problem \ref{pb:maxET1D}. The problem of improving bound 
\eqref{eq:maxEt1D} was debated at a number of scientific events, 
including the thematic program {\em Mathematical aspects of turbulence: 
where do we stand?} held at the Newton Institute in Cambridge, UK, 
during January--June 2022 and the workshop {\em Criticality and 
stochasticity in quasilinear fluid systems} which took place at the 
American Institute of Mathematics in San Jose, CA, in May 2022. A 
breakthrough came only recently when \citet{AlbrittonDeNitti2023} 
finally closed the gap by reducing the exponent in the upper bound in 
\eqref{eq:maxEt1D} down to 3/2, thus producing a sharp estimate 
matching the properties of the solutions of Problem \ref{pb:maxET1D} 
obtained by \citet{ap11a}.

Therefore, the problem is closed from the mathematical point of view: 
we have a rigorous upper bound which we know to be sharp and we also 
know the solutions of the Burgers system \eqref{eq:B} that saturate 
this bound. This successfully concludes steps S1--S2 stated in 
\S\,\ref{sec:bounds} and leaves us with step S3 where we want to 
identify the key physical mechanisms responsible for the extreme 
behavior. To address this point, we return to figure \ref{fig:maxET1D}c 
where we notice (see the inset) that, perhaps somewhat surprisingly, 
the enstrophy $\E\left(u(t;\tuuETm)\right)$ in the extreme flows is not 
a monotonously increasing function of time. In fact, in all cases shown 
in this figure, it first decreases before eventually beginning to 
rapidly grow. This behavior can be understood by referring to the 
energy equation \eqref{eq:dKdt} which shows that the rate of energy 
dissipation is proportional to the instantaneous enstrophy, such that 
large values of the enstrophy $\E(t)$ will result in a quick depletion 
of the energy $\K(t)$. Thus, when in Problem \ref{pb:maxET1D} the goal 
is to maximize the enstrophy $\E_{T}(u_{0})$ over "long" time windows 
$T$, initially keeping its rate of growth small or even negative turns 
out to be the best strategy as it allows one to conserve the energy. 
The ability to discover such nonintuitive solutions is a key advantage 
of framing the problem in terms of variational optimization.

The physics of extreme flow behavior can also be studied in terms of 
the Fourier-space representation with the goal to understanding of the 
fine structure of triadic interactions between modes. An effort in this 
direction was made by \citet{ProtasKangBustamante2024} who analyzed the 
extreme Burgers flows found as solutions of Problem \ref{pb:maxET1D} 
from this perspective. The main finding was that the fluxes sustaining 
this behavior were carried by only a handful of triads revealing a 
universal statistical distribution.

\subsection{On the Absence of Enstrophy Dissipation Anomaly in 2D Unforced Navier-Stokes Flows}
\label{sec:noanomaly}

In this section we consider a simpler and more tractable version of 
Question \ref{Q2} in the setting of unforced 2D Navier-Stokes flows. It 
is convenient to describe such flows in terms of the 2D vorticity system
\begin{subequations} \label{eq:2DNS}
\begin{alignat}{2} 
\partial_{t}\wn + \bnabla^{\perp} \psin \cdot \bnabla \wn &= \nu \Delta \wn &  \qquad &\text{in} \ (0,T] \times \Omega,  \label{eq:vort_eqn} \\
-\Delta \psin &= \wn & \qquad &\text{in} \ (0,T] \times \Omega, \label{eq:stream_vort} \\
\wn(0) &= \varphi &  &\text{in} \ \Omega,  \label{eq:wIC} 
\end{alignat} 
\end{subequations}
where $\bnabla^{\perp} := \left[\partial_{{x}_{1}}, -\partial_{{x}_{2}} 
\right]^{T}$, $\wn$ is the vorticity component perpendicular to the 
plane of motion (such that the vorticity vector is $\bomega = [0, 0, 
\wn]^{T}$) and $\psin$ the corresponding streamfunction, whereas 
$\varphi = \bnabla^{\perp}\cdot\u_{0}$ is the initial condition. Since 
we are particularly interested in how properties of solutions of 
\eqref{eq:2DNS} change with viscosity as $\nu \rightarrow 0$, this 
dependence is indicated with the subscript "$\nu$"; for some technical 
reasons we will assume here that $0 < \nu < 1$. The main difference 
between \eqref{eq:2DNS} and the complete 3D vorticity system is the 
absence of the term representing vortex stretching, cf.~the RHS in 
\eqref{eq:EuVorta}. As a consequence of this simplification, system 
\eqref{eq:2DNS} is known to be globally well-posed in the classical 
sense \citep{kl04}.

In addition to kinetic energy \eqref{eq:K} and enstrophy 
\eqref{eq:E}, 2D Navier-Stokes flows are also characterized by the 
palinstrophy 
\begin{equation}
\P(\wn(t)) := \frac{1}{2} \, \int_{\Omega} \left| \bnabla\wn(t, \x) \right|^2 \, d\x
\label{eq:P}
\end{equation}
and when there is no risk of confusion we will use the simplified 
notation $\P(t) := \P(\wn(t))$. The enstrophy and palinstrophy satisfy 
the equation
\begin{equation}
\frac{d\E(t)}{dt} = - 2 \nu \P(t)
\label{eq:dEdtP}
\end{equation}
obtained by multiplying \eqref{eq:vort_eqn} by $\wn$, integrating over 
the domain $\Omega$ and then performing integration by parts. It shows 
that in the absence of solid boundaries the enstrophy in 2D flows is a 
non-increasing function of time. Therefore, since it is controlled by 
the enstrophy via \eqref{eq:dKdt}, the rate of energy dissipation 
$d\K(t)/dt$ is in the 2D setting rather uninteresting. Given this 
observation and the fact that, phenomenologically, the forward cascade 
in 2D flows involves the enstrophy \citep{frisch1995turbulence}, the 
more relevant quantity in the present context is the rate of enstrophy 
dissipation $d\E(t)/dt$. We thus define our main quantity of interest 
here as the enstrophy dissipation per unit of time
\begin{equation} 
\label{eq:chi}
\chin(\varphi) := - \frac{1}{T} \int_0^T \frac{d\E(t)}{dt}\, dt = \frac{2 \nu}{T} \int_0^T \P(t)\, dt  
= {\frac{\nu}{T}} \, \int_{0}^{T} \intO \left| \bnabla \wn(t, \x; \varphi) \right|^2 \, d\x dt 
\end{equation} 
which is viewed as a function of the initial data $\varphi$. In analogy 
with Question \ref{Q2}, a natural question concerning this quantity is 
whether or not there exists initial data $\varphi_{\nu}$ possibly 
depending on $\nu$ and with a fixed $H^{1}$ norm such that 
$\chin(\varphi_{\nu}) \rightarrow C > 0$ as $\nu \rightarrow 0$, which 
would imply the enstrophy dissipation anomaly. 

The problem stated above has had an interesting history with Batchelor 
assuming an affirmative answer to this question in his statistical 
theory of 2D turbulence \citep{Batchelor1969}. However, 
\citet{Tran2006} argued that quantity \eqref{eq:chi} in fact vanishes 
in the inviscid limit such that unforced Navier-Stokes flows in 2D are 
not subject to dissipation anomaly. This result was also confirmed by 
rigorous mathematical analysis of the inviscid limit of 2D 
Navier-Stokes flows \citep{Lopes2006}. Even if there is no enstrophy 
dissipation anomaly in unforced 2D Navier-Stokes flows, in order to 
have a complete understanding of the problem, it is imperative to 
obtain sharp bounds on the dependence of quantity \eqref{eq:chi} on 
$\nu$. Using some assumptions about the form of the spectrum of 
solutions of \eqref{eq:2DNS}, \citet{Tran2006} conjectured that 
\eqref{eq:chi} is subject to the bound
\begin{equation} 
\label{eq:Tran}
\chin \leq C\,\left[-\ln(\nu)\right]^{-\frac{1}{2}},
\end{equation}
for some constant $C>0$ depending on the initial condition $\varphi$
and the length $T$ of the time window. 

The goal of \citet{MatharuProtasYoneda2022} was to obtain a 
mathematically rigorous upper bound on the rate of enstrophy 
dissipation and they discovered that such a bound is in fact  closely 
related to another problem which has recently received considerable 
attention, namely, the question of the convergence as $\nu \rightarrow 
0$ of Navier-Stokes flows to solutions of the inviscid Euler equations 
obtained by setting $\nu = 0$ in \eqref{eq:vort_eqn} and corresponding 
to the same initial condition $\varphi$. More specifically, noting 
\eqref{eq:dEdtP}, the fact that smooth solutions of the inviscid Euler 
system conserve enstrophy and using the reverse triangle inequality, we 
have
\begin{align}
\chin(\varphi)  &= {\frac{\nu}{T}} \, \int_{0}^{T} \left\| \bnabla\wn(T, \x; \varphi) \right\|^2_{L^2} \, dt = \frac{2 \nu}{T} \, \int_{0}^{T} \P(t) \, dt  \nonumber \\
&= \frac{1}{T} \, \left[\E(0) - \E(T)\right] = \frac{1}{T} \, \left[ \left\| \varphi \right\|^2_{L^2} - \left\| \wn(T, \x; \varphi) \right\|^2_{L^2} \right] \nonumber \\
&= \frac{1}{T} \, \left[ \left\| \omega(T, \x; \varphi) \right\|^2_{L^2} - \left\| \wn(T, \x; \varphi) \right\|^2_{L^2} \right] \nonumber \\
&\leq \frac{1}{T} \, \left[ \left\| \omega(T, \x; \varphi) \right\|_{L^2} + \left\| \wn(T, \x; \varphi) \right\|_{L^2} \right] \, \left\| \omega(T, \x; \varphi) - \wn(T, \x; \varphi) \right\|_{L^2} \nonumber \\
&\leq \frac{2}{T} \, \left\| \varphi \right\|_{L^2}  \, \left\| \omega(T, \x; \varphi) - \wn(T, \x; \varphi) \right\|_{L^2}, \label{eq:chibound}
\end{align} 
where $\omega(t,\x) := \omega_0(t,\x)$ denotes the vorticity in the 
inviscid Euler flow. The above inequality shows that the enstrophy 
dissipation over the time window $[0,T]$ can be bounded from above in 
terms of the difference of the vorticity fields in the viscous and 
inviscid flows obtained with the same initial data $\varphi$ at time $t 
= T$.  Quantifying this difference in terms of viscosity as $\nu 
\rightarrow 0$ has been the subject of several recent studies and arguably 
the best result of this type was obtained by \cite{Ciampa2021} who 
proved that 
\begin{equation}
\sup_{t \in [0,T]} \left\| \omega(\cdot, t) - \wn(\cdot,t) \right\|_{L^p} \le C\, M^{1-\frac{1}{p}}\max\left\{\phi_{\varphi, p, M}(C\, \nu^{\frac{e^{-CT}}{2}}), \left(C\, \nu^{\frac{e^{-CT}}{2}}\right)^{\frac{e^{-CT}}{2p}} \right\}, 
\label{eq:chiCCS}
\end{equation}
where $M := \|\varphi\|_{L^\infty}$ and now $C = C(T, M)$, whereas 
$\phi_{\varphi, p, M} \: : \: \RR^+ \to \RR^+$ is a continuous function 
such that $\phi_{\varphi, p, M}(0)=0$. In the light of relation 
\eqref{eq:chibound}, this inequality implies a viscosity-dependent 
upper bound on the enstrophy dissipation rate \eqref{eq:chi}. On the 
other hand a lower bound on the maximum enstrophy dissipation is also 
available and given by the following result
\begin{theorem}[\citet{Jeong2021}] 
\label{thm:lower}
There exists a family of initial data $\varphi_{\nu}$ in \eqref{eq:2DNS} 
such that the corresponding enstrophy dissipation is bounded below by
\begin{equation} \label{eq:chiJY}
\chin(\varphi_{\nu}) \ge {C} \nu \, \left[-\ln(\nu)\right]^{\frac{1}{2}} \qquad \text{as} \ \nu \rightarrow 0.
\end{equation}
\end{theorem}

\citet{MatharuProtasYoneda2022} showed that bound 
\eqref{eq:chibound}--\eqref{eq:chiCCS} is in fact sharp which was done 
by constructing families of optimal initial conditions $\phichk$ in 
\eqref{eq:wIC} such that the corresponding flows maximize the enstrophy 
dissipation $\chin(\phichk)$ for the given viscosity coefficient $\nu$ 
and the length $T$ of the time window. Such initial conditions were 
found by solving the following optimization problem for different $\nu$ 
and $T$
\begin{problem}
\label{pb:maxchi}  
Given {$\P_0,\nu, T > 0$} in system \eqref{eq:2DNS} and the
objective functional \eqref{eq:chi}, find
\begin{equation*}
{\phichk} = \underset{\varphi \in \mathcal{S}} {\argmax} \, \chin(\varphi), \quad \textrm{where} 
\quad \mathcal{S} := \left\{ \varphi \in H^1(\Omega) \: : \: \intO \varphi(\x) \, d\x = 0, \ \ \P(\varphi) = \P_0 \right\}. 
\end{equation*}
\end{problem}
The enstrophy dissipation is given in terms of a time integral of 
palinstrophy, cf.~\eqref{eq:chi}, and hence the objective functional 
$\chin(\varphi)$ can in principle be made arbitrarily large by allowing 
initial data $\varphi$ with large palinstrophy $\P(\varphi)$. To 
prevent this from happening, the constraint $\P(\varphi) = \P_0$ is 
included in the definition of the constraint manifold $\mathcal{S}$. 
Since the enstrophy dissipation is not affected by the mean vorticity
$(1/|\Omega|) \intO \wn(t, \x)\, d\x$, the zero-mean constraint serves 
as a normalization. Problem \ref{pb:maxchi} needs to be solved 
numerically and a state-of-the-art adjoint-based Riemannian gradient 
approach developed to solve this class of problems is presented in 
Appendix \ref{sec:solution}.

To probe the sharpness of estimate 
\eqref{eq:chibound}--\eqref{eq:chiCCS}, \citet{MatharuProtasYoneda2022} 
solved Problem \ref{pb:maxchi} for $\P_{0} = 1$ and with both $\nu$ and 
$T$ varying over a broad range of values. These results are summarized 
briefly below. Problem \ref{pb:maxchi} is nonconvex and as such admits 
multiple local maximizers at least for some values of $\nu$ and $T$. 
The results are organized in terms of ``branches'' defined as families 
of optimal initial conditions $\phichk$ obtained with fixed values of 
$\nu$ and varying $T$ such that the maximum enstrophy dissipation 
$\chin(\phichk)$ is a smooth function of the length $T$ of the time 
window. For each value of $\nu$ and each branch, the time windows are 
then chosen to capture the local maximum of $\chin(\phichk)$ and its 
neighbourhood. Information about the local maximizers found for $\nu = 
2.2361 \times 10^{-6}$ and $T = 0.1789$ on six distinct branches is 
collected in Table \ref{tab:branches} where we show the corresponding 
palinstrophy evolutions $\P(t)$, optimal initial conditions 
$\phichk(\x)$ and the vorticity fields realizing the maximum 
palinstrophy $\wn(\argmax_{0< t \le T}\P(t), \x)$. These branches were 
determined for the given value of $\nu$ using the continuation approach 
described in Appendix \ref{sec:cont} where $\chin$ is regarded as a 
smooth function of $T$. When searching for branches corresponding to 
different viscosity values, continuation with respect to $\nu$ with $T$ 
fixed was also used. The time evolution of the vorticity fields 
corresponding to all six branches is visualized in 
\href{https://youtu.be/4I_obQAgxUY}{movie 1}. This movie offers 
insights about the different physical mechanisms involving the 
stretching of thin vorticity filaments which are responsible for the 
growth of palinstrophy and hence also increased enstrophy dissipation. 
Additional comments about these different scenarios are provided at the 
end of this section.

\begin{landscape} \setlength\tabcolsep{0.5pt} 
\begin{table}
	\centering
	\vspace*{-2.0cm} \hspace*{-0.5cm}\begin{tabular}{ |c|c|c|c|c|c|c|} 
		\hline
		\rowcolor{Gray}
		Branch & 1 & 2 & 3 & 4 & 5 & 6  \\ 
		\hline &&&&&& \\ [-1.5em]
		\rotatebox{90}{\hspace{0.7cm}Palinstrophy} &  {\includegraphics[scale=0.2]{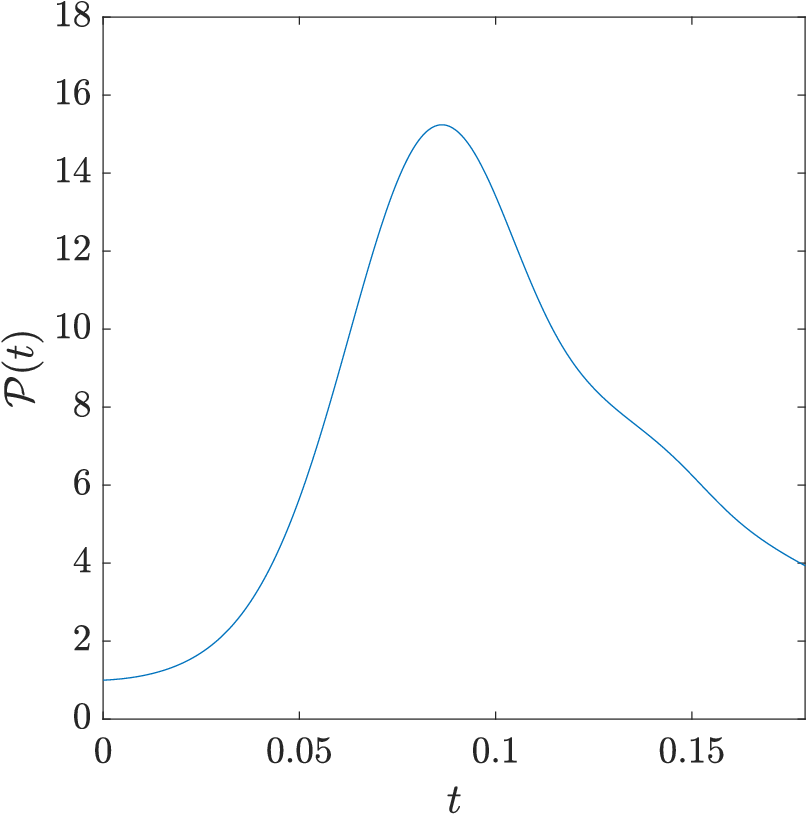}} & 
		{\includegraphics[scale=0.2]{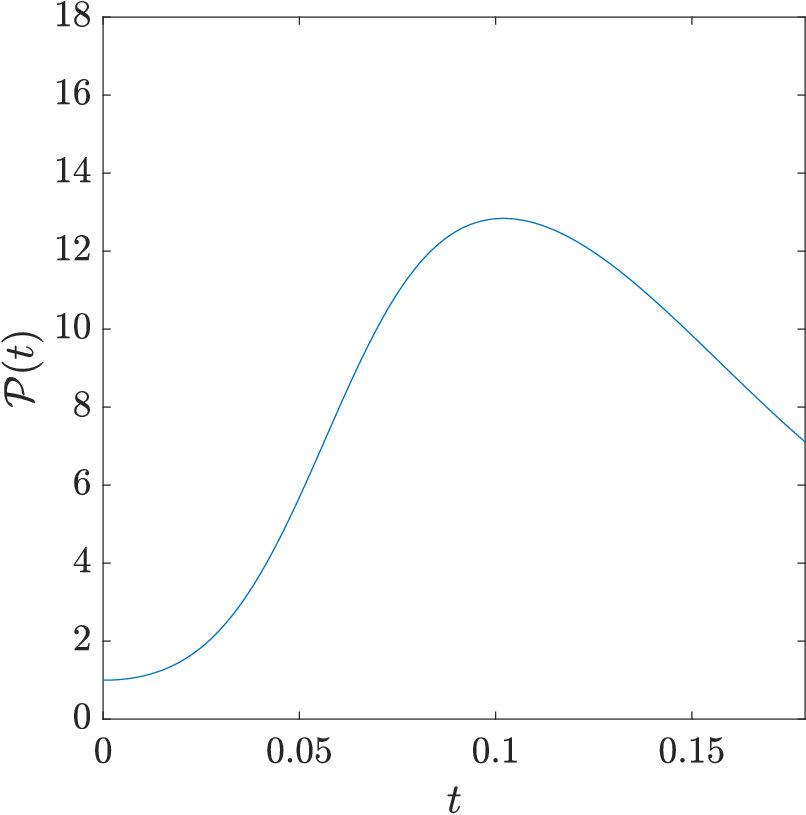}} &
		{\includegraphics[scale=0.2]{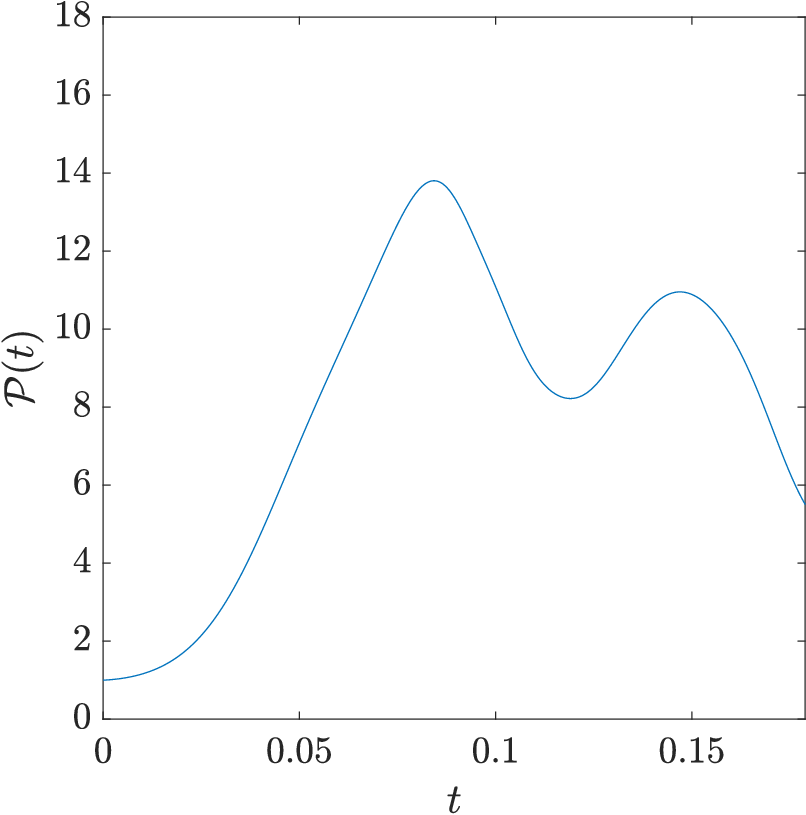}} &
		{\includegraphics[scale=0.2]{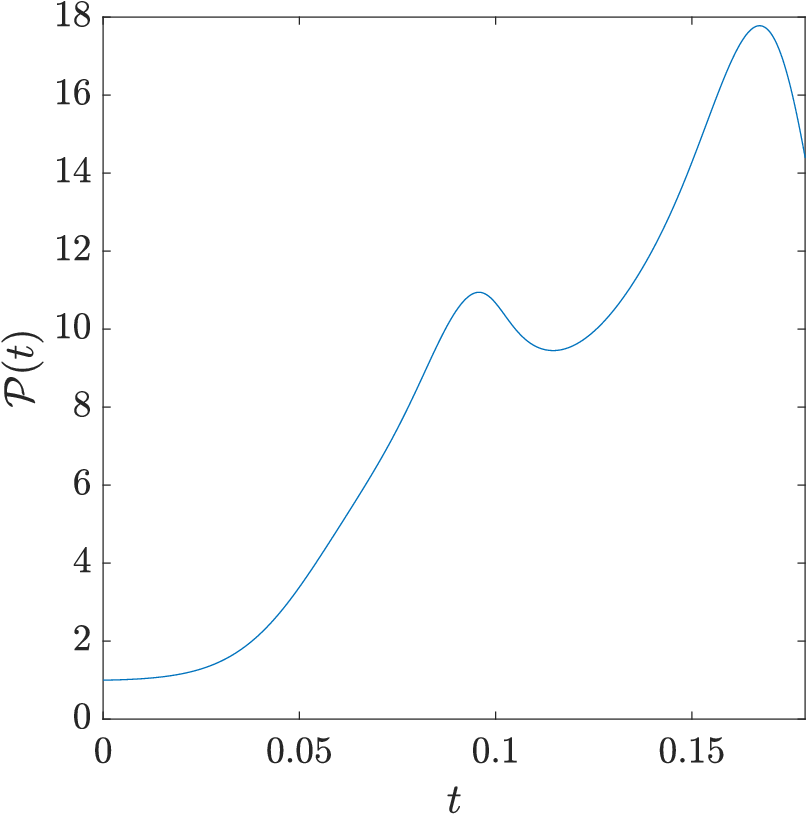}} & 
		{\includegraphics[scale=0.2]{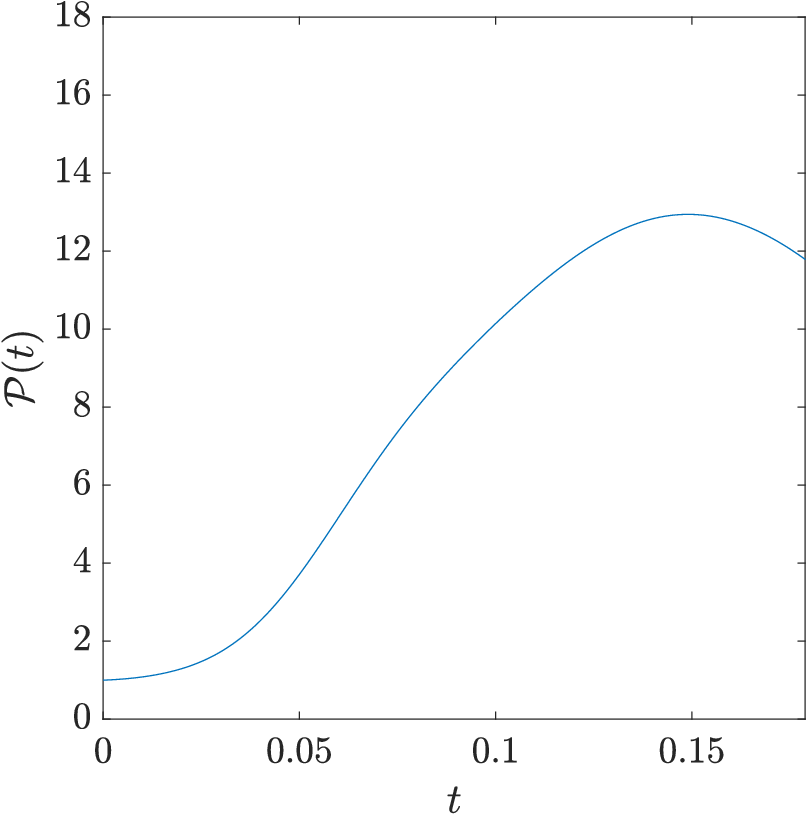}} &
		{\includegraphics[scale=0.2]{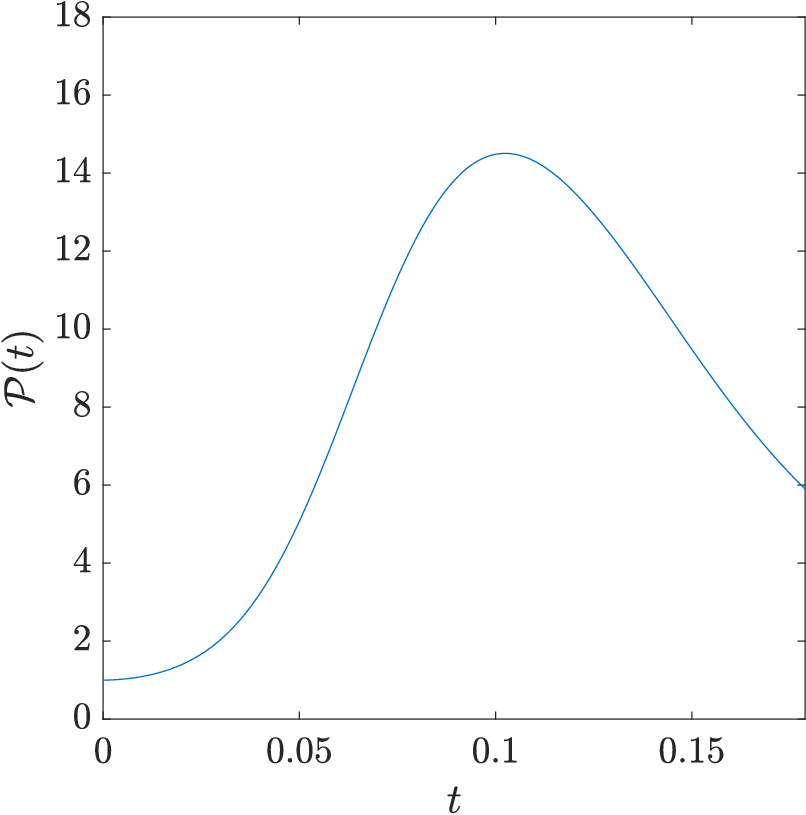}} 
		\\		
		\hline &&&&&& \\ [-1.5em]
		\rotatebox{90}{\hspace{0.6cm}Initial Condition} & {\includegraphics[scale=0.2]{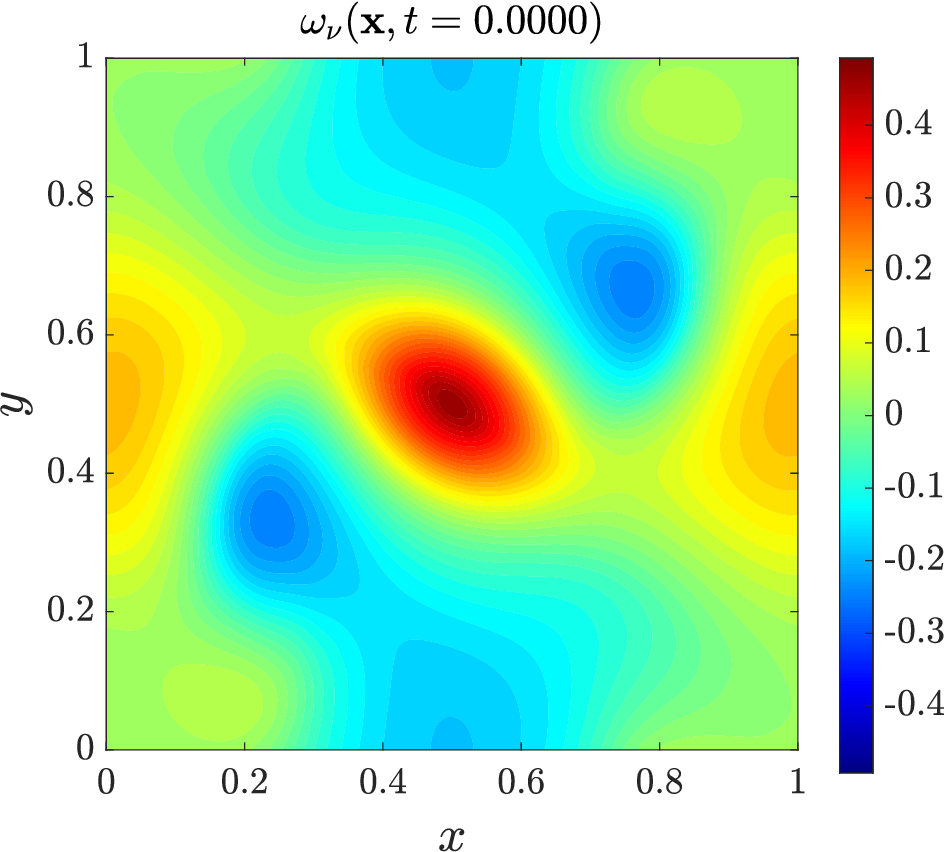}} & 
		{\includegraphics[scale=0.2]{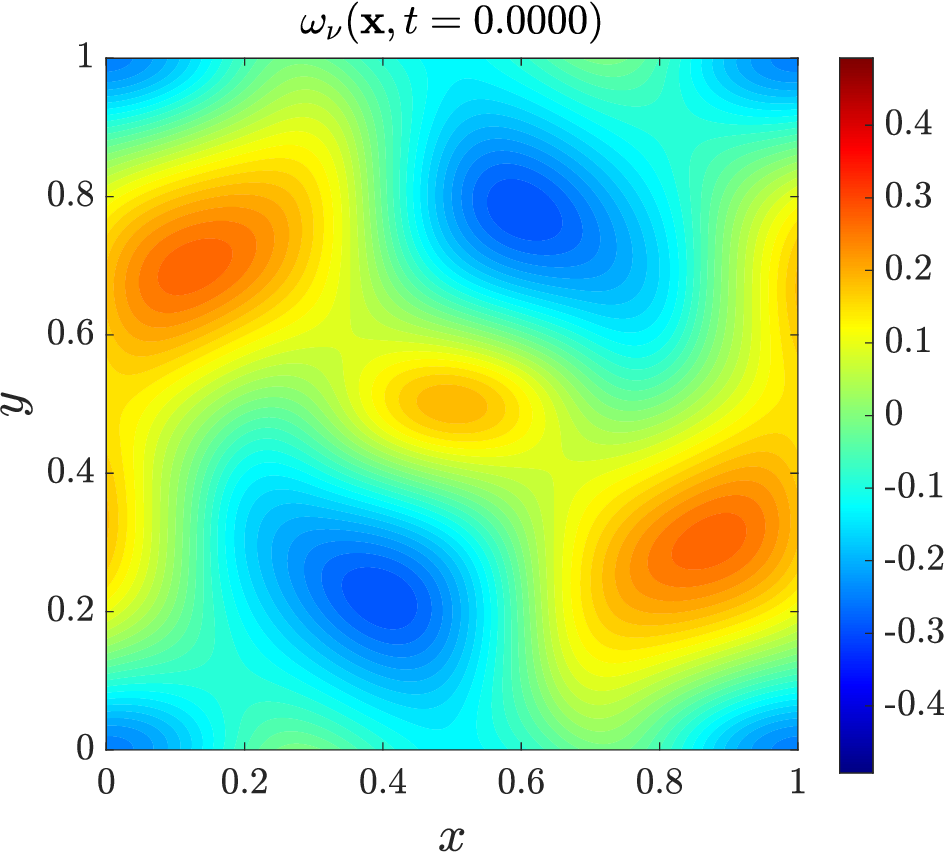}} & 
		{\includegraphics[scale=0.2]{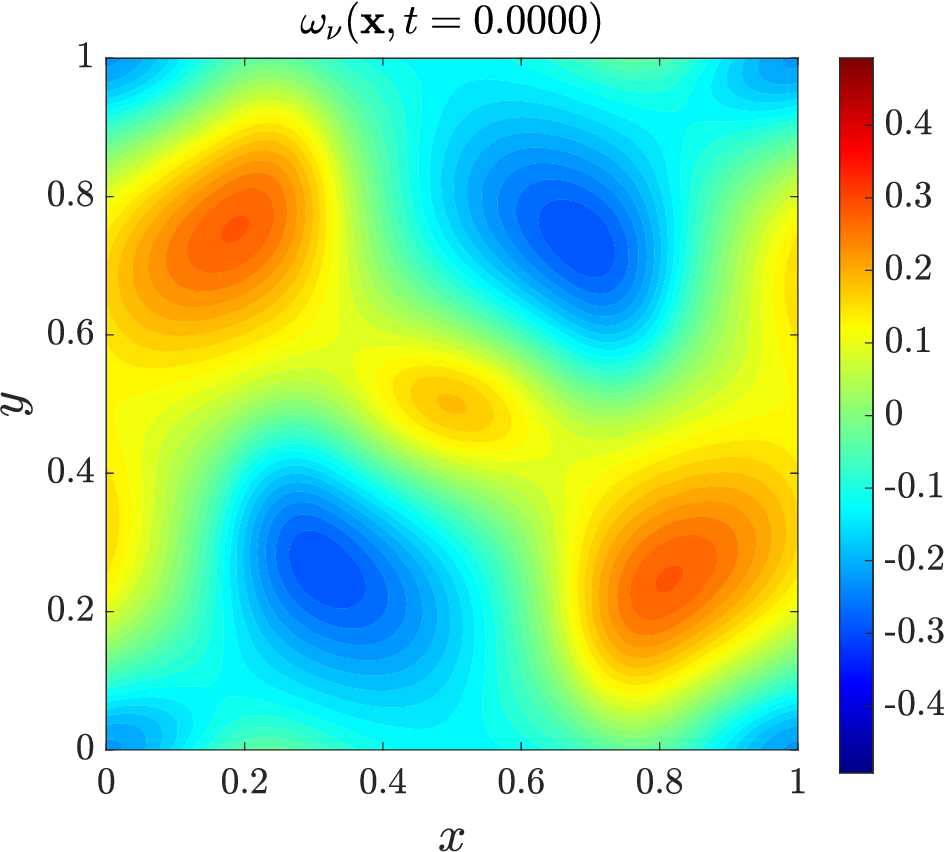}} &
		{\includegraphics[scale=0.2]{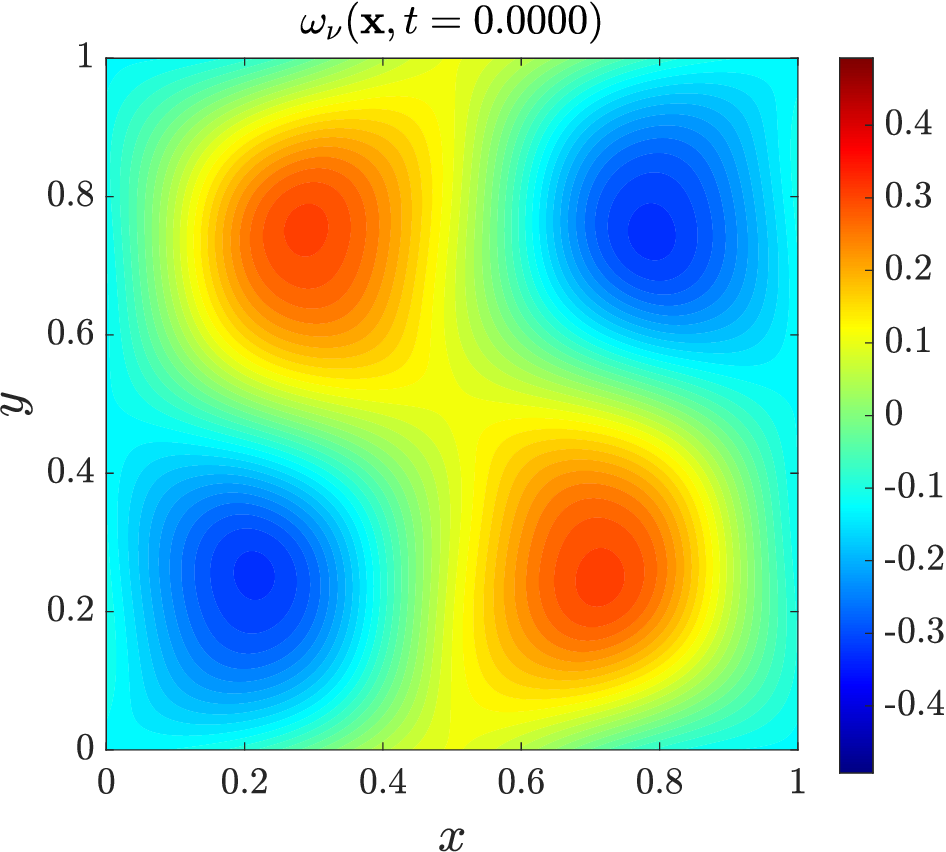}} &
		{\includegraphics[scale=0.2]{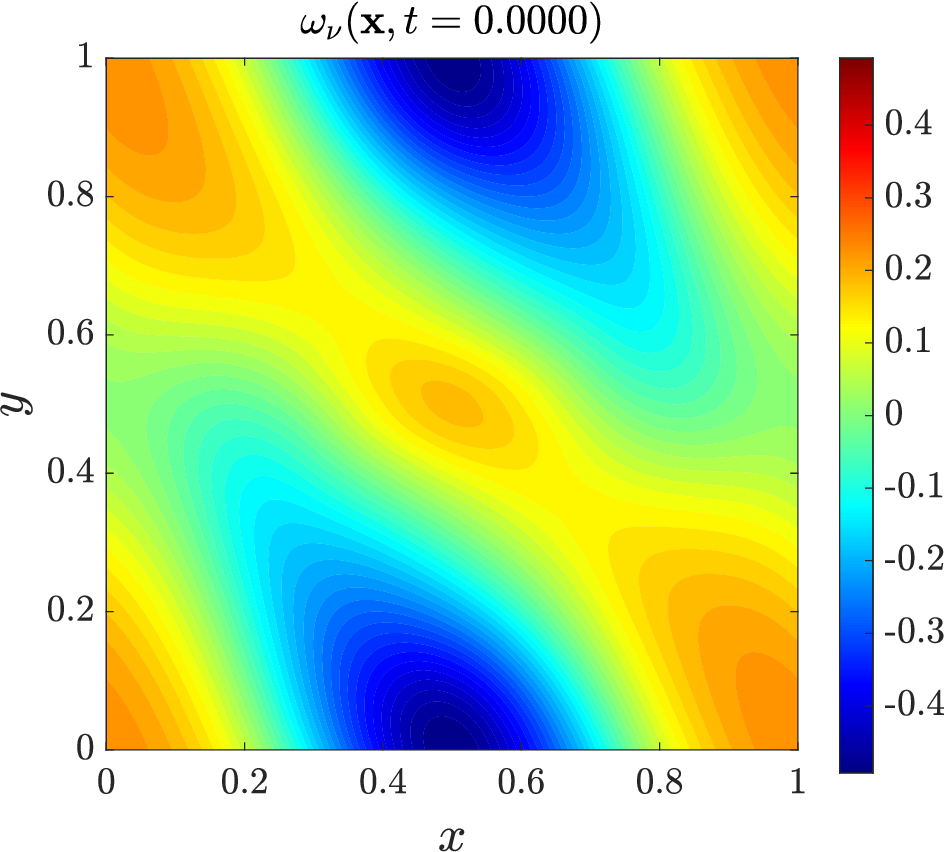}} &
		{\includegraphics[scale=0.2]{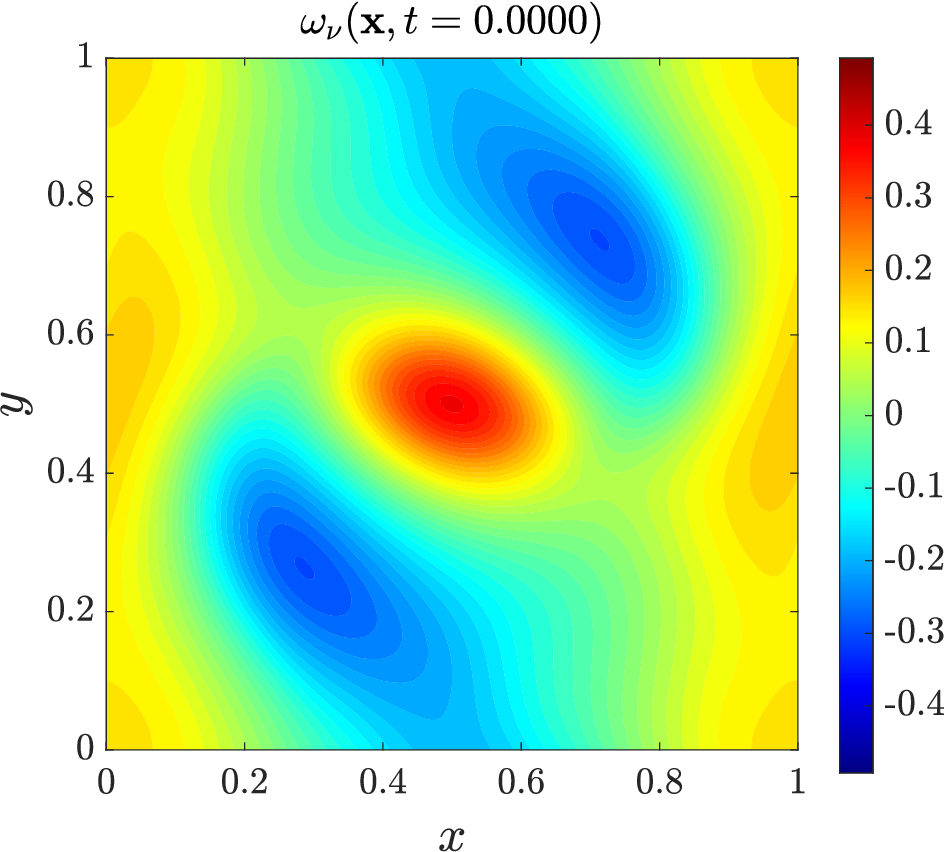}}  \\ 
		\hline &&&&&& \\ [-1.5em]
		\rotatebox{90}{\hspace{0.4cm}Palinstrophy Peak} & 		{\includegraphics[scale=0.2]{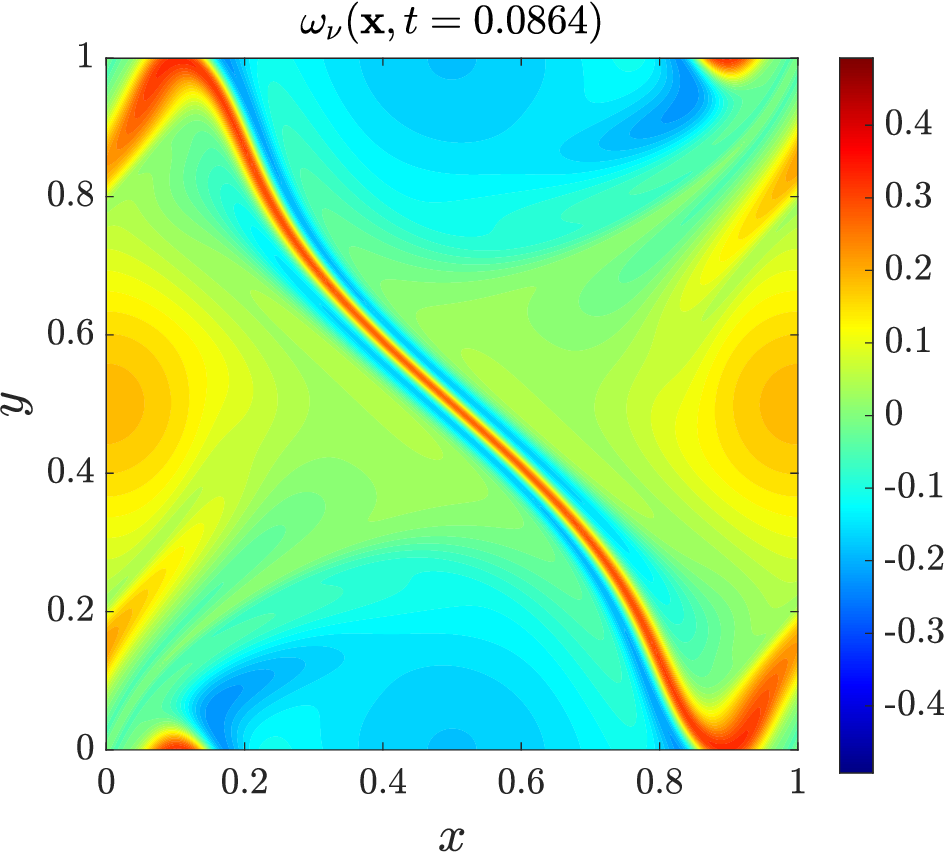}} & 
		{\includegraphics[scale=0.2]{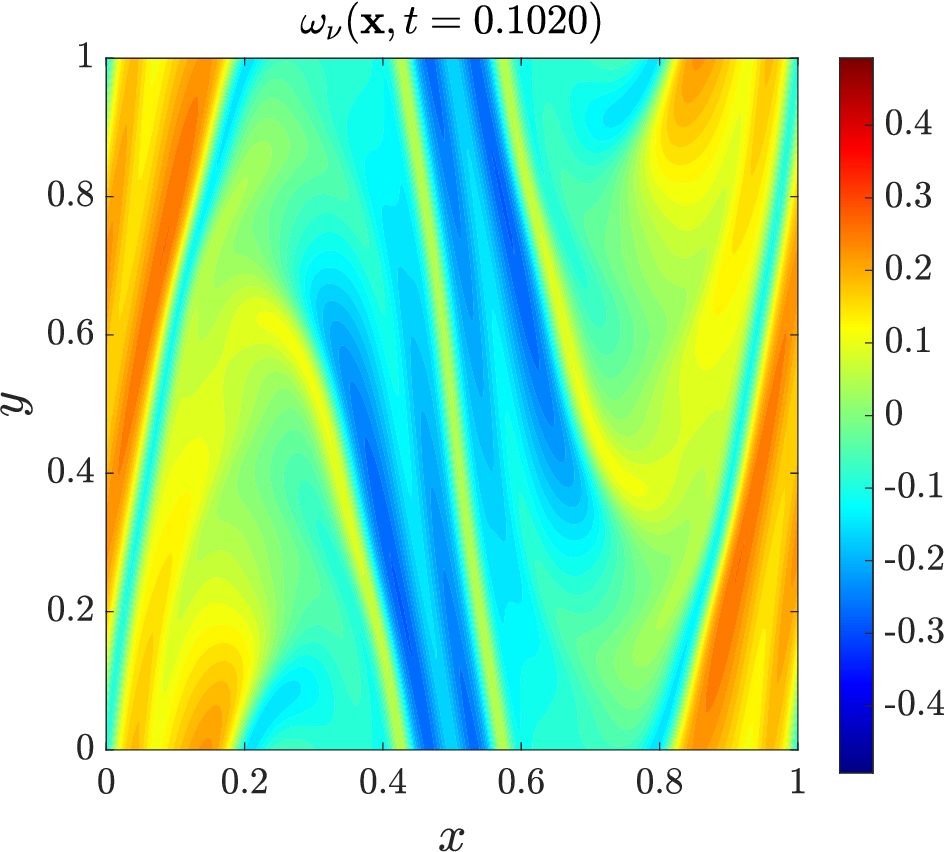}} &
		{\includegraphics[scale=0.2]{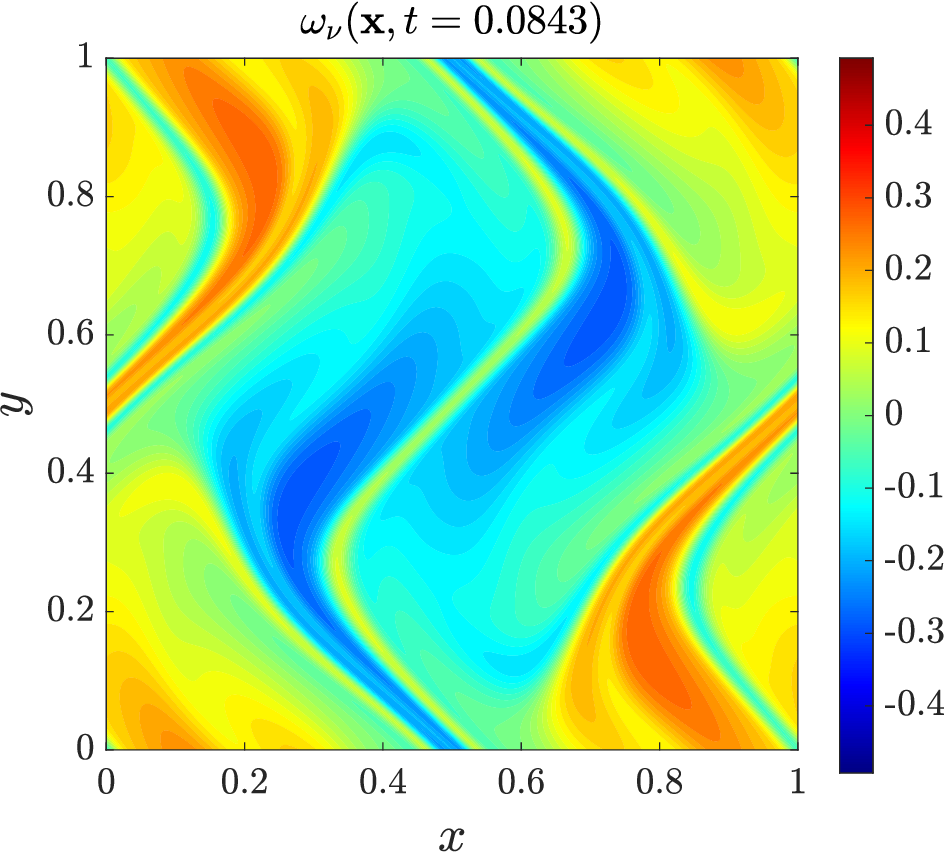}} &
		{\includegraphics[scale=0.2]{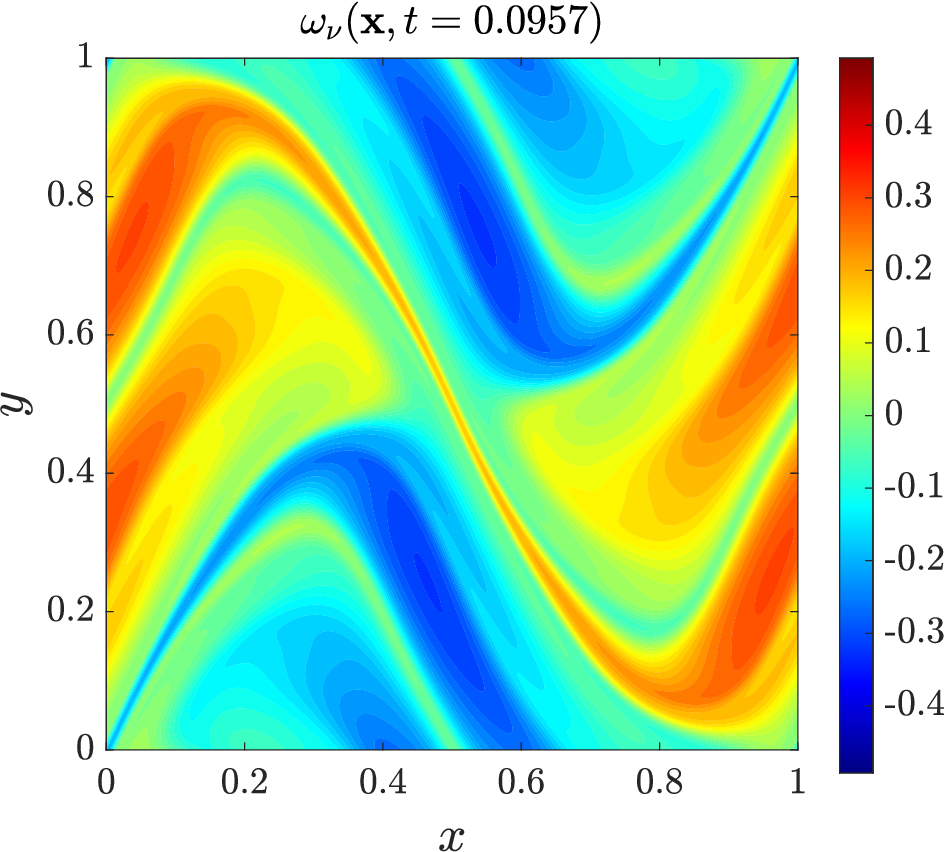}} & 
		{\includegraphics[scale=0.2]{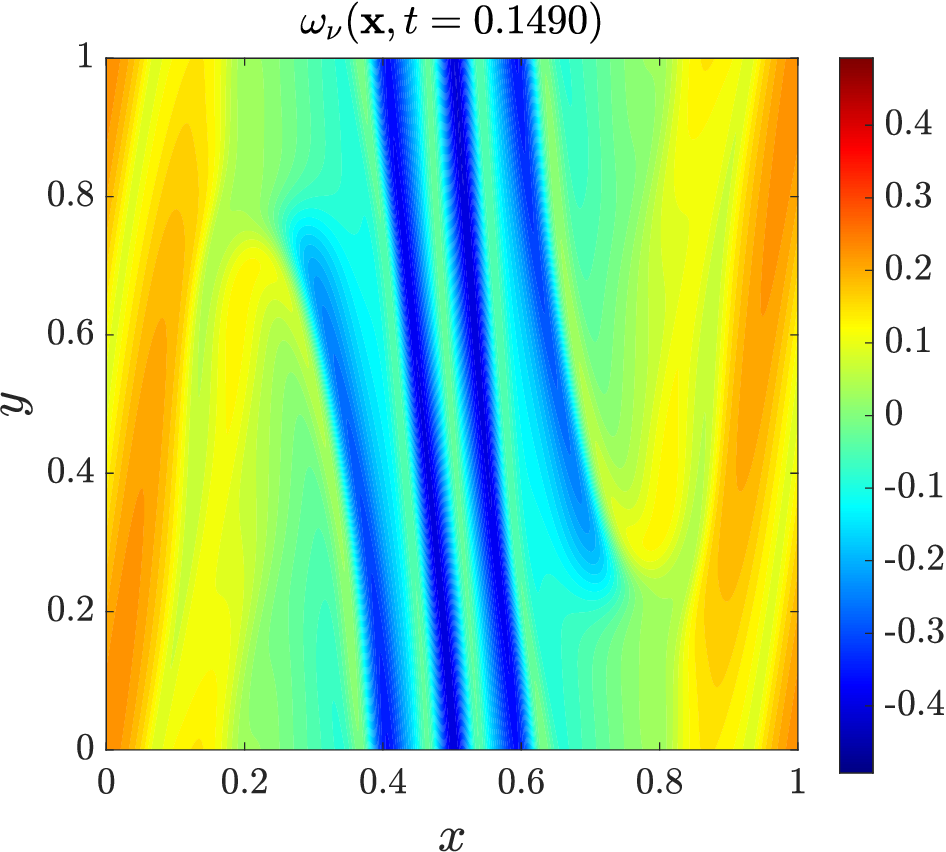}} &
		{\includegraphics[scale=0.2]{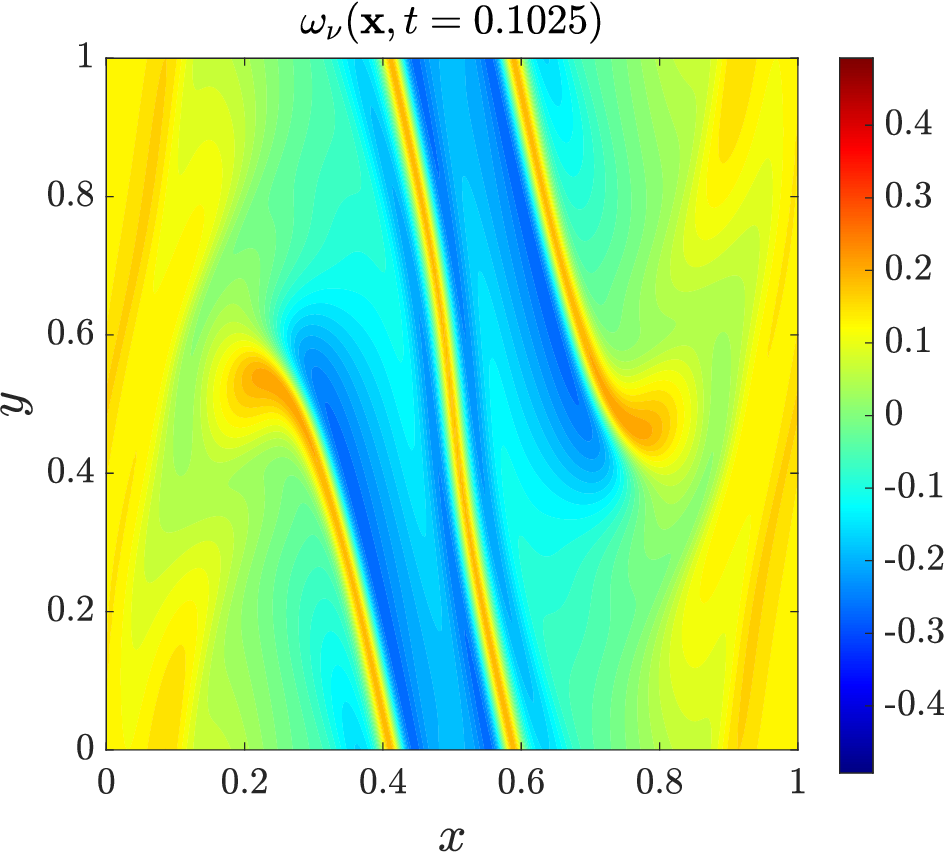}}  \\ 
		\hline &&&&&& \\ [-1.5em]
		\rotatebox{90}{\hspace{0.3cm}Palinstrophy Peak 2} & N/A & N/A &
		\includegraphics[scale=0.2]{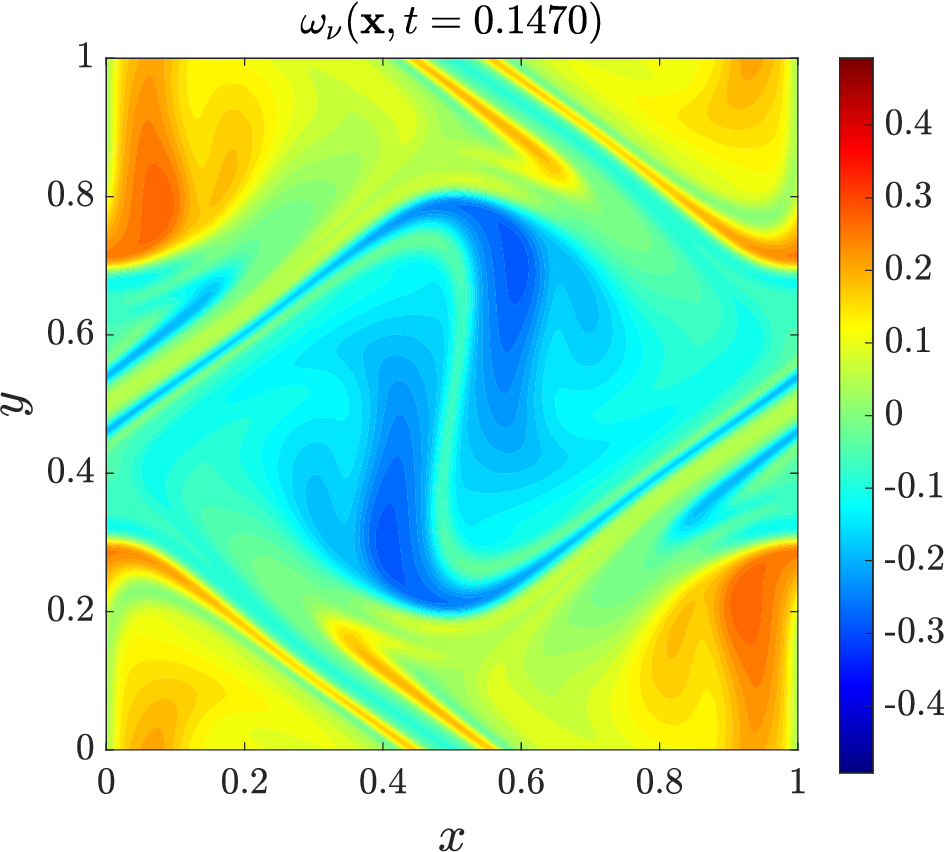} &
		\includegraphics[scale=0.2]{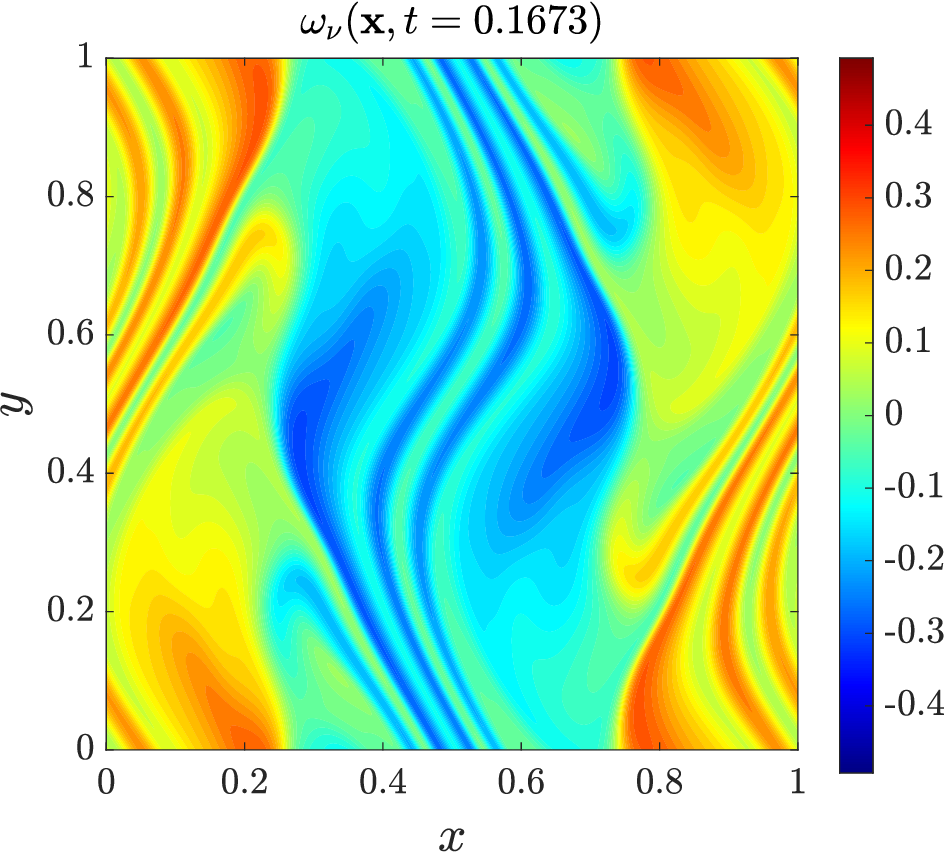} & 
		N/A & 
		N/A 
		\\ 
		\hline
	\end{tabular}	
	\caption{Summary information about the local maximizers
		  obtained by solving Problem \ref{pb:maxchi} with $\nu = 
		  2.2361 \times 10^{-6}$ and $T = 0.1789$. The time evolution 
		  of the vorticity fields is visualized in \href{https://youtu.be/4I_obQAgxUY}{movie 1}. 
		  ``N/A'' indicates that palinstrophy attains a single maximum only during the time 
		  evolution corresponding to the given branch.}	
	\label{tab:branches}
\end{table} 
\end{landscape}

\begin{figure}\centering
\mbox{
\hspace*{-0.5cm}
\subfigure[]
  {\raisebox{-2mm}{\includegraphics[scale=0.425]{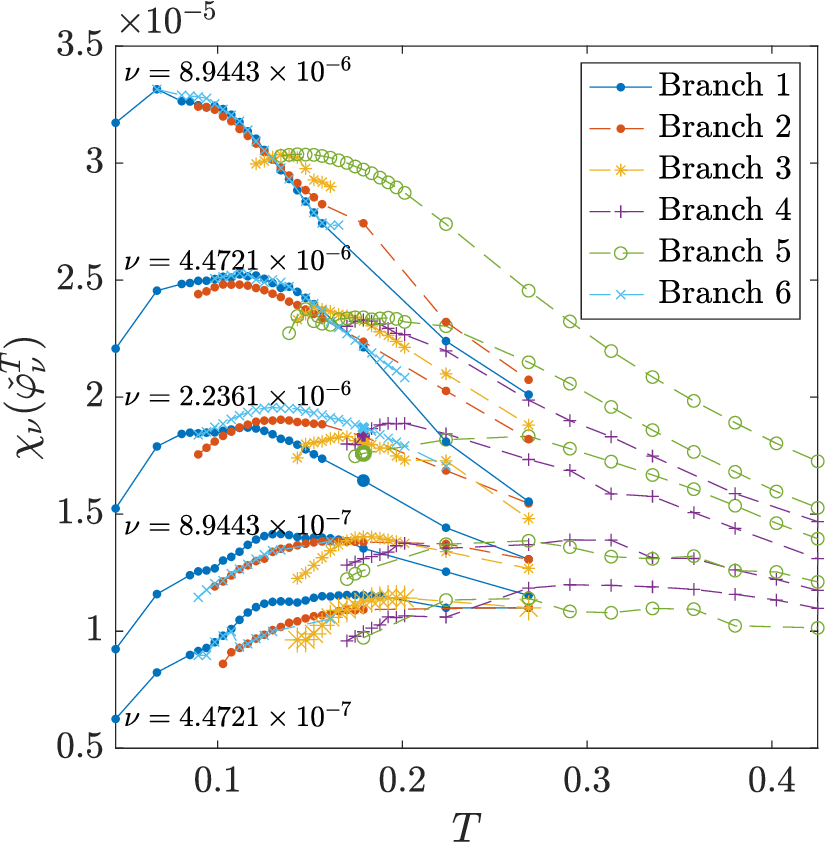}}}\qquad
  \subfigure[]
  {\includegraphics[width=0.495\textwidth]{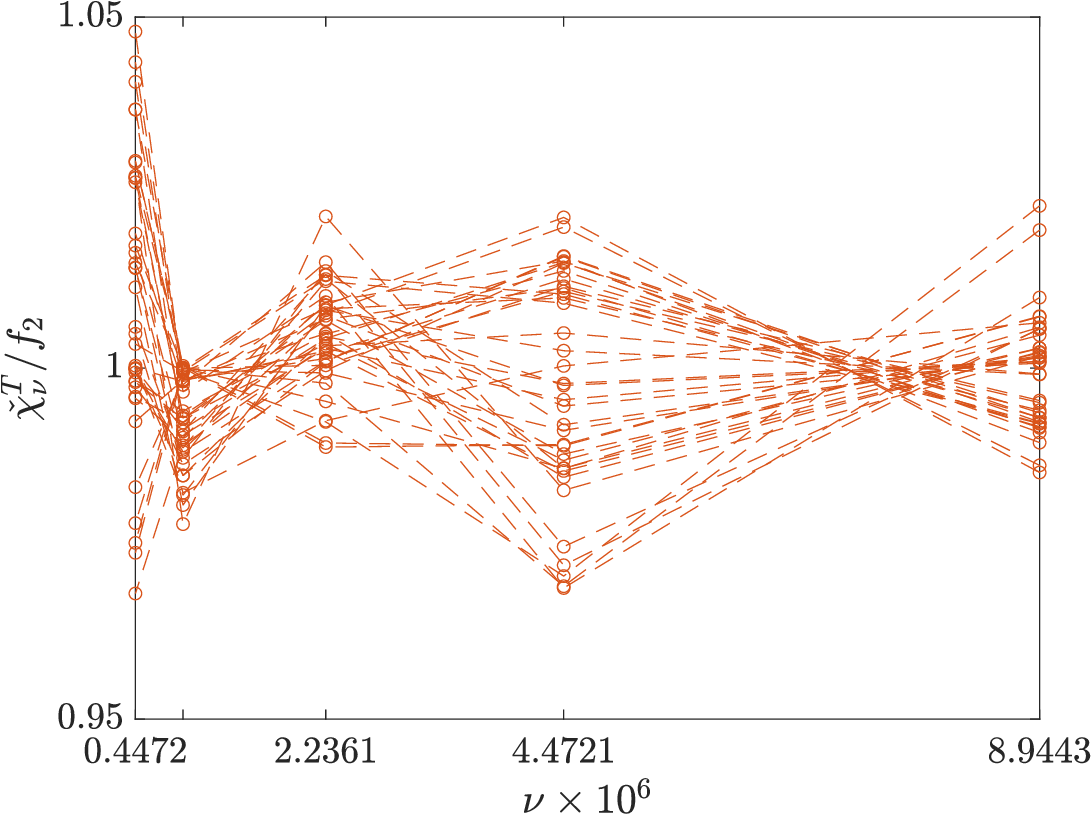}}}
\caption{(a) Dependence of the maximum enstrophy dissipation 
$\chin(\phichk)$ on $T$ for different indicated values of $\nu$ with distinct 
branches of local maximizers, cf.~Table \ref{tab:branches}; 
(b) dependence of $\chinck / f_2(\nu)$, cf.~\eqref{eq:CCSf}, with the optimal 
constants {$\tC = \tC(T)$} and exponents $\talpha = \talpha(T)$ on the 
viscosity $\nu$ for different $T$.} 
\label{fig:chin}
\end{figure}

The dependence of the maximum enstrophy dissipation $\chin(\phichk)$ on 
the length $T$ of the time window for five values of viscosity spanning 
more than one order of magnitude is shown in figure \ref{fig:chin}a, 
where we carefully distinguish branches of distinct local maximizers. 
We remark that for certain combinations of $\nu$ and $T$ only a subset 
of the local maximizers described in Table \ref{tab:branches} could be 
found. In figure \ref{fig:chin}a we observe that along each branch the 
maximum enstrophy dissipation $\chin(\phichk)$ admits a well-defined 
maximum with respect to $T$. However, for different values of $T$ the 
maximum of $\chin(\phichk)$ can be achieved on different branches. To 
analyze this data, we therefore introduce the function $\chinck := 
\max_{\textrm{branches}} \chin(\phichk)$ which, for each value of $\nu$ 
and $T$, represents the upper envelope of the curves shown in figure 
\ref{fig:chin}a. We add that the values of $\chin(\phichk)$ shown in 
figure \ref{fig:chin}a are for each value of $\nu$ at least an order of 
magnitude larger than the enstrophy dissipation corresponding to the 
initial conditions constructed by \citet{Jeong2021}, which realize the 
behavior given in \eqref{eq:chiJY}.

In order to identify quantitative connections between the data 
presented in figure \ref{fig:chin}a and estimates \eqref{eq:Tran}, 
\eqref{eq:chibound}--\eqref{eq:chiCCS} and \eqref{eq:chiJY}, 
\citet{MatharuProtasYoneda2022} introduced the ans{\"a}tze 
\begin{subequations}
  \label{eq:ff}
  \begin{align}
    f_1(\nu) &= C \, \left[-\ln(\nu)\right]^{-\frac{1}{2}}, \label{eq:Tranf} \\
    f_2(\nu) &= C \, \nu^{\alpha}, \label{eq:CCSf} \\
    f_3(\nu) &= C \, \nu\, \left[-\ln(\nu)\right]^{\frac{1}{2}} \label{eq:JYf}
  \end{align}
\end{subequations}
motivated by the structure of the different bounds.  More specifically, 
\eqref{eq:Tranf} is the expression from conjecture \eqref{eq:Tran}, 
\eqref{eq:CCSf} has the general form of the upper bound in 
\eqref{eq:chiCCS}, where only the second argument of the function 
$\max(\cdot)$ is considered since the function $\phi_{\varphi, p, M}$ 
appearing in the first argument is not given explicitly enough to allow 
for quantitative comparisons, whereas \eqref{eq:JYf} is the lower bound 
from Theorem \ref{thm:lower}, cf.~\eqref{eq:chiJY}.  The primary 
interest here is the dependence on the viscosity $\nu$ with the time 
window $T$ treated as a parameter (in enters explicitly only in 
estimate \eqref{eq:chibound}--\eqref{eq:chiCCS}).

To find out which of the functions \eqref{eq:Tranf}--\eqref{eq:JYf} 
best describes the actual dependence of the data shown in figure 
\ref{fig:chin}a on $\nu$, for each discrete value of $T$, 
\citet{MatharuProtasYoneda2022} determined the constant $C = C(T)$ in 
ansatz functions \eqref{eq:Tranf}--\eqref{eq:JYf} by solving the 
problem
\begin{equation}
{\tC(T) = \argmin_{C \in  \RR^+} \mu_i^T(C),} \qquad  i=1,2,3,
\label{eq:C}
\end{equation}
with the fitting error defined as $\mu_i^T(C) := \frac{1}{5} 
\sum_{j=1}^{5} \left|\widecheck{\chi}^T_{\nu_j} - f_i(\nu_j; 
C)\right|$, where $\nu_j \in \{ 8.9443 \times 10^{-6}, 4.4721 \times 
10^{-6}, 2.2361 \times 10^{-6}, 8.9443 \times 10^{-7}, 4.4721 \times 
10^{-7}\}$ are the considered values of the viscosity. The fitting 
error $\mu_i^T(C)$ measures the accuracy with which the different 
ansatz functions \eqref{eq:Tranf}--\eqref{eq:JYf} represent the data 
$\widecheck{\chi}^T_{\nu_j}$. In addition, we note that ansatz 
\eqref{eq:CCSf} also involves an a priori  undefined exponent $\alpha 
\in (0,1)$. To determine this additional parameter, problem 
\eqref{eq:C} was embedded in a bracketing procedure to find the 
exponent $\talpha = \talpha(T)$ producing the smallest fitting error 
$\mu_i^T(C)$ for a given value of $T$. This bracketing procedure was 
performed by first determining $\mu_i^T(\tC(T))$, by  solving problem 
\eqref{eq:C}, for a range of discrete values of $\alpha \in [0,1]$ and 
then using bisection to iteratively improve the approximation of 
$\talpha = \talpha(T)$ which produces the smallest fitting error. We emphasize 
that even though ans\"atze \eqref{eq:Tranf}--\eqref{eq:JYf} involve 
different numbers of parameters (one or two), they were all fitted to 
the data in figure \ref{fig:chin}a in the same way (i.e., by adjusting 
$C = C(T)$), which was done independently for different discrete 
exponents $\alpha$ in the case of ansatz \eqref{eq:CCSf}.

The most accurate fits were obtained with ansatz \eqref{eq:CCSf} and
the ratio $\chinck / f_2(\nu)$ is plotted as a function of $\nu$ for 
different $T$ in figure \ref{fig:chin}b using the values of $\tC = 
\tC(T)$ and $\talpha = \talpha(T)$ determined as described above. We 
see that for most values of $T$ it is close to unity over the entire 
range of $\nu$ indicating that ansatz function $f_2(\nu)$ accurately 
captures the dependence of $\chinck$ on $\nu$. On the other hand, we 
note that relations $f_1(\nu)$ and $f_3(\nu)$, respectively, 
overestimate and underestimate the actual dependence of $\chinck$ on 
$\nu$ (the plots of $\chinck / f_i(\nu)$, $i=1,3$ are not shown here 
for brevity). This observation is consistent with the fact that 
\eqref{eq:Tranf} represents estimate \eqref{eq:Tran}, which is more 
conservative than bound \eqref{eq:chibound}--\eqref{eq:chiCCS}, whereas 
\eqref{eq:JYf} has the form of the lower bound \eqref{eq:chiJY}.

\begin{figure}\centering
    \includegraphics[scale=0.4]{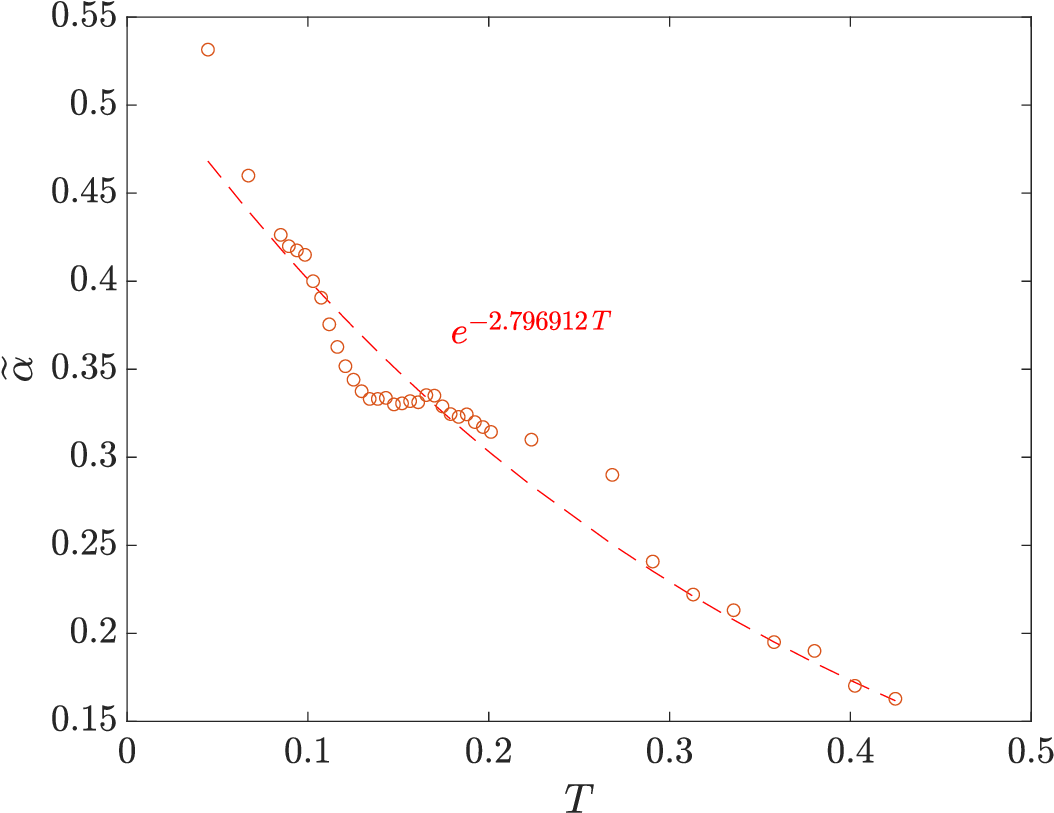}
\caption{Dependence of the optimal exponents $\talpha = \talpha(T)$ in 
ansatz $f_2(\nu)$  on the length $T$ of the time window.  The dashed 
line represents the exponential fit, in the form indicated, to the values 
of $\talpha = \talpha(T)$.}
 \label{fig:talpha}
\end{figure}

Finally, the optimal exponents $\talpha = \talpha(T)$ determined for 
ansatz  \eqref{eq:CCSf} are shown in figure \ref{fig:talpha} where an 
overall decreasing trend with $T$ is evident. As regards the ``dip'' 
occurring for $0.0894 \lessapprox T \lessapprox 0.1342$, we speculate 
that it may be the result of some branches not being captured in the 
results shown in figure \ref{fig:chin}a. We note that, remarkably, the 
dependence of the exponent $\talpha$ on $T$ reveals an approximately 
exponential form consistent with the structure of the upper bound in 
\eqref{eq:chiCCS}, more specifically, the exponential dependence of the 
exponents of $\nu$ in this bound on $T$.

We conclude that the dependence of the maximum enstrophy dissipation 
$\chinck$ in the extreme flows found by \citet{MatharuProtasYoneda2022} 
on $\nu$ with fixed $T$  is quantitatively consistent with the upper 
bound \eqref{eq:chibound}--\eqref{eq:chiCCS}, cf.~figure 
\ref{fig:chin}b, which is also the sharpest estimate available to date. 
We note that it depends on the quantity $\| \phichk \|_{L^\infty}$ (via 
the constant $M$). Since the optimal initial conditions were sought in 
the space $H^1(\Omega)$, one does not have an a priori control over 
this quantity, however, in the computations reported here there was no 
evidence of $\| \phichk \|_{L^\infty}$ becoming large. Thus, these 
caveats notwithstanding, we conclude that the combined estimate 
\eqref{eq:chibound}--\eqref{eq:chiCCS} is sharp and does not offer any 
room for improvement, other than perhaps a logarithmic correction. 
Remarkably, the exponential dependence of the exponent in this upper 
bound on $T$ is also quantitatively consistent with these results, 
cf.~figure \ref{fig:talpha}.

These findings accomplish steps S1--S2 stated in \S\,\ref{sec:bounds} 
and demonstrate that the problem discussed in this section  can also be 
considered closed from the mathematical point of view. As regards step 
S3 concerning key physical mechanisms realizing the observed extreme 
behavior, \citet{MatharuProtasYoneda2022} have found six distinct 
branches of local maximizers, each associated with a different scenario 
for palinstrophy amplification, cf.~Table \ref{tab:branches}. As is 
evident from \href{https://youtu.be/4I_obQAgxUY}{movie 1}, while in all 
cases the maximal palinstrophy amplification involves stretching of 
thin vorticity filaments, there are multiple ways to arrange these 
structures in the periodic domain $\Omega$ and which of these different 
mechanisms produces the largest enstrophy dissipation depends on the 
value of viscosity $\nu$ and the length $T$ of the time window, 
cf.~figure \ref{fig:chin}a. It is noteworthy that all these flow 
evolutions feature very thin filaments which however do not undergo the 
Kelvin-Helmholtz instability as they are stabilized by the shear 
induced by the large vortices also present in the flow field.  The 
flows on branches 3 and 4, which feature multiple palinstrophy maxima, 
employ a mechanism reminiscent of the continuous baker's map to amplify 
the palinstrophy.  Moreover, we see that, interestingly, in some cases 
seemingly very similar optimal initial conditions $\phichk$ give rise 
to quite different flow evolutions featuring different numbers of local 
palinstrophy maxima (one or two) in the considered time window $[0,T]$, 
see, e.g., the maximizers from Branches 2 and 3 in Table 
\ref{tab:branches}. To close this discussion, we note that the question 
about enstrophy dissipation anomaly in forced 2D Navier-Stokes flows 
(with $\f \not\equiv \0$ and $d = 2$ in \eqref{eq:NS}) is still open 
and we return to this topic in \S\,\ref{sec:anomalyopen}.

\FloatBarrier

\section{Systematic Search for Singularities in 3D Navier-Stokes and Euler Flows}
\label{sec:search}

In this section we come to consider Question \ref{Q1} which is arguably 
one of the most important outstanding problems in theoretical 
fluid mechanics. In fact, from the chronological point of view, this 
question has inspired most of the research program surveyed in this 
essay. While the blow-up problem is fundamentally a question in 
mathematical analysis of PDEs, a lot of computational studies have been 
carried out since the mid-1980s in order to shed light on the 
hydrodynamic mechanisms which might lead to singularity formation in 
finite time. Given that such flows evolving near the edge of regularity 
involve formation of very fine structures, these computations 
typically require the use of state-of-the-art computational resources 
available at a given time. Some of the computational studies focused on 
the possibility of finite-time blow-up in the 3D Navier-Stokes and/or 
Euler system were conducted by \cite{bmonmu83,k93,p01,
  gbk08,h09,bb12,CampolinaMailybaev2018}, all of whom
considered problems defined on domains periodic in all three 
dimensions. A simplified semi-analytic model of vortex reconnection was 
developed and analyzed based on the Biot-Savart law and asymptotic 
techniques by \cite{MoffattKimura2019a}. We also mention the studies by 
\cite{sc09} along with the references found therein, in which various 
complexified forms of the Euler equation were investigated. The idea of 
this approach is that, since the solutions to complexified equations 
have singularities in the complex plane, singularity formation in the 
real-valued problem is manifested by the collapse of the complex-plane 
singularities onto the real axis. Overall, the outcome of these 
investigations is rather inconclusive: while for the Navier-Stokes 
system most of the recent computations do not offer support for 
finite-time blow-up, the evidence appears split in the case of the 
Euler system. Some of the investigations \citep{bb12} hinted at the 
possibility of singularity formation in a finite time. In this 
connection we also highlight the {computational} investigations of 
\citet{lh14b} in which blow-up was {documented} in axisymmetric Euler 
flows on a bounded cylindrical domain. Key to this scenario is the 
interaction of the flow with the solid boundary and the underlying 
physical mechanisms in this so-called "teacup" flow were elucidated by 
\citet{Barkley2020}. A significant effort ensued aimed at justifying 
these results mathematically and we refer the reader to 
\citet{DrivasElgindi2023,Elgindi2025} for surveys of the latest 
developments. We also mention an investigation by \citet{Hou:22:Euler} 
who provided evidence for blow-up in axisymmetric Euler flows on 
bounded domains in which the singularity occurs away from the boundary. 
In contrast to most other studies, the works of 
\citet{lh14b,Hou:22:Euler} relied on adaptive mesh refinement employed 
to resolve fine structures in flows at the edge of regularity.

In all of the aforementioned investigations, the initial conditions 
$\u_{0}$ in \eqref{eq:NS} and \eqref{eq:Eu} were chosen in some ad-hoc, 
albeit physically justified, manner. On the other hand, our efforts 
described here have followed a fundamentally different approach where 
the initial conditions are found systematically through the solution of 
a variational optimization problem defined  such that the corresponding 
Navier-Stokes or Euler flows locally maximize certain quantities 
characterizing the regularity of the solutions. Here we first focus on 
the questions concerning extreme, possibly singular, behavior in 
Navier-Stokes flows before turning our attention to similar questions 
in the context of inviscid Euler flows. In both cases we begin by 
reviewing well-known results concerning the existence of classical 
solutions of systems \eqref{eq:NS} and \eqref{eq:Eu} together with the 
associated a priori bounds. These bounds control how much different 
quantities serving as regularity indicators for the solutions can grow 
and come in two flavours, namely, instantaneous bounds characterizing 
the rates of growth, and finite-time bounds related to growth over 
finite time intervals. Then, we will use these results to formulate 
variational optimization problems aimed at finding initial data 
$\u_{0}$ for systems \eqref{eq:NS} and \eqref{eq:Eu} that might 
possibly lead to a singularity. Finally, we will discuss properties of 
the different extreme flows obtained in this way.

\subsection{Extreme Behavior in Navier-Stokes Flows}
\label{sec:extremeNS}

In the context of Navier-Stokes flows, the mathematical discussion of 
Question \ref{Q1} revolves around the so-called "conditional regularity 
results". These are easy to verify conditions which need to be 
fulfilled by a Leray-Hopf weak solution for it to also satisfy system 
\eqref{eq:NS} in the classical sense, i.e., pointwise in $(0,T] \times 
\Omega$. Such solutions will then also be infinitely smooth 
(real-analytic) \citep{RobinsonRodrigoSadowski2016}. However, it is not 
a priori known if these conditions are true. Conditional regularity 
results are often accompanied by a priori estimates involving some 
related quantities and also applicable to weak solutions.  

Arguably, the best known conditional regularity result is the enstrophy 
condition \citep{RobinsonRodrigoSadowski2016} asserting that $\u(t)$ 
is a smooth solution of system \eqref{eq:NS} on the time interval 
$[0,T]$ if and only if its enstrophy \eqref{eq:E} remains bounded, 
i.e.,
\begin{equation}
\label{eq:supE}
\mathop{\sup}_{0 \leq t \leq T} \E(\u(t))  < \infty.
\end{equation}
While it is not known whether \eqref{eq:supE} is true for all initial 
data $\u_0$ and arbitrarily large $T$, Leray-Hopf weak solutions 
satisfy $\int_0^T \E(\u(t)) \, dt < \infty$, cf.~\eqref{eq:Kt}. Condition 
\eqref{eq:supE} implies that should a singularity form in a classical 
solution $\u(t)$ of the Navier-Stokes system \eqref{eq:NS} at some finite 
time $0 < t_0 < \infty$, then necessarily
\begin{equation}
\lim_{t \rightarrow t_0} \E(\u(t)) = \infty.
\label{eq:Etblowup}
\end{equation}

\begin{figure}[t]
\mbox{
\subfigure[]{\includegraphics[width=0.49\textwidth]{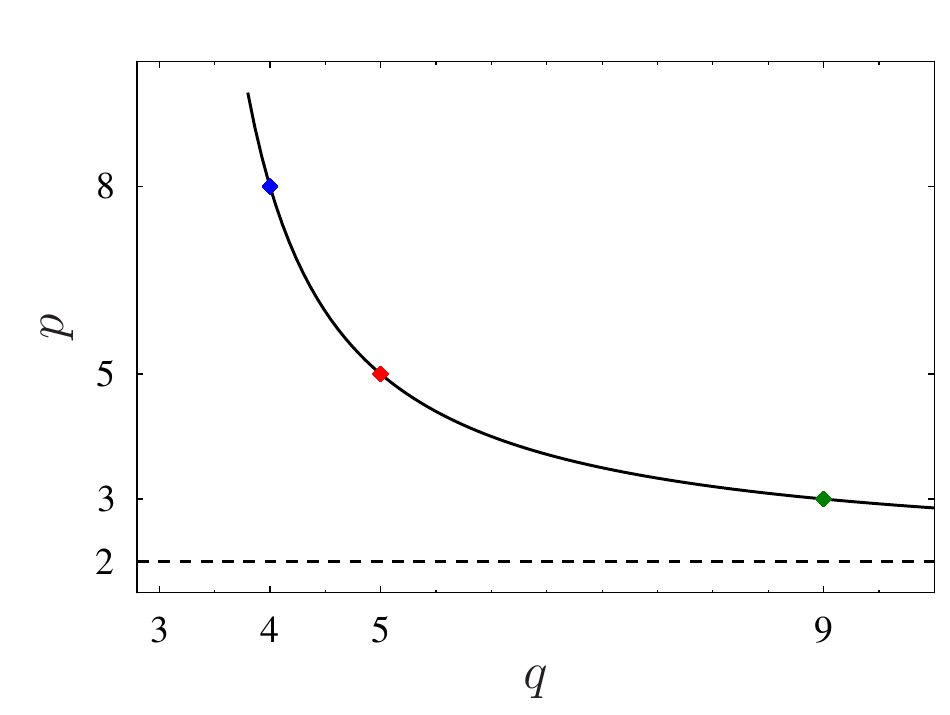}\label{fig:LPSq}}
\subfigure[]{\includegraphics[width=0.49\textwidth]{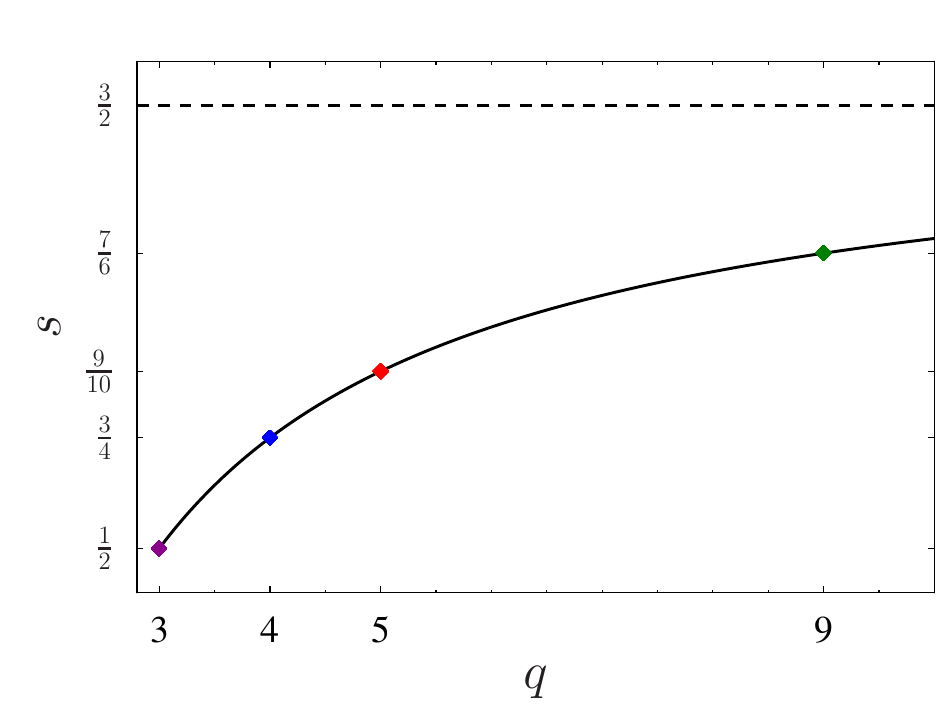}\label{fig:LPSs}}
}
\caption{
Dependence of \subref{fig:LPSq} $p$ in \eqref{eq:LPS} and 
\subref{fig:LPSs} $s$ in \eqref{eq:SobEmbb} on the index $q$ with solid 
symbols representing the values considered in this essay. The dashed 
horizontal lines correspond to the limiting values of $p$ and $s$ obtained 
when $q \rightarrow \infty$.
}
\label{fig:LPSq_s}
\end{figure}

Another important conditional regularity result is the family
of the Ladyzhenskaya-Prodi-Serrin conditions asserting that
Navier-Stokes flows $\u(t)$ are smooth and satisfy system
\eqref{eq:NS} in the classical sense provided that
\citep{KisLad57,Prodi1959,Serrin1962}
\begin{equation}
\u \in L^p([0,T];L^q(\Omega)), \quad 2/p+3/q = 1, \quad q > 3.
\label{eq:LPS}
\end{equation}
We thus have $p = 2q / (3-q)$ and the values of this exponent are shown 
as a function of $q$ in figure \ref{fig:LPSq_s}a. These conditions were 
generalized by \citet{Gibbon2018} to include norms of the 
derivatives of the velocity field. As regards the limiting case with $q 
= 3$, the corresponding condition was established by
\cite{Escauriaza2003}
\begin{equation}
\u \in L^{\infty}([0,T];L^3(\Omega)).
\label{eq:LPS3}
\end{equation}
Condition \eqref{eq:LPS} implies that should a singularity form in a 
classical solution $\u(t)$ of the Navier-Stokes system \eqref{eq:NS} at 
some finite time $0 < t_0 < \infty$, then necessarily
\begin{equation}
\lim_{t \rightarrow t_0} \int_0^t \| \u(\tau) \|_{L^q}^p \, d\tau = \infty, \quad 2/p+3/q = 1, \quad q > 3.
\label{eq:LPSblowup}
\end{equation}
At the same time, the time evolution of the solution norm $\| \u(t) 
\|_{L^q(\Omega)}$ on the time interval $[0,T]$ is subject to a priori 
bounds valid also for Leray-Hopf weak solutions, which might involve 
singularities.  An estimate of this type was known earlier and was 
rederived with an upper bound explicitly depending on the initial data 
by \cite{KangProtas2021}
\begin{equation}
\int_0^T \| \u(\tau) \|_{L^q}^{\frac{4q}{3(q-2)}} \, d\tau \le  C \, \K_0^{\frac{2q}{3(q-2)}}, \qquad 2 \le q \le 6.
\label{eq:LPSbound}
\end{equation}
We note that the integrals in \eqref{eq:LPSblowup} and 
\eqref{eq:LPSbound} differ in the exponent in the integrand expressions 
which is smaller in the latter case.

Condition \eqref{eq:LPSblowup} implies that, if blow-up occurs at $t = 
t_{0}$, then $\lim_{t \rightarrow t_{0}} \| \u(t) \|_{L^{q}} = \infty$. 
However, $\| \u(t) \|_{L^{q}}$ cannot diverge too rapidly as $t_{0}$ is 
approached since in that case the integral in \eqref{eq:LPSblowup} 
would remain finite (in other words, it would exist as an improper 
integral). Inequalities \eqref{eq:Kt} and \eqref{eq:LPSbound} provide 
further constraints on how a hypothetical blow-up could occur as 
described by \eqref{eq:Etblowup} and \eqref{eq:LPSblowup}. These 
relations show that potential singularity formation in Navier-Stokes 
flows is a subtle phenomenon and the range of admissible scenarios in 
terms of the rate of divergence of the quantities of interest is 
relatively narrow.

\subsubsection{Instantaneous and Finite-Time Bounds}
\label{sec:boundsNS}

In order to obtain insights about realizability of the enstrophy condition 
\eqref{eq:supE}, we assume here the Navier-Stokes system \eqref{eq:NS} 
admits a smooth classical solution $\u(t)$ for times $t \in [0,T]$, 
where $T$ is sufficiently small, which is guaranteed by local existence 
theorems \citep{RobinsonRodrigoSadowski2016}. We then consider the 
equation for the evolution of the enstrophy \eqref{eq:E} obtained 
multiplying \eqref{eq:NSa} by $\Delta\u$, integrating over $\Omega$ 
and performing integrations by parts (these operations are justified 
for $t \in [0,T]$ since the solution $\u(t)$ is smooth there)
\begin{equation}
\frac{d\E(\u(t))}{dt} = -\nu\int_\Omega|\Delta\u|^2\,d\x + \int_{\Omega} \u\cdot\nabla\u\cdot\Delta\u\, d\x =: \R_{\E_{0}}(\u).
\label{eq:dEdt} 
\end{equation}
As shown by \cite{ld08}, by following steps analogous to 
\eqref{eq:cubic1D}--\eqref{eq:dEdt1D}, relation \eqref{eq:dEdt} can be 
used to obtain the following upper bound on the rate of growth of 
enstrophy
\begin{equation}
\frac{d\E}{dt} \leq \frac{27}{8\,\pi^4\,\nu^3} \E^3. 
\label{eq:dEdt_estimate_E}
\end{equation} 
By simply integrating this differential inequality in time, we obtain 
the finite-time bound
\begin{equation}
\E(\u(t)) \leq \frac{\E_0}{\sqrt{1 - \frac{27}{4\,\pi^4\,\nu^3}\,\E_0^2\, t}}, 
\label{eq:Et_estimate_E0}
\end{equation}
which becomes infinite at time $t_0 = 4\,\pi^4\,\nu^3 / (27\, \E_0^2)$. 
Thus, based on inequality \eqref{eq:Et_estimate_E0}, which is the best 
estimate of this type available to date, it is not possible to establish 
the boundedness of the enstrophy $\E(\u(t))$ required in condition 
\eqref{eq:supE} and hence also the regularity of Navier-Stokes flows globally 
in time. However, boundedness of enstrophy and hence existence of smooth 
solutions can be established for arbitrarily long times provided the 
initial data $\u_0$ is ``small'', more precisely, when $\K_0 \E_0 = 
\O(\nu^4)$ \citep{ld08}. On the other hand, assuming that in an extreme 
flow we have $(d/dt)\E(t) \sim \E(t)^{\alpha}$  and invoking the same 
argument  based on the energy equation and Gr{\"o}nwall's lemma as in 
\S\,\ref{sec:Burgers}, cf.~\eqref{eq:Kt}--\eqref{eq:maxEtG}, we 
conclude that blow-up can occur only when $\alpha \in (2,3]$ and this 
amplification rate is sustained sufficiently long (longer for smaller 
$\alpha$). These considerations will be useful in analyzing how close 
the extreme Navier-Stokes flows discussed in \S\,\ref{sec:resultsNS} 
come to actually forming a singularity.

In a similar vein, the Ladyzhenskaya-Prodi-Serrin condition
\eqref{eq:LPSblowup} can be studied by considering the rate of growth
of the $L^q$ norm of the velocity field, for which an upper bound was
already known to Leray, see also \citet{RobinsonRodrigoSadowski2016},
\begin{equation}
\frac{1}{q} \frac{d}{dt} \| \u(t) \|_{L^q}^q \le  C \nu^{\frac{q+3}{q-3}} \| \u(t) \|_{L^q}^{p+1} 
= C \nu^{\frac{q+3}{q-3}} \| \u(t) \|_{L^q}^{\frac{q(q-1)}{q-3}} 
, \qquad q > 3.
\label{eq:dLqdt}
\end{equation}
We note that the exponent on the RHS becomes unbounded as $q 
\rightarrow 3^{+}$ and we are not aware of any estimate of this type 
applicable when $q = 3$. Integrating \eqref{eq:dLqdt} with respect to 
time, we obtain
\begin{equation}
\|\u(t)\|_{L^q}\leq \frac{1}{\left(\|\u_{0}\|_{L^q}^{-p}-p\,C\,t\right)^{1/p}},
\label{eq:tLPS}
\end{equation}
which becomes unbounded as $t \rightarrow t_{0} = 1 / 
(p\,C\,\|\u_{0}\|_{L^q}^{p})$. Together with \eqref{eq:LPS}, this 
represents a local existence result akin to \eqref{eq:Et_estimate_E0}. 
In analogy to the argument invoked above, we add that if, 
hypothetically, the norm $\|\u(t)\|_{L^{q}}$ were to be amplified at a 
rate faster than the RHS in \eqref{eq:dLqdt}, i.e., with 
$(d/dt)\|\u(t)\|_{L^{q}} \sim \|\u(t)\|_{L^{q}}^{\alpha}$ and some 
$\alpha > (q-1) / (q - 3)$, then $\|\u(t)\|_{L^{q}}$  would become 
unbounded without \eqref{eq:LPSblowup} taking place. Thus, somewhat 
counterintuitively, for a blow-up to occur in a Navier-Stokes flow, the 
norm $\|\u(t)\|_{L^{q}}$  must not grow too rapidly. Leveraging a 
priori estimates \eqref{eq:LPSbound} and Gr{\"o}nwall's lemma 
\eqref{eq:Gro}, as was done in \S\,\ref{sec:Burgers}, 
cf.~\eqref{eq:Kt}--\eqref{eq:maxEtG}, and above for the enstrophy 
condition, we arrive at upper bounds on the rate of growth of the 
$L^{q}$ norm that ensure there is no blow-up in finite time
\begin{equation}
\frac{d}{dt}\|\u(t)\|_{L^q}\leq
\begin{cases}
C\|\u(t)\|_{L^q}^{\frac{7q-6}{3(q-2)}}, \qquad & 2\leq q \leq6, \\
\\
C\|\u(t)\|_{L^q}^{\frac{2q-3}{q-3}}, & q>6. \\   
\end{cases} 
\label{eq:noblowup_ub}
\end{equation}
Together with \eqref{eq:dLqdt}, these inequalities demarcate the range 
of growth rates given in terms of the exponent of $\|\u(t)\|_{L^{q}}$ 
consistent with singularity formation in finite time. These 
considerations will be useful in analyzing how close the extreme 
Navier-Stokes flows discussed in \S\,\ref{sec:resultsNS} come to 
actually forming a singularity.

Since there are no a priori bounds on the enstrophy $\E(T)$ and the 
integral $\int_0^T \| \u(\tau) \|_{L^q(\Omega)}^p \, d\tau$ valid for 
arbitrarily large times $T$, we would like to know how much these 
quantities can grow in Navier-Stokes flows under the worst-case 
scenarios and, in particular, if they can become unbounded in finite 
time as this would signal singularity formation, 
cf.~\eqref{eq:Etblowup} and \eqref{eq:LPSblowup}. Since the rates of 
growth of these quantities are subject to estimates 
\eqref{eq:dEdt_estimate_E} and \eqref{eq:dLqdt}, as a first step one 
would like to verify whether or not these estimates are sharp. If 
\eqref{eq:dEdt_estimate_E} is not sharp, then necessarily the 
corresponding finite-time bound \eqref{eq:Et_estimate_E0} must 
represent an overestimate; likewise for 
\eqref{eq:dLqdt}--\eqref{eq:tLPS}. These questions will be addressed by 
formulating and solving a number of variational optimization problems 
discussed below.

\subsubsection{Optimization Problems}
\label{sec:optNS}

We begin by considering the instantaneous bounds and the first 
optimization problem was formulated by \citet{ld08} to probe the 
sharpness of estimate \eqref{eq:dEdt_estimate_E}.
\begin{problem}
\label{pb:maxdEdt3D}
  Given $\E_0\in\RR_+$ and the objective functional $\R_{\E_{0}}(\u)$, cf.~\eqref{eq:dEdt},
find
\begin{align*}
\tuE  & =  \mathop{\arg\max}_{\u \in \mathcal{S}_{\E_0}} \, \R_{\E_{0}}(\u), \qquad \text{where} \\
\mathcal{S}_{\E_0} & :=  \left\{\u\in H^2(\Omega)\,\colon \; \bnabla\cdot\u =0, \ \int_{\Omega} \u \, d\x = \0, \ \E(\u) = \E_0 \right\},
\end{align*} 
\end{problem}
\noindent{}
where the functional setting, $\u \in H^{2}(\Omega)$, represents the 
minimum regularity of the argument needed for the functional to be well 
defined. This problem is the 3D counterpart of Problem 
\ref{pb:maxdEdt1D} considered in the context of the 1D Burgers system 
in \S\,\ref{sec:Burgers}.

In order to probe the sharpness of the companion bound 
\eqref{eq:dLqdt}, the following objective functional was introduced by 
\citet{BleitnerProtas2026}
\begin{align}
    \frac{1}{q}\frac{d}{dt}\|\u\|_q^q &= \int_\Omega |\u|^{q-2} \u\cdot \bnabla p \, d\x + \nu \int_{\Omega} |\u|^{q-2}\u\cdot \Delta \u \, d\x
    \label{sakfndsj}
    \\
    &= - (q-2)\int_\Omega |\u|^{q-4} \u\cdot (\u\cdot \bnabla) \u\, \Delta^{-1}(\bnabla \u\colon \bnabla \u^T)\, d\x \nonumber
    \\
    &\qquad - \nu  \int_{\Omega} |\u|^{q-2}|\bnabla \u|^2 \, d\x - \frac{4(q-2)\nu}{q^2}\int_{\Omega} \left|\bnabla |\u|^{\frac{q}{2}}\right|^2d\x \nonumber
    \\
    & =: \R_{B}^q(\u), \qquad q \ge 3.
    \label{eq:dLqdt2}    
\end{align}
It was obtained by testing \eqref{eq:NSa} with $|\u|^{q-2} \u$, using 
\eqref{eq:p} as well as the identity $\int_{\Omega} |\u|^{q-2} \u \cdot 
(\u\cdot \bnabla) \u \, d\x = 0$. As shown by 
\citet{BleitnerProtas2026}, it is well defined when $\u \in 
W^{1,\frac{3q}{q+1}}(\Omega)$. Since this Sobolev space is a general 
Banach space and is not equipped with the Hilbert structure 
representable in terms of an inner product, for reasons discussed in 
Appendix \ref{sec:grad}, solution of an optimization problem formulated 
in such a functional setting is complicated. More specifically, an 
inner product is required for the usual definition of the gradient of 
the objective functional. To get around this difficulty, 
\citet{BleitnerProtas2026} maximized functional \eqref{eq:dLqdt2} over 
the largest Sobolev space with the Hilbert structure $H^{s}(\Omega)$, 
cf.~\eqref{eq:Hs}, embedded in $W^{1,\frac{3q}{q+1}}(\Omega)$, as 
determined by the Sobolev embedding \citep{af05}
\begin{equation}
    H^{\frac{3}{2}-\frac{1}{q}}(\Omega) \hookrightarrow W^{1,\frac{3q}{q+1}}(\Omega), \qquad q \ge 3.
    \label{eq:HsWp}
\end{equation}
This then leads to the optimization problem
\begin{problem}
\label{pb:maxdLqdt}
  Given $q \ge 3$, $B\in\RR_+$ and the objective functional $\R_{B}^q(\u)$, cf.~\eqref{eq:dLqdt2},
find
\begin{align*}
\tuB  & =  \mathop{\arg\max}_{\u \in \mathcal{X}_{B}} \, \R_{B}^q(\u), \qquad \text{where}  \\
\mathcal{X}_{B}  & :=  \left\{\u\in H^{\frac{3}{2}-\frac{1}{q}}(\Omega) \,\colon \; \bnabla\cdot\u =0, \ \int_{\Omega} \u \, d\x = \0, \ \|\u\|_{L^{q}} = B \right\}.
\end{align*} 
\end{problem}
We will come back to the question of solving optimization problems in 
functional spaces without Hilbert structure below.
We now move on to state optimization problems aimed at probing 
conditions \eqref{eq:Etblowup} and \eqref{eq:LPSblowup} defined over 
finite time windows $[0,T]$. For fixed magnitudes of the initial data 
$\u_{0}$ and lengths $T$ of the time horizon, they define 
locally-optimal initial data for extreme Navier-Stokes flows evolving on these 
time windows. In regard to the enstrophy condition \eqref{eq:Etblowup}, 
we thus have 
\begin{problem}\label{pb:maxET}
  Given $\E_0, T \in\mathbb{R}_+$ and the objective functional $\E_{T}(\u_{0}) := \E(\u(T;\u_{0}))$, find
\begin{align*}
\tuET & =  \mathop{\arg\max}_{\u_0 \in {\Q}_{\E_0}} \, \E_{T}\left(\u_0\right), \quad \text{where} \\
{\Q}_{\E_0} & :=  \left\{\u_0\in H^1(\Omega)\,\colon\,\bnabla\cdot\u_0 = 0, \; \E(\u_0) = \E_0 \right\}
\end{align*} 
\end{problem}
\noindent{}
which was studied by \citet{KangYunProtas2020}. It is the 3D counterpart 
of Problem \ref{pb:maxET1D} considered in the context of the 1D Burgers 
system in \S\,\ref{sec:Burgers}.

Motivated by the Ladyzhenskaya-Prodi-Serrin conditions 
\eqref{eq:LPS}--\eqref{eq:LPS3}, \citet{RamirezProtas2026} defined the 
following two objective functionals (see also \citet{KangProtas2021})
\begin{align}
\Phi_T^{q}(\u_0)  & := \frac{1}{T} \int_0^T \| \u(\tau) \|_{L^q}^{p} \, d\tau, 
\qquad p = \frac{2q}{q-3}, \quad q > 3, \label{eq:Phi} \\
\Psi_T(\u_0)  & :=  \| \u(T) \|_{L^3}^{3} \label{eq:Psi}
\end{align}
for a given $T > 0$. The first one coincides with the integral in 
\eqref{eq:LPSblowup}, except for the prefactor $T^{-1}$ which offsets 
the increase of the integral as the interval $[0,T]$ becomes large. 
Definition \eqref{eq:Psi} is motivated  by the observation that the 
norm $\| \cdot \|_{L^{\infty}}$ in \eqref{eq:LPS3} is not a 
differentiable function of its argument, cf.~\eqref{eq:Lq}, hence its 
use in the objective functional would result in a non-smooth 
optimization problem that would be much harder to solve. This 
difficulty is circumvented by considering $\| \u(t) \|_{L^3}$ at 
different fixed times $t = T$, rather than over the interval $[0,T]$, 
in the objective functional \eqref{eq:Psi}.

The form of the functionals \eqref{eq:Phi}--\eqref{eq:Psi} 
involving Lebesgue norms $\| \u(t) \|_{L^q}$, $q \ge 3$, suggests that 
optimization should be performed over these spaces. However, as already 
discussed in the context of Problem \ref{pb:maxdLqdt}, the difficulty is 
that these spaces are not endowed with the Hilbert structure which is 
the preferred setting for PDE optimization problems, 
cf.~Appendix \ref{sec:grad}.  Therefore, for each of the two 
objective functionals \eqref{eq:Phi}--\eqref{eq:Psi}, 
\citet{RamirezProtas2026} considered two distinct formulations of the 
optimization problem:
\begin{itemize}
\item in the first, optimization is performed in the largest Sobolev 
space with the Hilbert structure $H^{s}(\Omega)$ embedded in the given 
Lebesgue space $L^q(\Omega)$, which lends itself to solution 
using standard methods of numerical PDE optimization,
\item in the second, maximization is performed directly in the 
Lebesgue space $L^q(\Omega)$ leading to variational optimization  
problems with a nonstandard structure; in particular, in the absence of 
an inner product, this necessitates the introduction of a different 
notion of the gradient of the objective functional, namely, the metric 
gradient.
\end{itemize}
In regard to the first formulation, the relevant Hilbert-Sobolev space 
$H^{s}(\Omega)$  is determined based on the Sobolev embedding theorem 
in 3D \citep{af05}
\begin{equation}
 H^{s}(\Omega)\hookrightarrow L^{q}(\Omega), \quad \mbox{if}\quad s\ge\frac{3}{2}-\frac{3}{q}. 
 \label{eq:SobEmbb}
\end{equation}
In other words,  $H^{s}(\Omega)$ with $s = 3 / 2 - 3 / q$ is the 
"largest" $L^{2}$-based Sobolev space embedded in $L^{q}(\Omega)$ and 
hence serves as a proxy for the letter in the first formulation. The 
values of the index $s$ in \eqref{eq:SobEmbb} are shown as function of 
$q$ in Figure \ref{fig:LPSq_s}b.

This thus leads to the following four optimization problems:
\begin{problem}\label{pb:PhiHs}
 Given $B, T \in\mathbb{R}_+$, $q>3$, $s=3/2-3/q$ and the objective functional $\Phi_T^q(\u_0)$ from
 \eqref{eq:Phi}, find
\begin{align}
\tuBT & =  \mathop{\arg\max}_{\u_0 \in {\M}_{B}} \, \Phi_T^q(\u_0), \quad \text{where} \\
 {\M}_{B} & :=  \left\{\u_0\in H^{s}(\Omega)\,\colon\,\bnabla\cdot\u_0 = 0, \; \int_{\Omega} \u_0 \, d\x = 0,  \; \|\u_0\|_{L^q} = B \right\}.\label{manifold:pb1}
\end{align} 
\end{problem}
\begin{problem}\label{pb:PhiLq}
Given $B, T \in\mathbb{R}_+$, $q>3$ and the objective functional $\Phi_T^q(\u_0)$ from
\eqref{eq:Phi}, find
\begin{align}
\tuBT & =  \mathop{\arg\max}_{\u_0 \in \mathcal{L}_{B}} \, \Phi_T^q(\u_0), \quad \text{where} \\
 {\mathcal{L}}_{B} & :=  \left\{\u_0\in L^q(\Omega)\,\colon\,\bnabla\cdot\u_0 = 0, \; \int_{\Omega} \u_0 \, d\x = 0,  \; \|\u_0\|_{L^q} = B \right\}. \label{manifold:pb2}
\end{align} 
\end{problem}
\begin{problem}\label{pb:PsiHs}
Given $B, T \in\mathbb{R}_+$ and the objective functional $\Psi_T(\u_0)$ from
\eqref{eq:Psi}, find
\begin{align}
\tuBT & =  \mathop{\arg\max}_{\u_0 \in {\N}_{B}} \, \Psi_T(\u_0), \quad \text{where} \\
 {\N}_{B} & :=  \left\{\u_0\in H^{1/2}(\Omega)\,\colon\,\bnabla\cdot\u_0 = 0, \; \int_{\Omega} \u_0 \, d\x = 0,  \; \|\u_0\|_{L^3} = B \right\}.
\end{align} 
\end{problem}
\begin{problem}\label{pb:PsiLq}
Given $B, T \in\mathbb{R}_+$ and the objective functional $\Psi_T(\u_0)$ from
\eqref{eq:Psi}, find
\begin{align}
\tuBT & =  \mathop{\arg\max}_{\u_0 \in {\mathcal{S}}_{B}} \, \Psi_T(\u_0), \quad \text{where} \\
 {\mathcal{Q}}_{B} & :=  \left\{\u_0\in L^3(\Omega)\,\colon\,\bnabla\cdot\u_0 = 0, \; \int_{\Omega} \u_0 \, d\x = 0,  \; \|\u_0\|_{L^3} = B \right\}.
\end{align} 
\end{problem}
\noindent Problems \ref{pb:PhiHs} and \ref{pb:PsiHs} represent the 
first formulation mentioned above, whereas  Problems \ref{pb:PhiLq} and 
\ref{pb:PsiLq} represent the second. The reason why we consider 
Problems \ref{pb:PhiLq} and \ref{pb:PsiLq}, while 
\citet{BleitnerProtas2026} did not maximize functional $\R_{B}^{q}(\u)$ 
over the space $W^{1,\frac{3q}{q+1}}(\Omega)$ is because the presence 
of derivatives in the definition of this function space makes 
computation of the gradient with respect to this topology intractable. 
On the other hand, gradients in the Lebesgue spaces $L^{q}(\Omega)$, $q 
\ge 3$, can be computed reliably \citep{RamirezProtas2026}.

Problems \ref{pb:maxdEdt3D}--\ref{pb:PsiLq} all share the property of 
being Riemannian, in the sense that their local maximizers are sought 
over constraint manifolds with a Riemannian structure. A 
state-of-the-art approach to the numerical solution of such problems 
which exploits this structure is presented in Appendix 
\ref{sec:solution}. Insights about the bounds reviewed in 
\S\;\ref{sec:boundsNS} obtained by solving these problems are discussed 
next.

\subsubsection{Results}
\label{sec:resultsNS} 

\begin{figure}
\begin{center}
\subfigure[$|\k|^2 = 1$]{\includegraphics[width=0.32\textwidth]{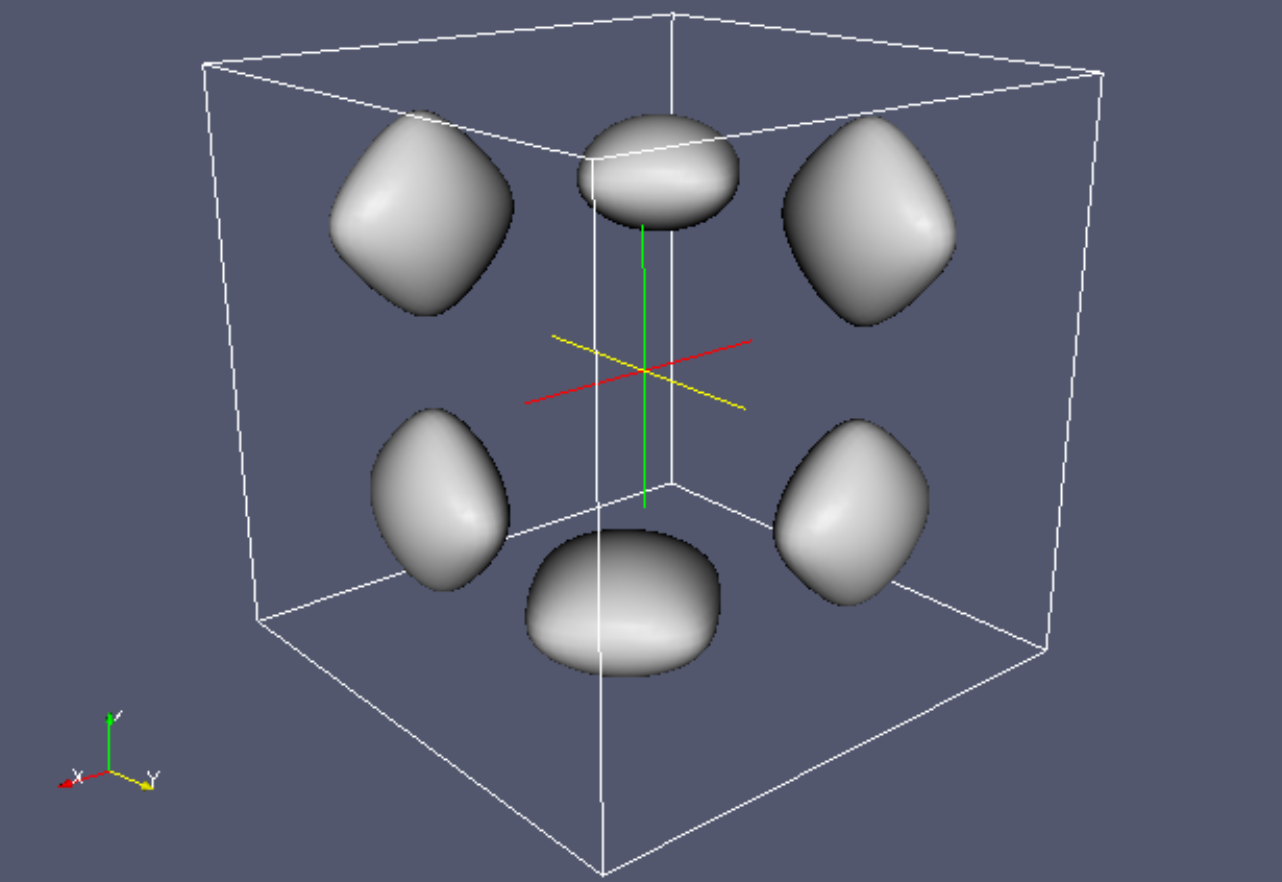}}
\subfigure[$|\k|^2 = 2$]{\includegraphics[width=0.32\textwidth]{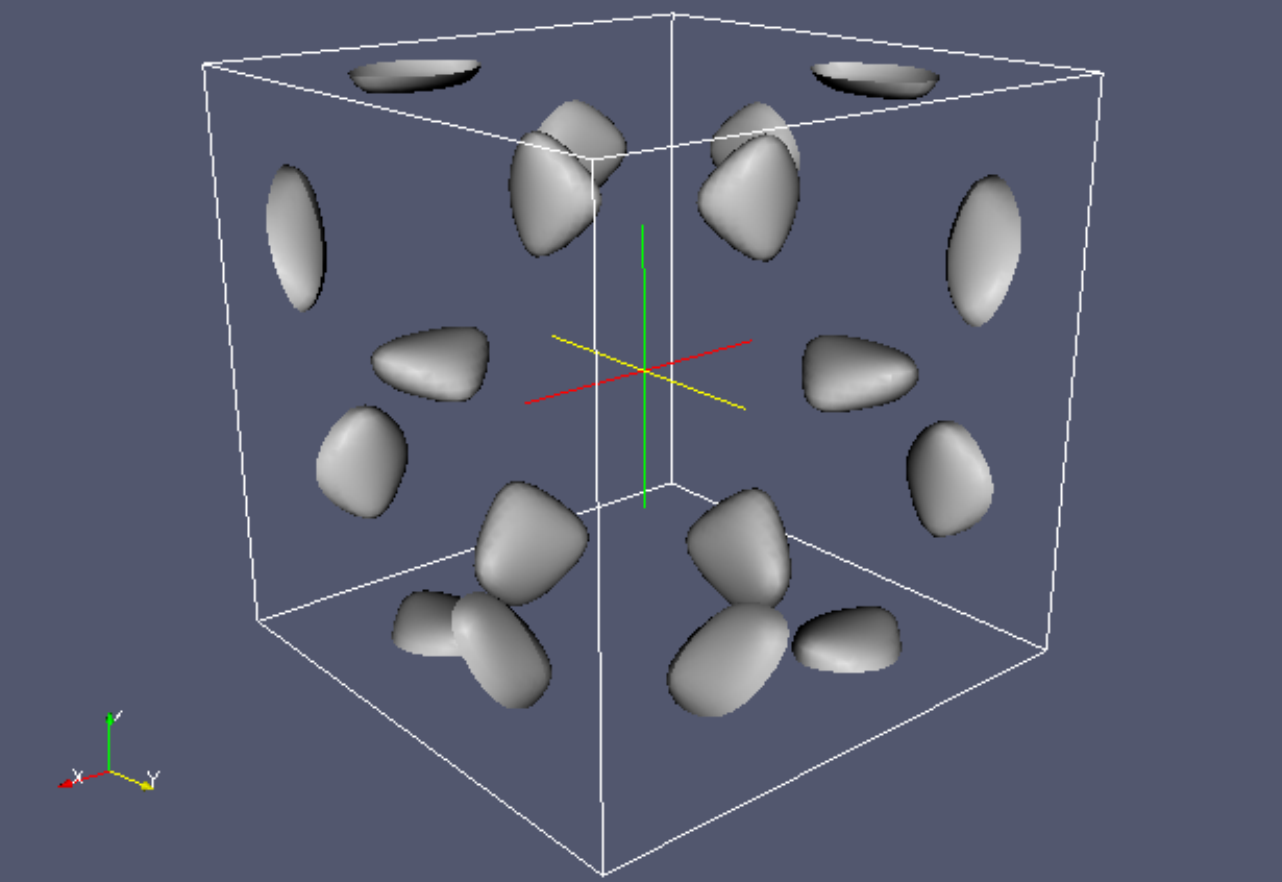}}
\subfigure[$|\k|^2 = 3$]{\includegraphics[width=0.32\textwidth]{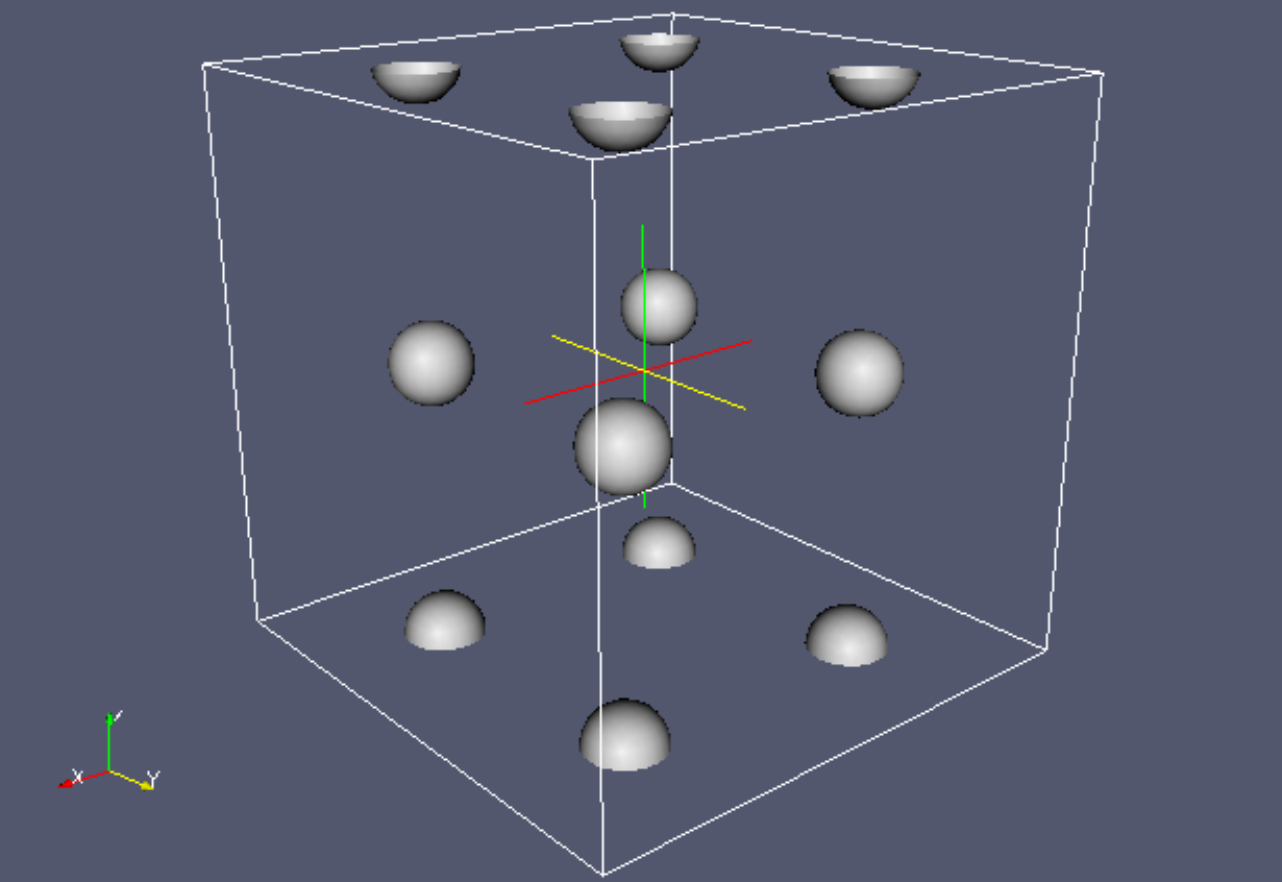}}
\subfigure[staggered ABC flow]{\includegraphics[width=0.32\textwidth]{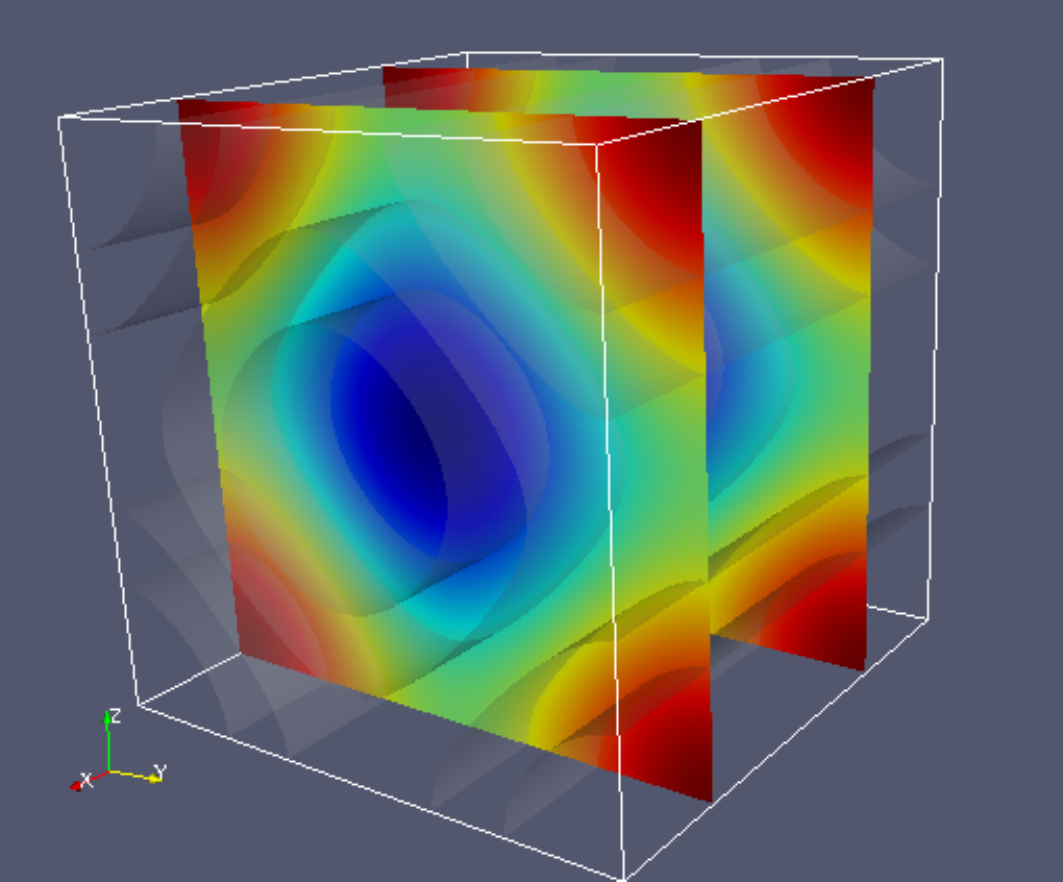}}
\subfigure[aligned ABC flow]{\includegraphics[width=0.32\textwidth]{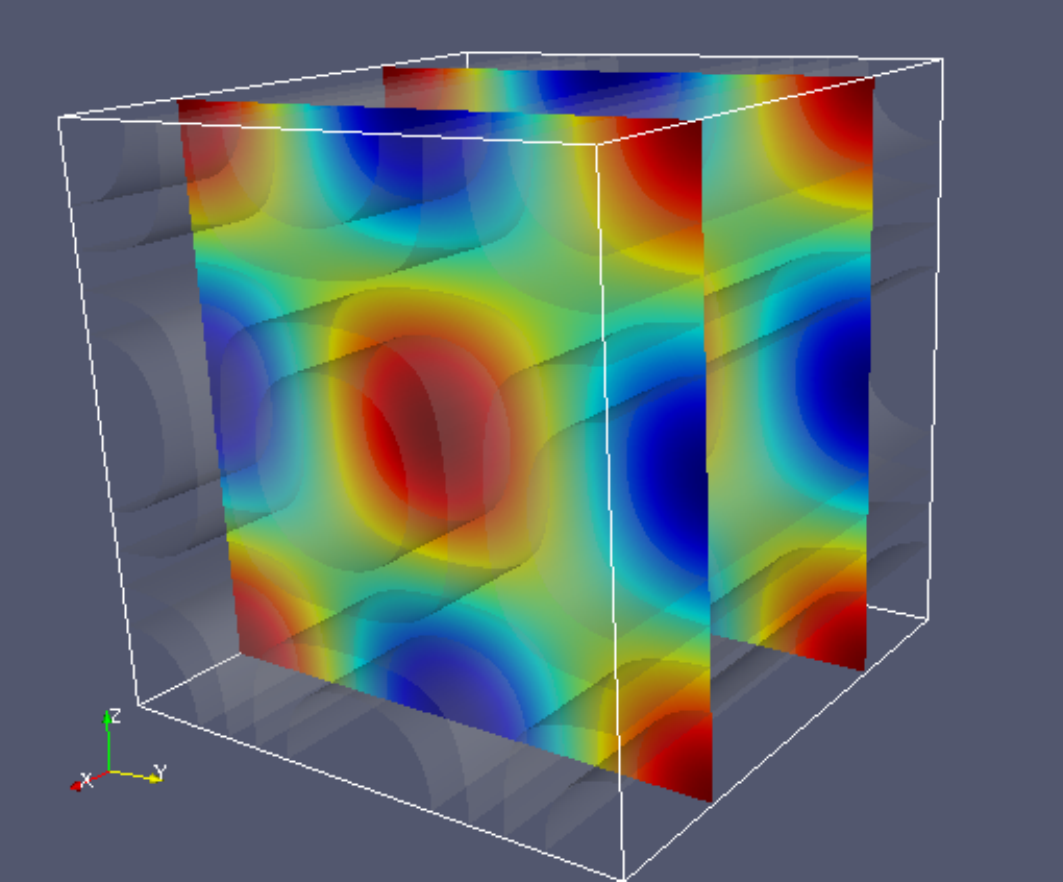}}
\subfigure[Taylor-Green flow]{\includegraphics[width=0.32\textwidth]{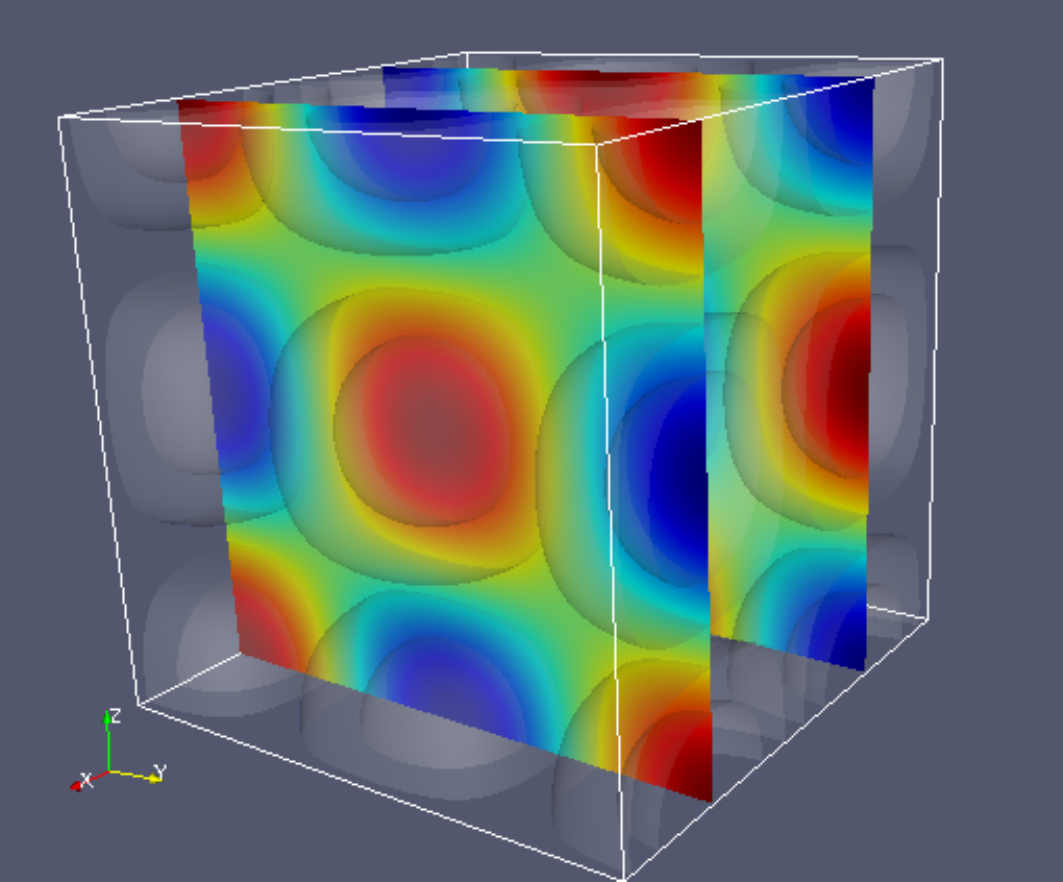}}
\caption{Local maximizers of Problem \eqref{pb:maxdEdt3D} in the limit 
 $\E_0 \to 0$  given by the eigenfunctions $\u_{1}$ of \eqref{eq:u1} 
 with (a,d) $|\k|^2 = 1$, (b,e) $|\k|^2 = 2$ and (c,f) $|\k|^2 = 3$. 
 Figures (a--c) represent the isosurfaces defined by the the relation 
 $|\bnabla\times\u_1|(\x) =  0.95||\bnabla\times\u_1||_{L^\infty}$, 
 whereas figures (d--f) depict the isosurfaces and cross-sectional 
 distributions in the $y-z$ plane of the $x_1$ component of the field $\u_1$ 
 \citep{ap16}.}
\label{fig:maxdEdt_vortexCells}
\end{center}
\end{figure}

We begin by reviewing the results obtained by solving the instantaneous 
optimization problems. Problem \ref{pb:maxdEdt3D} was solved by 
\citet{ld08} and later revisited by \citet{ap16}, whereas Problem 
\ref{pb:maxdLqdt} was recently studied by \citet{BleitnerProtas2026}. 
In both cases, the main interest was to solve these problems as the 
constraint parameters, respectively, $\E_{0}$ and $B$, increase in 
order to probe the sharpness of estimates \eqref{eq:dEdt_estimate_E} 
and \eqref{eq:dLqdt}. On the other hand, Problems \ref{pb:maxdEdt3D} 
and \ref{pb:maxdLqdt} are analytically solvable in the small-data 
limit, i.e., as $\E_{0} \rightarrow 0$ and $B \rightarrow 0$, using  
asymptotic techniques \citep{ap16}. To fix attention, we focus here on 
the first problem and, to simplify the notation, will drop the 
subscript $\E_0$ when referring to the optimal field. The 
Euler-Lagrange system representing the first-order optimality 
conditions for Problem \ref{pb:maxdEdt3D} is given by \citep{l69}
\begin{subequations}\label{eq:KKT_E}
\begin{align} 
\B(\tu,\tu) - 2\nu\Delta^2\tu - \lambda\Delta\tu - \bnabla q & = 0 \qquad\mbox{in}\,\,\Omega , \label{eq:KKT_E_gradR}\\
\nabla\cdot\tu & = 0 \qquad\mbox{in}\,\,\Omega , \label{eq:KKT_E_divConstr}\\
\E(\tu) - \E_0 & = 0, \label{eq:KKT_E_E0Constr}
\end{align}
\end{subequations}
where $\lambda\in\mathbb{R}$ and $q:\Omega\to\mathbb{R}$ are the Lagrange 
multipliers associated with the constraints defining the manifold 
$\mathcal{S}_{\E_0}$ and $\B(\u,\v)$ is a bilinear form given by
\begin{displaymath}
\B(\u,\v) :=  \Delta\left( \u\cdot\bnabla\v \right) + (\bnabla\u)^T\Delta\v - 
\u\cdot\bnabla(\Delta\v).
\end{displaymath}
Using the formal series expansions with $\beta > 0$
\begin{subequations}\label{eq:series3D}
\begin{align}
\tu & = \u_0 + \E_0^{\beta}\u_1 + \E_0^{2\beta}\u_2 + \ldots, \\
\lambda & = \lambda_0 + \E_0^{\beta}\lambda_1 + \E_0^{2\beta}\lambda_2 + \ldots, \\
q & = q_0 + \E_0^{\beta}q_1 + \E_0^{2\beta}q_2 + \ldots
\end{align} 
\end{subequations}
in \eqref{eq:KKT_E} and collecting terms proportional to different 
powers of $\E_0^{\beta}$, we obtain at the leading order $\u_0 \equiv 
0$, $\beta = 1/2$ and $\E(\u_1) = 1$. The first-order correction then 
satisfies the eigenvalue problem \citep{ap16}
\begin{equation}
\label{eq:u1}
2\nu\Delta\u_1 + \lambda_0\u_1 = 0, \qquad \bnabla\cdot\u_{1} = 0
\end{equation}
with $q_1 \equiv 0$. Using expansions \eqref{eq:series3D} and relation
\eqref{eq:u1} in \eqref{eq:dEdt} gives the approximation
\begin{equation}\label{eq:R03D}
\R_{\E_{0}}(\tu) = - \nu\E_0\int_{\Omega} \left| \Delta \u_1 \right|^2 \, d\x + \O(\E_0^{3/2}) \approx - \lambda_0\E_0
\end{equation}
valid for $\E_{0} \rightarrow 0$. The eigenfunctions of problem 
\eqref{eq:u1} corresponding to the eigenvalues $\lambda_0/(2 \nu) = 
(2\pi)^2|\k|^2$ with $|\k| =: k = 1,2,3$ are visualized in figure 
\ref{fig:maxdEdt_vortexCells}. They all reveal a cellular structure 
and, as discussed by \citet{ap16}, correspond to the well-known 
Arnold-Beltrami-Childress (ABC) flows and the Taylor-Green vortex 
\citep{mb02}. We remark that the Taylor-Green vortex has been employed 
as the initial data in a number of studies aimed at triggering singular 
behavior in Euler flows \citep{bmonmu83,bb12}. It is therefore 
interesting that it arises in the variational formulation considered 
here as a local maximizer of functional \eqref{eq:dEdt} in the limit 
$\E_0 \rightarrow 0$. Eigenfunctions of \eqref{eq:u1} are also local 
maximizers of Problem \ref{pb:maxdLqdt} in the limit $B \rightarrow 0$. 
However, this problem has a richer structure and in fact any 
divergence-free vector field $\u_{1}$ is also a critical point of 
functional \eqref{eq:dLqdt2} for $B \rightarrow 0$ 
\citep{BleitnerProtas2026}. 

\begin{figure}
\linespread{1.1}
\setcounter{subfigure}{0}
\begin{center}
\Bmp{\textwidth}
\subfigure[]{\includegraphics[width=0.48\textwidth]{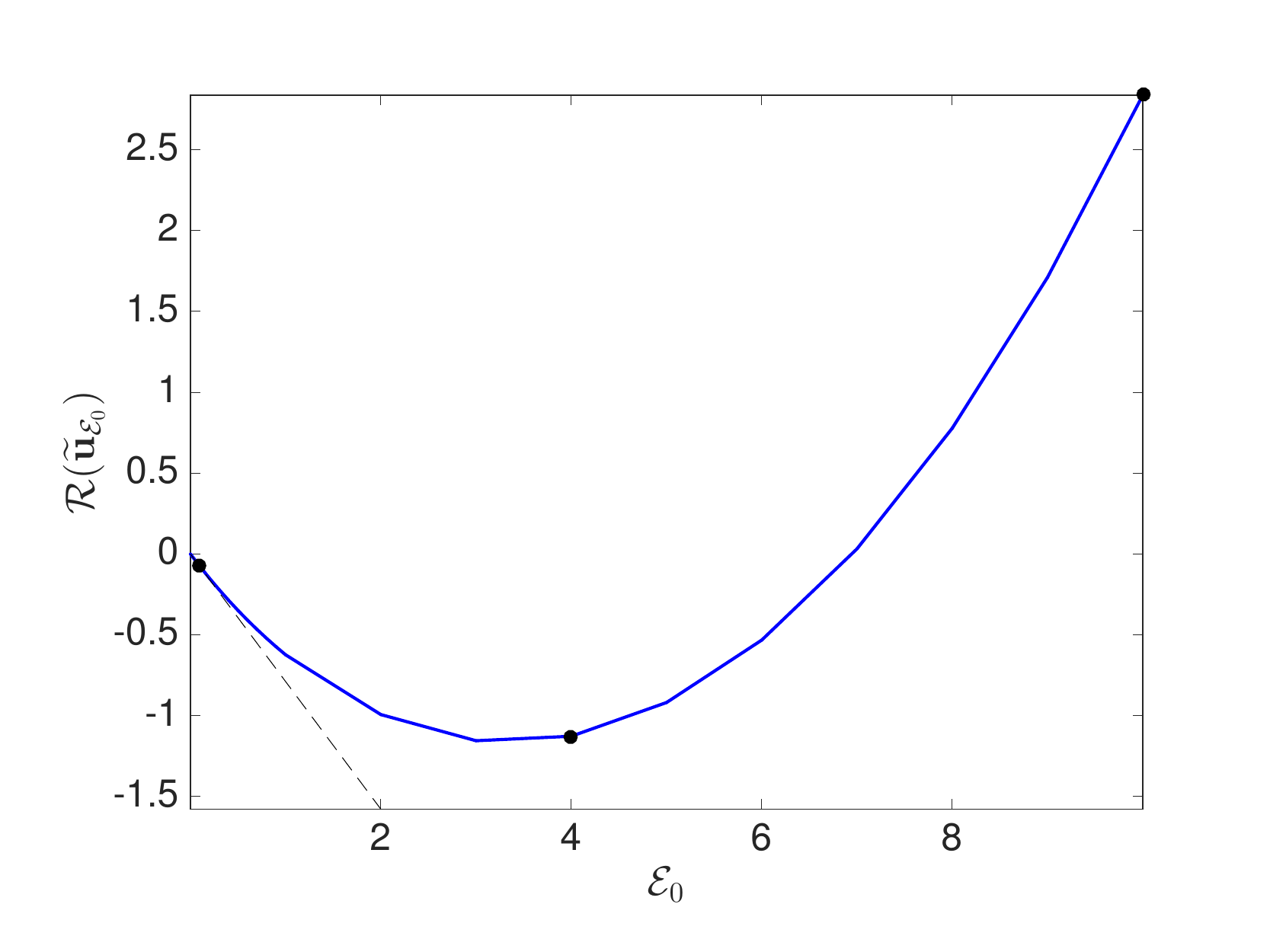}}
\subfigure[]{\includegraphics[width=0.47\textwidth]{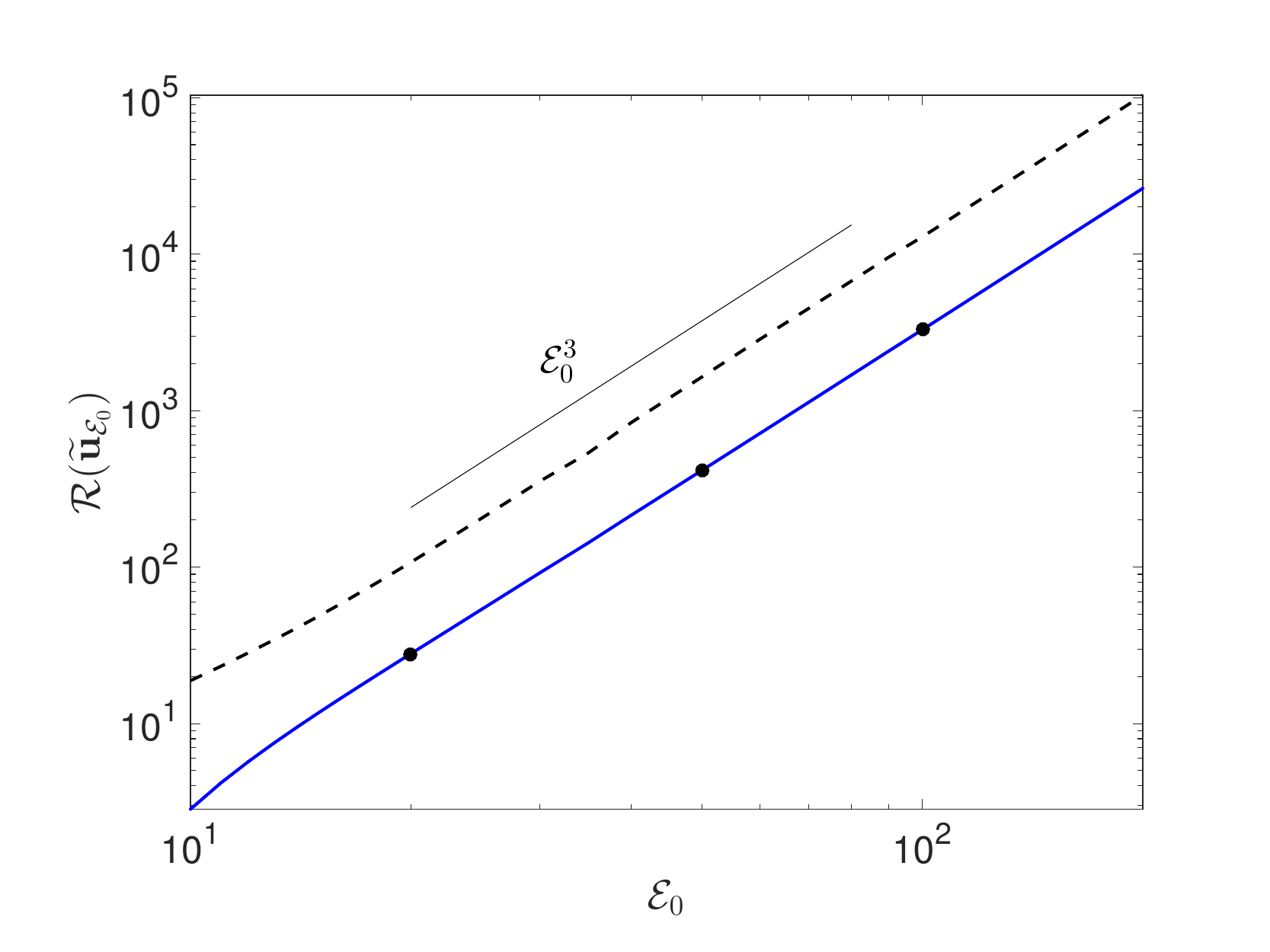}}
\vspace*{-0,3cm}
\Emp
\subfigure[]{\includegraphics[width=0.49\textwidth]{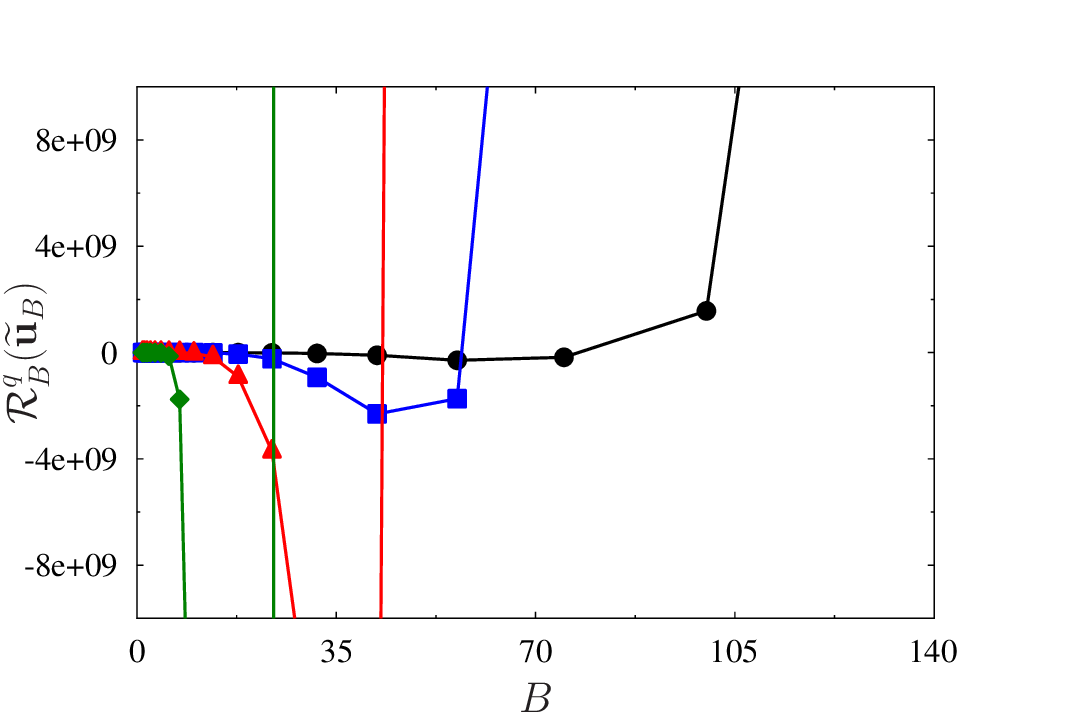}}
\subfigure[]{\includegraphics[width=0.49\textwidth]{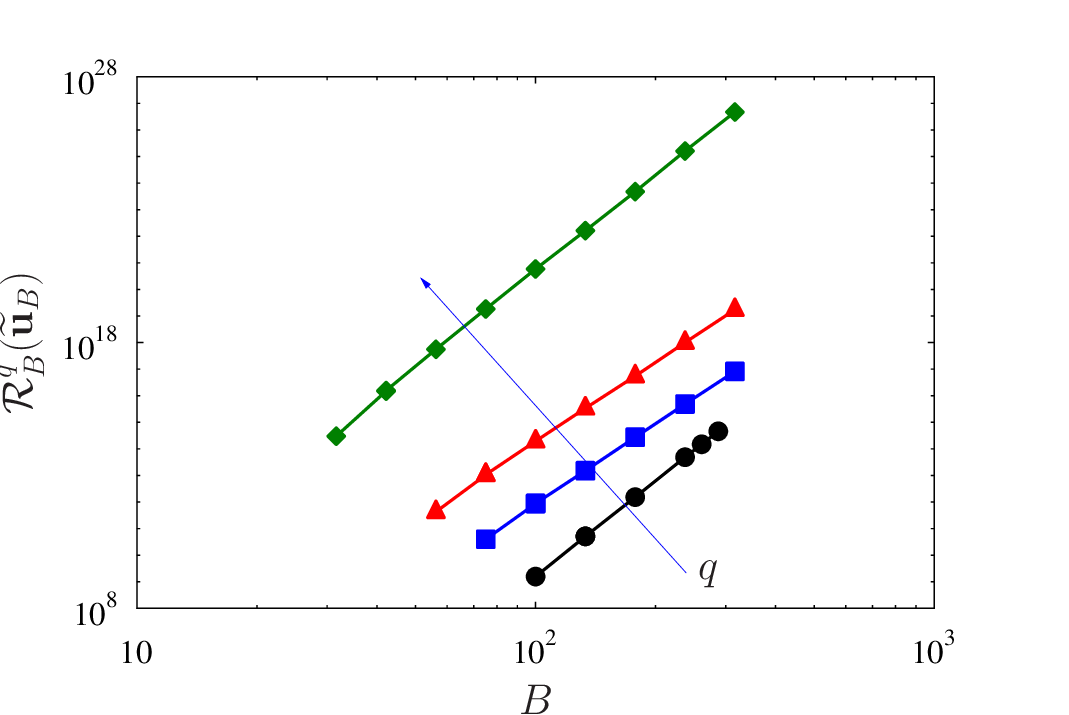}}
\caption{
(a,b) The dependence of the maximum rate of growth of enstrophy 
  $\R_{\E_{0}}(\tuE)$ on $\E_{0}$ for (a) small $\E_{0}$ and (b) large 
  $\E_{0}$ in solutions of Problem \ref{pb:maxdEdt3D}; the dashed lines 
  in panels (a) and (b) represent, respectively, 
  the asymptotic relation \eqref{eq:R03D} and the RHS in estimate 
  \eqref{eq:dEdt_estimate_E}. (c,d)  The dependence of the maximum rate 
  of growth of the $L^{q}$ norm $\R_{B}^{q}(\tuB)$ on $B$ for (c) 
  small $B$ and (d) large $B$ in solutions of Problem 
\ref{pb:maxdLqdt}; (black circles) $q = 4$, (blue squares) 
  $q = 5$, (red triangles) $q = 6$,  (green diamonds) $q = 9$; the 
  arrow in panel (d) indicates the tend with the increase of $q$.  
}
\label{fig:RvsE0}
\end{center}
\end{figure}

In analogy with the terminology employed in \S\,\ref{sec:noanomaly}, 
here "branches" refer to families of solutions $\tuE$ and $\tuB$ of 
Problems \ref{pb:maxdEdt3D} and \ref{pb:maxdLqdt} parameterized by, 
respectively, $\E_{0}$ and $B$. The local maximizers obtained in the 
small-data limit are useful since they serve as "seeds" of such 
branches as they are computed for finite values of $\E_{0}$ and $B$ 
using the continuation approach described in Appendix \ref{sec:cont}. 
These branches are illustrated in figures \ref{fig:RvsE0}a,b and 
\ref{fig:RvsE0}c,d where we show the dependence of, respectively, 
$\R_{\E_{0}}(\tuE)$ on $\E_{0}$ and $\R_{B}^{q}(\tuB)$ on $B$ for small 
and large values of the constraint parameters; for Problem 
\ref{pb:maxdLqdt} we focus on $q = 4,5,6,9$. The values of the 
viscosity coefficient used  by \citet{ld08,ap16} in the solution of 
Problem \ref{pb:maxdEdt3D}  and  by \citet{BleitnerProtas2026} in the 
solution of Problem \ref{pb:maxdLqdt} were different and, respectively, 
$\nu = 0.01$ and $\nu = 1$. This discrepancy, however, does not pose 
problems here since these results are not directly compared and are 
used to probe the sharpness of different estimates. The objective 
functionals \eqref{eq:dEdt} and \eqref{eq:dLqdt2} involve a balance 
between negative-definite terms representing viscous (dissipative) 
effects and nonlinear advection effects represented by the 
sign-indefinite pressure term in the latter case. As is evident from 
figures \ref{fig:RvsE0}a,c, for small values of the constraint 
parameters the viscous effects dominate, in the sense that there exist 
$\overline{\E}_{0} > 0$ and $\overline{B} > 0$ such that 
$\R_{\E_{0}}(\tuE) < 0$ for $\E_{0} \in (0,\overline{\E}_{0})$ and 
$\R_{B}^{q}(\tuB) < 0$ for $B \in (0,\overline{B}_{0})$. In figure 
\ref{fig:RvsE0}a we see that the behavior of $\R_{\E_{0}}(\tuE)$ in the 
limit $\E_{0} \rightarrow 0$ is indeed well approximated by the 
asymptotic relation \eqref{eq:R03D}. On the other hand, for larger 
values of the constraint parameters, the nonlinear advection effects 
dominate and we have $\R_{\E_{0}}(\tuE) > 0$ for $\E_{0} > 
\overline{\E}_{0}$ and $\R_{B}^{q}(\tuB) > 0$ for $B > 
\overline{B}_{0}$. The transition between these two regimes is much 
sharper in the case of solutions of Problem \ref{pb:maxdLqdt} and 
becomes sharper for larger values of $q$. Figures \ref{fig:RvsE0}b,d 
indicate that both $\R_{\E_{0}}(\tuE)$ and $\R_{B}^{q}(\tuB)$ exhibit a 
power-law dependence on the constraint parameters $\E_{0}$ and $B$ for 
sufficiently large values of these parameters.  To quantify this, a 
least-square error fit reveals that \citep{ld08,ap16}
\begin{equation}\label{eq:RvsE0fit}
\R_{\E_0}(\tuE) = C_1\E_0^{\,\alpha_1}, \qquad C_1 = 3.72 \times 10^{-3} , \ \alpha_1 = 2.97 \pm 0.02 
\end{equation}
for solutions of Problem \ref{pb:maxdEdt3D} in the limit $\E_{0} 
\rightarrow \infty$. As regards solutions of Problem \ref{pb:maxdLqdt} 
for different $q$ and $B \rightarrow \infty$, we have 
\begin{equation}
\R_{B}^q(\tuB) = C_{2} \| \tuB \|_{L^q}^{\alpha_{2}}, \qquad C_{2}, \ \alpha_{2} > 0
\label{eq:RvsBfit}
\end{equation}
with the prefactors and exponents given in Table \ref{tab:RvsBfit}. 
These results allow us to conclude that both estimates 
\eqref{eq:dEdt_estimate_E} and \eqref{eq:dLqdt} are in fact sharp in 
terms of the exponent. Since the prefactor in the former bound is given 
explicitly, we can evaluate it using $\nu = 0.01$ to give $C_1 = 
27/(8\pi^4\nu^3) \approx 3.465 \times 10^4$. Therefore, the solutions 
of Problem \ref{pb:maxdEdt3D} underestimate this prefactor by about 
seven orders of magnitude. Since the prefactor in estimate 
\eqref{eq:dLqdt} is not given explicitly, such a comparison cannot be 
made for this bound.

In addition to sharpness under the worst-case conditions, another 
question pertaining to inequality \eqref{eq:dEdt_estimate_E} is whether 
the upper bound on its RHS can also be realized under generic 
conditions in turbulent flows. This problem was studied by 
\citet{Schumacher2010} who demonstrated that in turbulent flows the 
rate of change of ensemble-averaged squared vorticity grows at most as 
$\frac{d}{dt}\langle \bomega^2 \rangle \sim \langle \bomega^2 
\rangle^{3/2}$, where $\langle \cdot \rangle$ denotes ensemble 
averaging.

\begin{table}
    \centering
    \begin{tabular}{c | c c c }
         $q$ & exponent in \eqref{eq:dLqdt} & \multicolumn{1}{c}{fitted exponent $\alpha_{2}$} & fitted prefactor $C_{2}$ \\
         \hline
         $4$ & $12$ & $11.88           \pm 0.03$ & $2.9\cdot10^{-15}$\\
         $5$ & $10$ & $\phantom{0}9.96 \pm 0.02$ & $1.1\cdot10^{-8}\phantom{^0}$\\
         $6$ & $10$ & $\phantom{0}9.92 \pm 0.02$ & $2.8\cdot10^{-6}\phantom{^0}$\\
         $9$ & $12$ & $11.94           \pm 0.04$ & $7.0\cdot10^{-4}\phantom{^0}$
    \end{tabular}
 \caption{Exponents in the upper bound \eqref{eq:dLqdt} and the parameters 
 $C_{2}$, $\alpha_{2}$  in ansatz \eqref{eq:RvsBfit} fitted to the 
 data in figure \ref{fig:RvsE0}d for different $q$ \citep{BleitnerProtas2026}.}
    \label{tab:RvsBfit}
\end{table}

\begin{figure}
\centering
\mbox{
\subfigure[]{\includegraphics[width=0.35\textwidth]{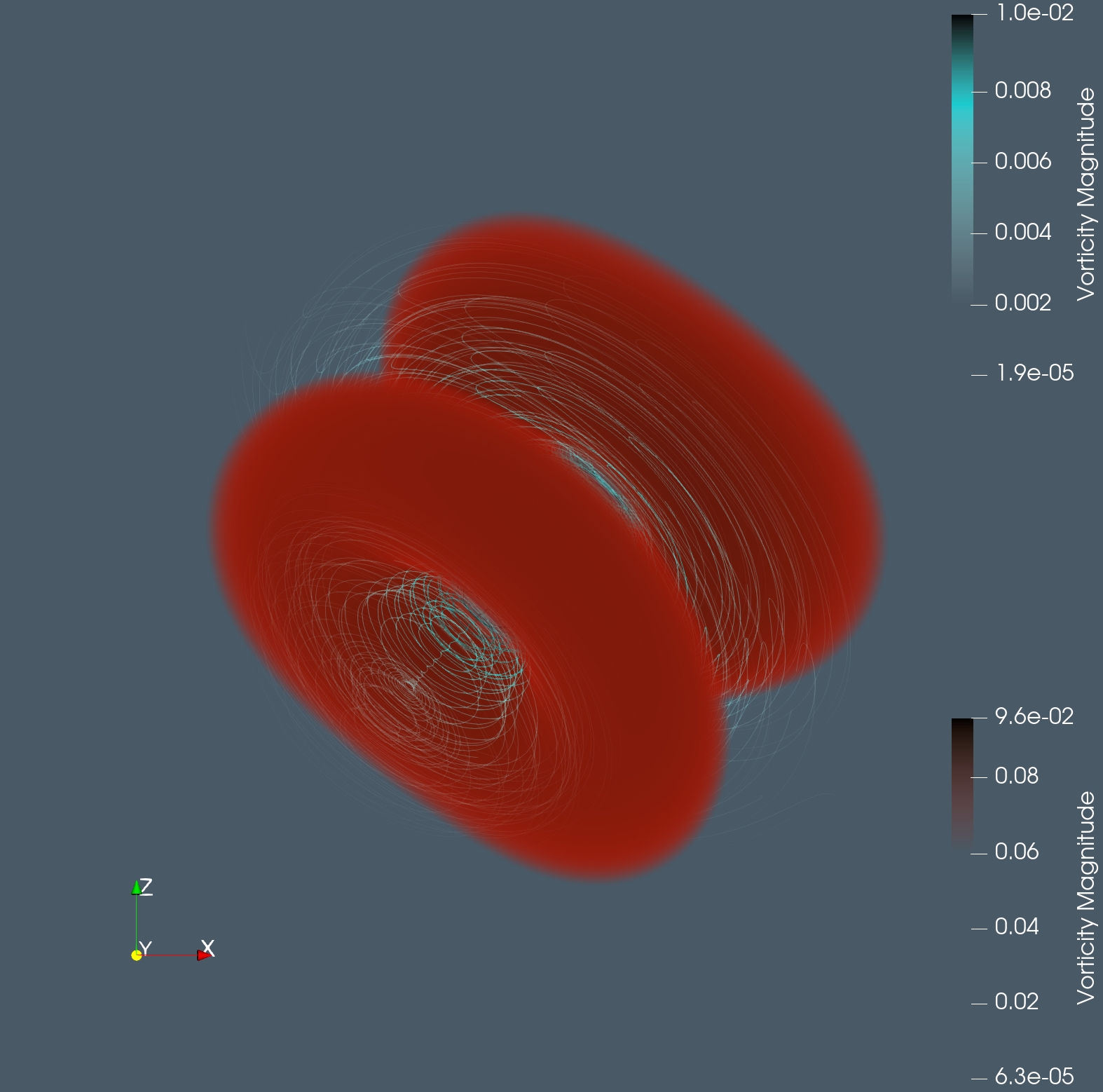}} \qquad\qquad
\subfigure[]{\includegraphics[width=0.35\textwidth]{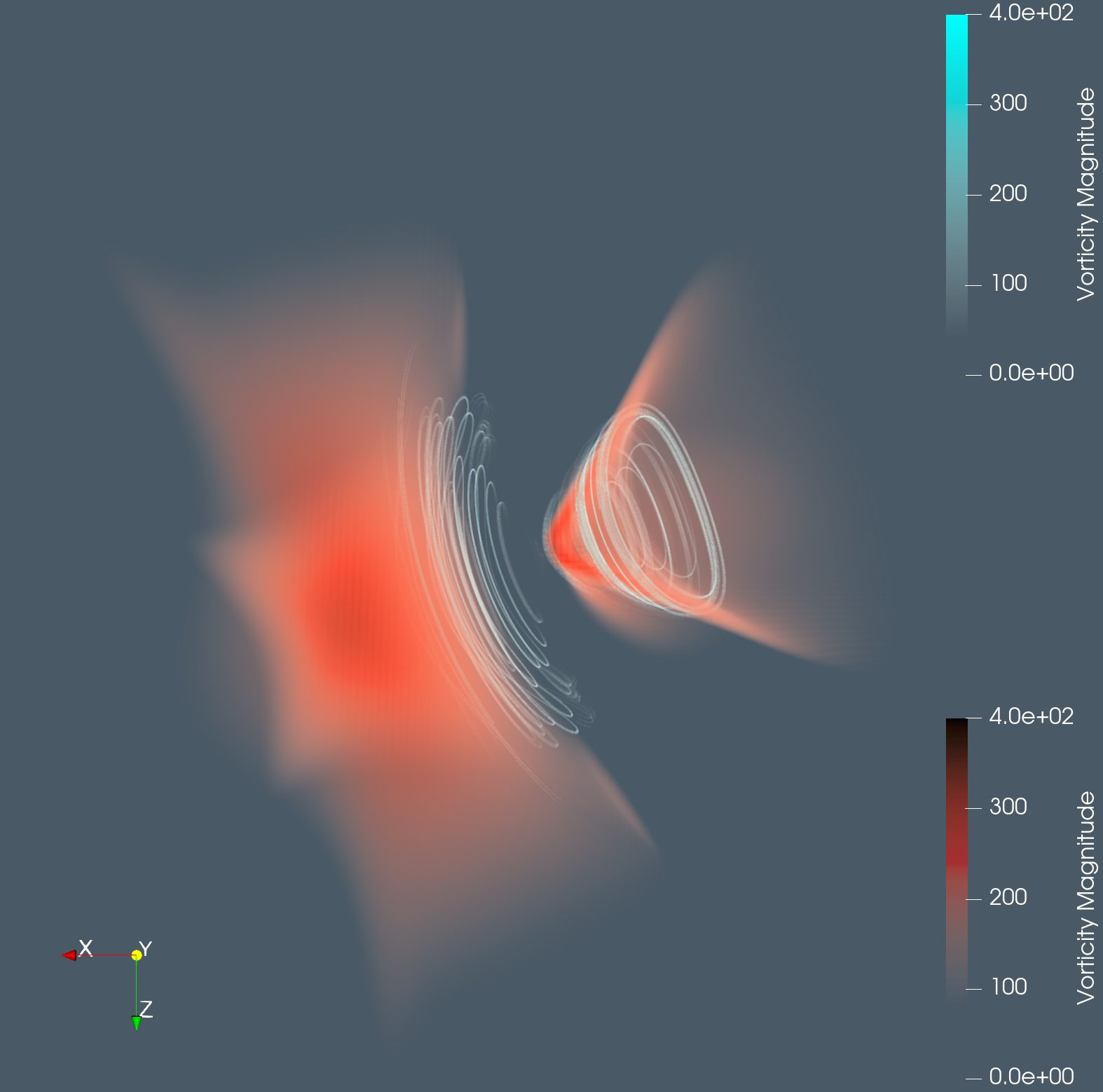}}
}
\caption{Maximizers (a) $\tuE$ of Problem \ref{pb:maxdEdt3D} obtained 
for $\E_{0} = 20$ and (b) $\tuB$ of Problem \ref{pb:maxdLqdt} obtained 
for $q = 5$ and $B = 177.8$. Both figures show the volume rendering of 
the vorticity magnitude $|\bomega(\x)|$ (in red) and selected vortex lines 
(in blue, with the color intensity proportional to the local value of 
$|\bomega(\x)|$).}
\label{fig:tuEB}
\end{figure}

Local maximizers $\tuE$ and $\tuB$ of Problems \ref{pb:maxdEdt3D} and 
\ref{pb:maxdLqdt} obtained for $\E_{0} > \overline{\E}_{0}$ and $B > 
\overline{B}_{0}$ are shown in figures \ref{fig:tuEB}a,b. We add that 
as $\E_{0}, B \rightarrow \infty$ the structure of these fields remains 
essentially unchanged, but they become increasingly localized in the 
sense that their characteristic length scales vanish relative to the 
size of the domain $\Omega$. As already observed by \citet{ld08}, the 
structure of the maximizers $\tuE$ of Problem \ref{pb:maxdEdt3D} is 
quite clear and involves a pair of nearly axisymmetric colliding vortex 
rings, cf.~figure \ref{fig:tuEB}a. Problem  \ref{pb:maxdEdt3D} also 
admits multiple branches of suboptimal local maximizers that for a 
given value of $\E_{0}$ correspond to smaller values of the objective 
functional $\R_{\E_{0}}(\tuE)$ than shown in figures 
\ref{fig:RvsE0}a,b. In the limit $\E_{0} \rightarrow 0$, these 
suboptimal branches approach the limiting states shown in figures 
\ref{fig:maxdEdt_vortexCells}b,e and \ref{fig:maxdEdt_vortexCells}c,f, 
and other eigenfunctions $\u_{1}$ of the Laplacian with $|\k| > 3$, 
cf.~\eqref{eq:u1}. For $\E_{0} > \overline{\E}_{0}$, these suboptimal 
maximizers also have the form of two nearly axisymmetric vortex rings 
similar to those shown in figure \ref{fig:tuEB}a, but aligned with a 
different symmetry plane of the domain $\Omega$. On the other hand, the 
structure of the local maximizer $\tuB$ of Problem \ref{pb:maxdLqdt} in 
figure \ref{fig:tuEB}b is less clear and does not appear reducible to a 
simple flow. However, it does reveal the presence of two localized 
objects, both with an axisymmetric structure typical of vortex rings, 
although these objects are quite different from each other.

It is an interesting question whether there is any quantitative 
relation between the maximizers $\tuE$ and $\tuB$ of Problems 
\ref{pb:maxdEdt3D} and \ref{pb:maxdLqdt}. By performing direct 
evaluation for different $q$ and large $\E_0$, $B$, we note that 
$\R_{\E_{0}}(\tuB) < 0$ and $\R_{B}^q(\tuE) < 0$, which means that the 
states maximizing $(d/dt) \| \u(t) \|_{L^{q}}$ can only produce 
$d\E(t)/dt < 0$ and vice versa. We therefore conclude that, at least 
with the local maximizers of Problems \ref{pb:maxdEdt3D} and 
\ref{pb:maxdLqdt} found by \cite{ld08,ap16,BleitnerProtas2026}, it is 
not possible to simultaneously saturate both estimates 
\eqref{eq:dEdt_estimate_E} and \eqref{eq:dLqdt} with the same family of 
states.

We now move on to discuss solutions of Problems 
\ref{pb:maxET}--\ref{pb:PsiLq} defined on finite time windows. We focus 
on the main findings and to fix attention consider Problems 
\ref{pb:PhiHs}--\ref{pb:PhiLq} for $q = 9$ only; the reader is referred 
to \citet{KangYunProtas2020,KangProtas2021,RamirezProtas2026} for 
further details. Each of these problems now depends on two parameters: 
a measure of the magnitude of the initial condition $\u_{0}$, i.e., 
$\E_{0}$ or $B$, and the length $T$ of the time window. For each value 
of $\E_{0}$ and $B$ the general approach is to first consider time 
windows $T$ that are sufficiently short such that smooth classical 
solutions of the Navier-Stokes system \eqref{eq:NS} are guaranteed to 
exist \citep{RobinsonRodrigoSadowski2016}. The optimization problems 
are then solved on progressively longer time windows that may eventually 
become longer than the time of local existence (whose numerical 
value is unknown). Should a singularity form at some time $t_0$, this 
has to be manifested in the unbounded growth of the optimized 
quantities $\E(\u(T))$, $\| \u(T) \|_{L^3}$ and $\frac{1}{T} \int_0^T 
\| \u(\tau) \|_{L^q}^{p} \, d\tau$ as the optimization window is 
extended $T \rightarrow t_0$, cf.~\eqref{eq:Etblowup} and 
\eqref{eq:LPSblowup}. In this way, a hypothetical singularity can be 
detected by studying classical solutions of system \eqref{eq:NS} only 
which are smooth and can be accurately approximated using standard 
numerical techniques, cf.~Appendix \ref{sec:numer}, without the need to 
compute singular solutions that satisfy system \eqref{eq:NS} in a weak 
sense only. Flows obtained as solutions of Problems 
\ref{pb:maxET}--\ref{pb:PsiLq} did not reveal an unbounded growth of 
the quantities of interest which would signal singularity formation. 
However, these quantities do exhibit a significant transient growth 
which we quantify below as the most extreme behavior realizable under 
the Navier-Stokes dynamics.

\begin{figure}
\begin{center}\vspace*{-3.7cm}
\mbox{\hspace*{-0.35cm}
\Bmp{0.5\textwidth}
\subfigure[]{\includegraphics[width=1.0\textwidth]{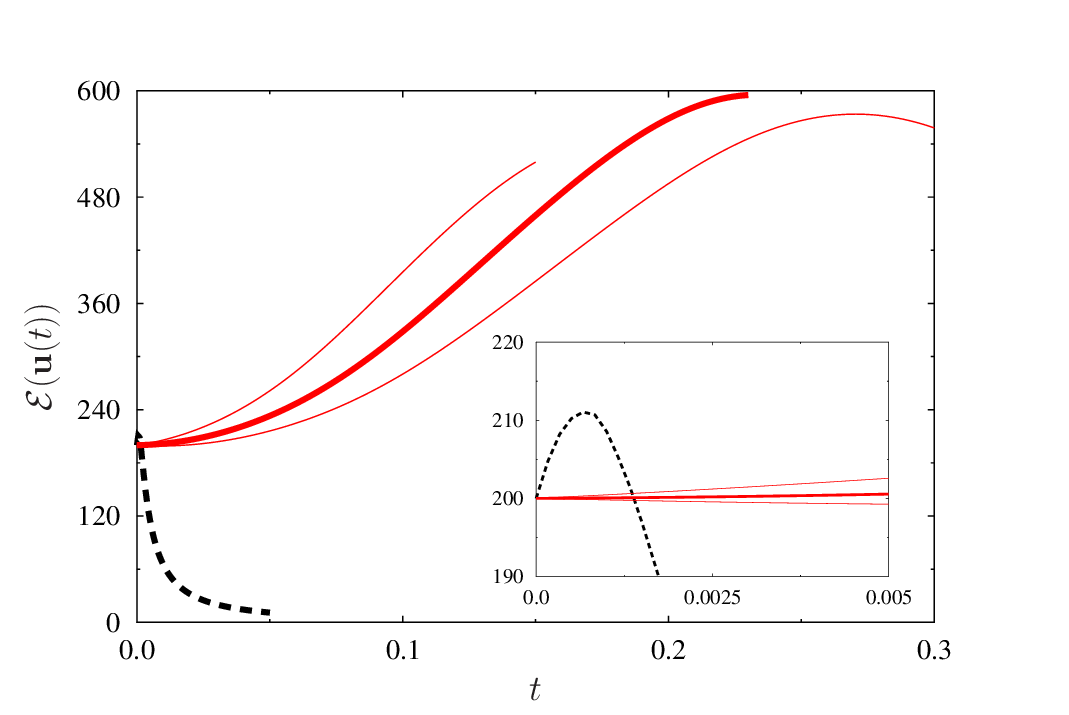}}
\Emp
\Bmp{0.5\textwidth}\hspace*{-1.0cm}
\subfigure[]{\includegraphics[width=1.05\textwidth]{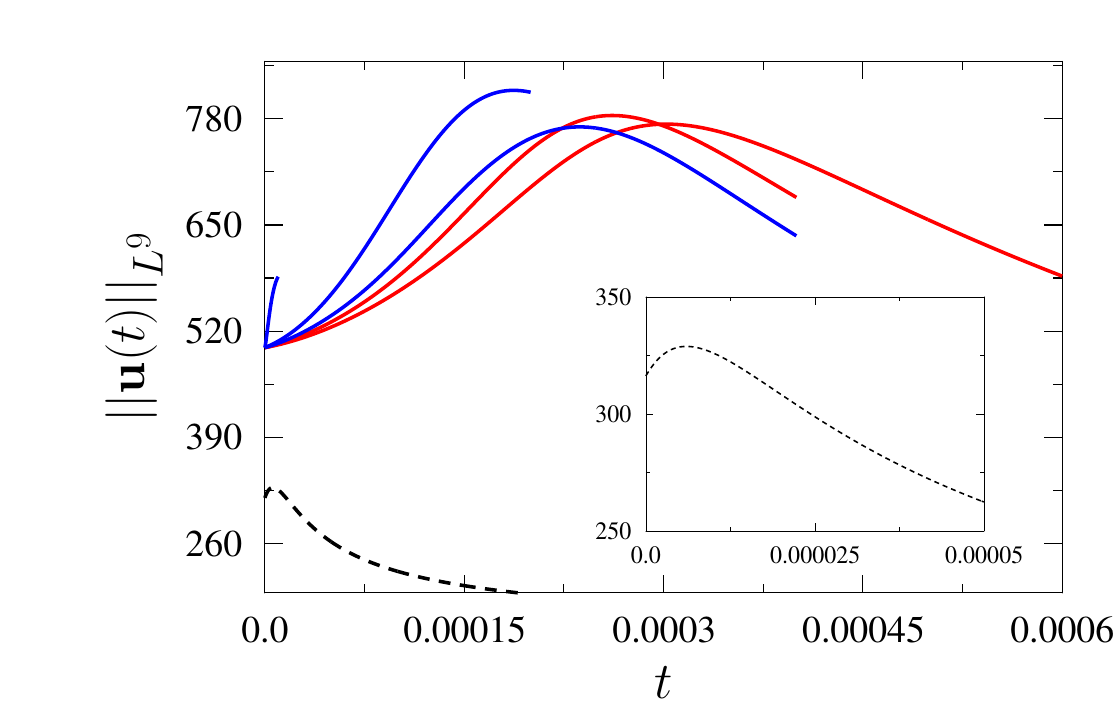}}
\Emp
} \\ \vspace*{-0.475cm}
\mbox{\hspace*{-0.35cm}\vspace*{-0.35cm}
\Bmp{0.5\textwidth}
\subfigure[]{\includegraphics[width=1.0\textwidth]{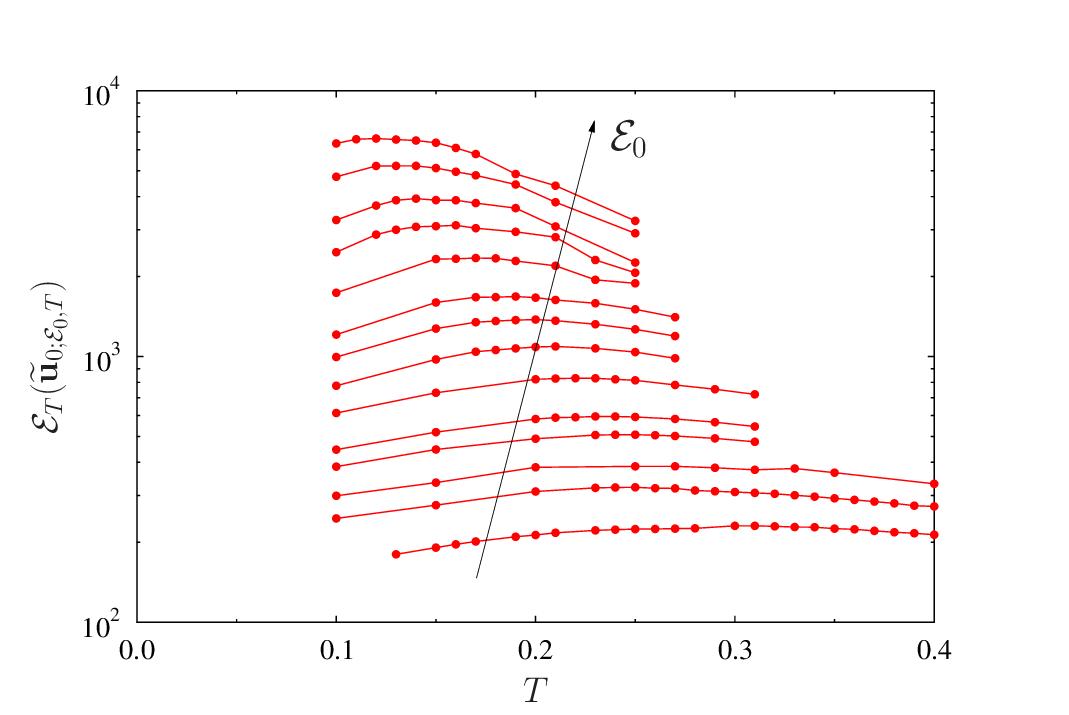}}
\Emp
\Bmp{0.5\textwidth}\hspace*{-0.6cm}\vspace*{-0.3cm}
\subfigure[]{\includegraphics[width=1.0\textwidth]{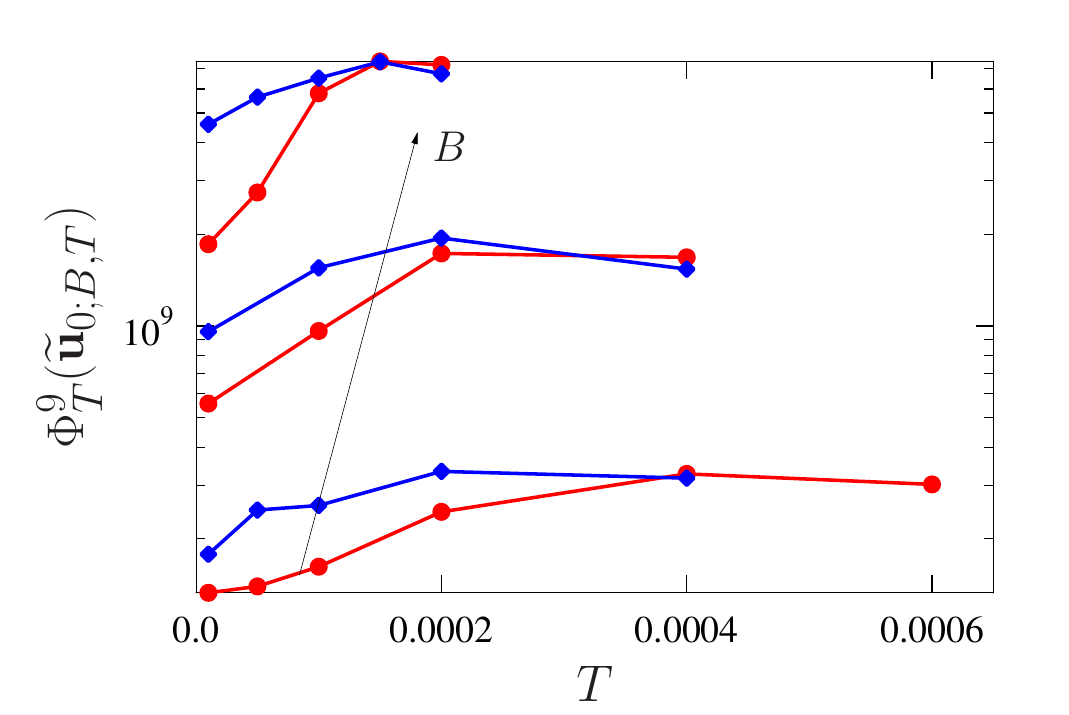}}
\Emp
}\\ \vspace*{-0.475cm}
\mbox{\hspace*{-0.35cm}
\Bmp{0.5\textwidth}
\subfigure[]{\includegraphics[width=1.0\textwidth]{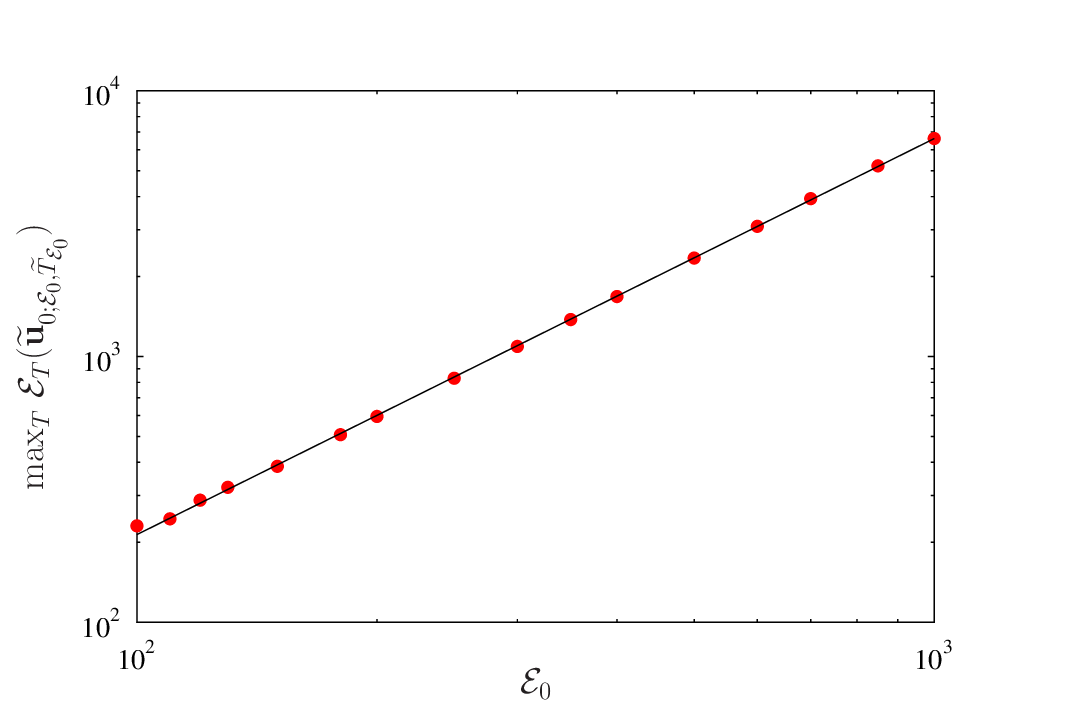}}
\Emp
\Bmp{0.5\textwidth}\hspace*{-0.75cm}
\subfigure[]{\includegraphics[width=1.05\textwidth]{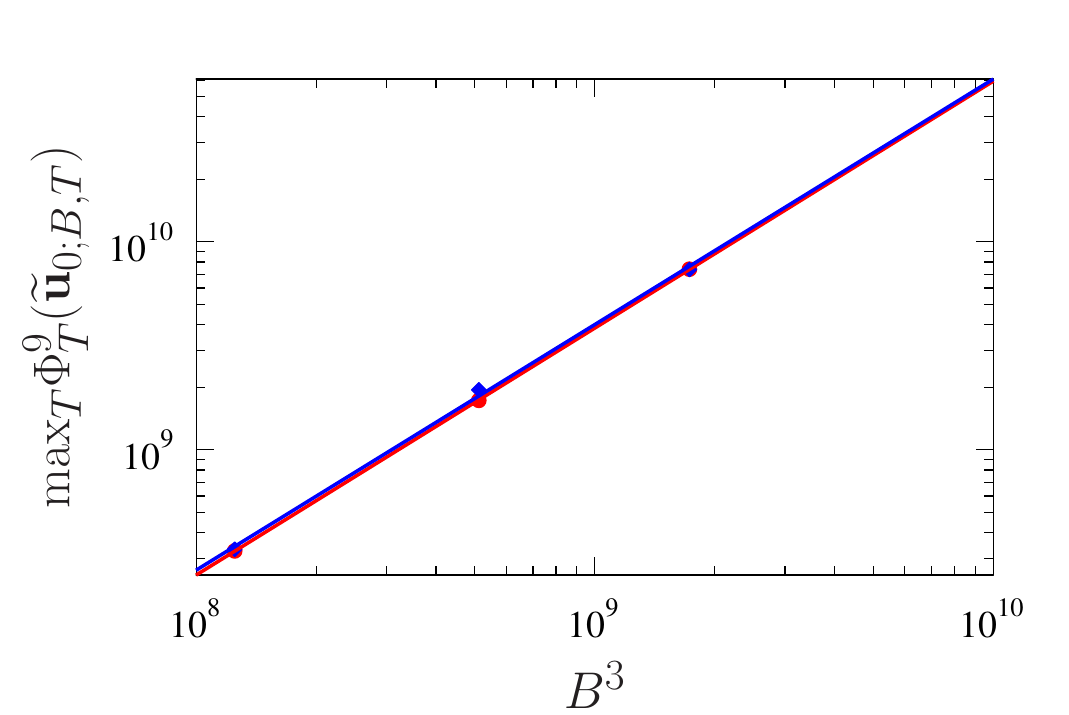}}
\Emp
}\\ \vspace*{-0.35cm}
\caption{\small (a) Time evolution of the enstrophy $\E(\u(t))$ in the Navier-Stokes 
flows with the 
  initial condition given by (black dashed line) the maximizer $\tuE$ 
  of Problem \ref{pb:maxdEdt3D} and (red solid lines) the maximizers 
  $\tuET$ of Problem \ref{pb:maxET} for $\E_0 = 200$ and $T = 
  0.15,0.23,0.3$ (the curve corresponding to the optimal length of the 
  time window $\tTE = 0.23$ is marked with a thick line whereas the 
  inset represents magnification of the initial stages of evolution). 
  (b) Time evolution of $||\u(t)||_{L^9}$ in the Navier-Stokes flows 
  with the initial condition given by (black dashed line) the maximizer 
  $\tuB$ of Problem \ref{pb:maxdLqdt} with $B = 316$ and the
   optimal initial conditions $\tuBT$ obtained by solving Problem 
  \ref{pb:PhiHs} (blue) and Problem \ref{pb:PhiLq} (red) with $q = 9$ 
  and $B = 500$; three representative times $T$ are considered 
  including $\tTB$ which gives the largest value of 
  $\Phi^{9}_{T}(\tuBT)$ for a given value of $B$. (c) Maximum attained 
  enstrophy $\E_T(\tuET)$ as a function of the optimization time $T$ in 
  Problem \ref{pb:maxET} for initial enstrophies $100 \le \E_0 \le 
  1000$. (d) Maximum attained values of the objective functional 
  $\Phi^{9}_{T}(\tuBT)$ as functions of the optimization time $T$ in 
  Problems \ref{pb:PhiHs} (blue) and \ref{pb:PhiLq} (red) for different 
  values of the constraint $B=500 ,800, 1200$; in panels (c,d) each 
  symbols represents a solution of the optimization problem obtained 
  for the indicated values of $T$ with the constraint parameters 
  remaining constant on each curve; the trend with the increase of 
  $\E_{0}$ and $B$ is indicated with arrows. (e) Dependence of the 
  maximum enstrophy growth $\max_{T > 0} \, \E_T(\tuET)$ on the initial 
  enstrophy $\E_{0}$ with the solid line representing the fit 
  \eqref{eq:maxTET}. (f) Dependence of 
  $\max_{T}\,\Phi^{9}_{T}\left(\widetilde{\u}_{0;B,T}\right)$ on 
  $B^3=||\widetilde{\u}_{0;B,T}||_{L^9}^3$ with solid lines 
  representing the fits \eqref{eq:sqSb_q9}--\eqref{eq:sqLG_q9}.}
\label{fig:maxE3D}
\end{center}
\end{figure}

In figures \ref{fig:maxE3D}a,b we show the time evolution of $\E(t)$ 
and $\| \u(t) \|_{L^{9}}$ in the Navier-Stokes flows corresponding to 
the optimal initial conditions $\tuET$ and $\tuBT$ found by solving 
Problems \ref{pb:maxET} and \ref{pb:PhiHs}--\ref{pb:PhiLq} for some 
representative parameter values. For comparison, we also include data 
for the flows with the initial conditions $\tuE$ and $\tuB$ obtained by 
solving the instantaneous optimization problems, Problems 
\ref{pb:maxdEdt3D} and \ref{pb:maxdLqdt} (in the latter case, the 
values of the parameter $B$ used in the solution of the instantaneous 
and finite-time problems are a bit different). We see that, in the 
flows obtained by solving the finite-time optimization problems, 
$\E(t)$ and $\| \u(t) \|_{L^{9}}$ exhibit a significant transient 
growth. On the other hand, in the flows obtained by solving the 
instantaneous optimization problems, these quantities grow very rapidly 
initially only, at $t = 0^{+}$, but then their growth is depleted such 
that overall they increase very little. Interestingly, in one case 
shown in figure \ref{fig:maxE3D}a, the enstrophy $\E(t)$ initially 
decreases, an effect already discussed in \S\,\ref{sec:Burgers} in the 
context of extreme Burgers flows in 1D obtained by solving an analogous 
optimization problem, cf.~Problem \ref{pb:maxET1D}. It is interesting 
whether in the limit $\E_{0} \rightarrow \infty$ solutions of Problem 
\ref{pb:maxET} would exhibit a self-similar structure analogous to the 
one evident in the solutions of the 1D Burgers system, cf.~figures 
\ref{fig:maxET1D}b,c. The behavior of the flows obtained by solving the 
two problems motivated by the Ladyzhenskaya-Prodi-Serrin conditions, 
i.e., Problems \ref{pb:PhiHs} and \ref{pb:PhiLq}, is similar.

In order to quantify the maximum transient growth in the Navier-Stokes 
flows obtained by solving Problems \ref{pb:maxET} and 
\ref{pb:PhiHs}--\ref{pb:PhiLq}, the latter with $q = 9$, with different 
parameters, in figures \ref{fig:maxE3D}c,d  we show, respectively, 
$\E_{T}(\tuET)$ and $\Phi_{T}^{9}(\tuBT)$ as functions of the length 
$T$ of the optimization window where each branch corresponds to 
different values of the constraint parameters $\E_{0}$ and $B$. We see 
that, for each value of these parameters, $\E_{T}(\tuET)$ and 
$\Phi_{T}^{9}(\tuBT)$ have well-defined maxima with respect to $T$ 
which shift towards smaller values of $T$ as $\E_{0}$ and $B$ increase. 
The times when these maxima are attained are denoted $\tTE$ and $\tTB$.

\begin{figure}[t]
\centering
\mbox{
\subfigure[]{\includegraphics[width=0.49\textwidth]{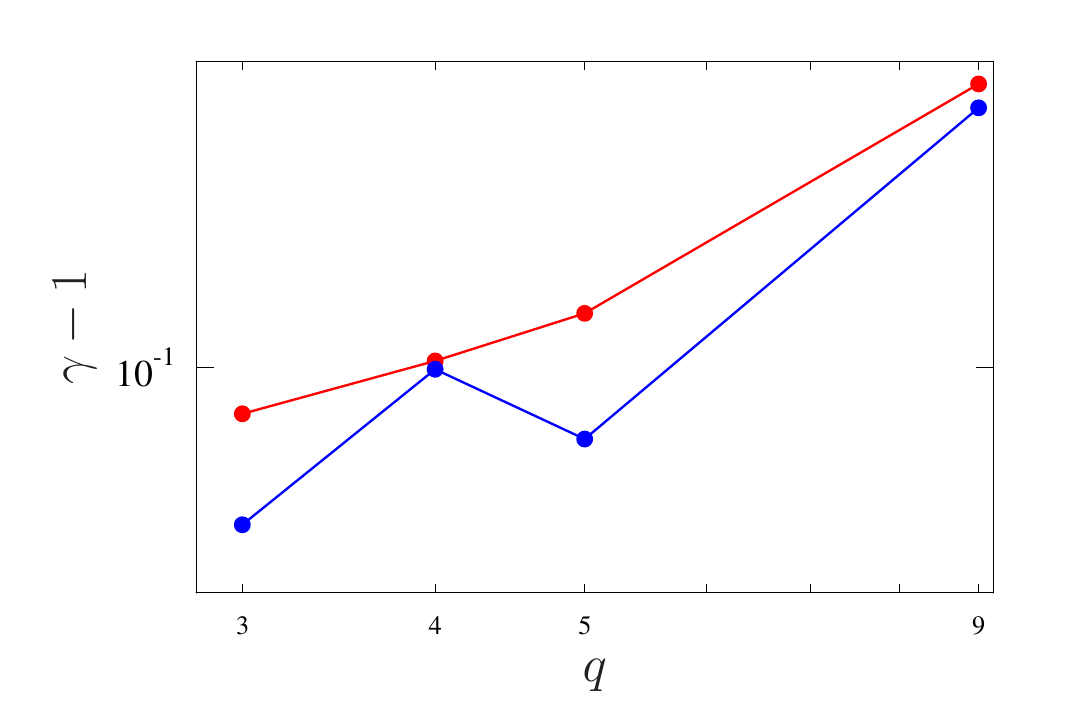}\label{fig:gamma}}
\quad
\subfigure[]{\includegraphics[width=0.49\textwidth]{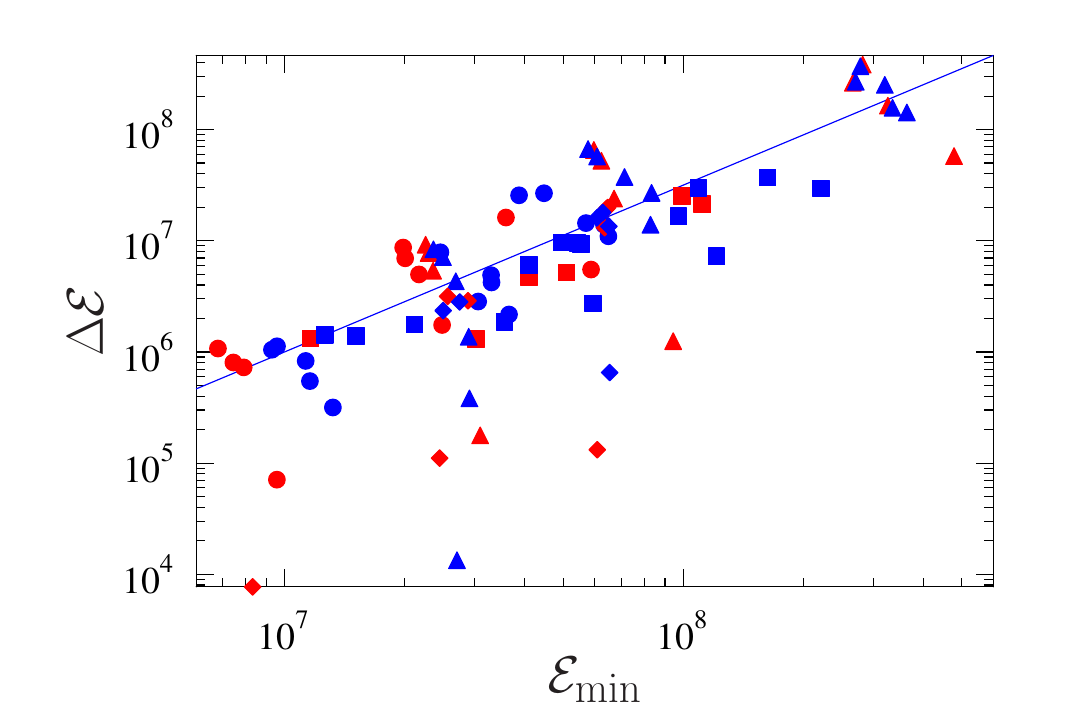}\label{fig:deltaE}}
}
\caption{\subref{fig:gamma} Dependence of the exponent $\gamma$ in 
expressions \eqref{eq:gamma} on  $q$  for Navier-Stokes flows with the optimal 
initial conditions found by solving (blue) Problems \ref{pb:PhiHs} and 
\ref{pb:PsiHs} and (red) Problems \ref{pb:PhiLq} and \ref{pb:PsiLq}. 
\subref{fig:deltaE} Dependence of $\Delta\E$, cf.~\eqref{eq:deltaE}, on 
the minimum enstrophy  $\E_{\text{min}}$ in Navier-Stokes flows with the optimal 
initial conditions found by solving (blue) Problems \ref{pb:PhiHs} and 
\ref{pb:PsiHs}, and (red) Problems \ref{pb:PhiLq} and \ref{pb:PsiLq} for different $B$ and $T$. 
In \subref{fig:deltaE}, each symbol corresponds to a different value of 
$q$: (squares) $q=3$, (circles) $q=4$, (triangles) $q=5$ and (diamonds) 
$q=9$. The solid blue line represents the relation $\Delta \E \sim C 
\E_{\text{min}}^{3/2}$ for some $C>0$ \citep{RamirezProtas2026}. }
\label{fig:compq}
\end{figure}

It is interesting to understand how the largest values of 
$\E_{T}(\tuET)$ and $\Phi_{T}^{9}(\tuBT)$ attained on each branch scale 
with the "size" of the initial data and to quantify this we plot the 
maxima over branches, $\max_{T} \E_{T}(\tuET)$ and $\max_{T} 
\Phi_{T}^{9}(\tuBT)$, as functions of $\E_{0}$ and $B$ in figures 
\ref{fig:maxE3D}e,f. In both cases, a power-law dependence is evident 
and by performing least-square error fits we obtain 
\citep{KangYunProtas2020,RamirezProtas2026}
\begin{subequations}
\begin{alignat}{2}
& \text{Problem \ref{pb:maxET}:}& \qquad
& \max_{T}\,\E_{T}(\tuET)\approx 0.224\, 
\E_{0}^{1.490},
\label{eq:maxTET} \\	
& \text{Problem \ref{pb:PhiHs}:}& \qquad
& \max_{T}\,\Phi_{T}^9\left(\widetilde{\u}_{0;B,T}\right)\approx 0.09\, 
\left(B^{3}\right)^{1.18},
\label{eq:sqSb_q9} \\
& \text{Problem \ref{pb:PhiLq}:}& \qquad
& \max_{T}\,\Phi_{T}^9\left(\widetilde{\u}_{0;B,T}\right)\approx 0.07\, 
\left(B^{3}\right)^{1.19},
\label{eq:sqLG_q9}
\end{alignat}
\end{subequations}
where the quantity $B^{3}$ was chosen as the independent variable in 
the last two relations recognizing the fact that with $q = 9$ the 
objective functional \eqref{eq:Phi} scales with the third power of the 
velocity. It is intriguing to note that the exponent in 
\eqref{eq:maxTET} is essentially equal to 3/2 and is therefore 
effectively the same as has been established in an analogous setting in 
the context of the 1D Burgers equation in \S\,\ref{sec:Burgers}, 
cf.~\eqref{eq:maxETap}. Extreme Navier-Stokes flows with initial 
conditions found by solving Problems \ref{pb:PhiHs}--\ref{pb:PhiLq} for 
$q = 4,5$ and Problems \ref{pb:PsiHs}--\ref{pb:PsiLq} produced results 
qualitatively similar to what is reported in figures 
\ref{fig:maxE3D}b,d,f \citep{RamirezProtas2026}. By performing 
fits as in \eqref{eq:sqSb_q9}--\eqref{eq:sqLG_q9} for the power-law 
relations 
\begin{equation}
\max_{T}\,\Psi_{T}\left(\tuBT\right)\sim \left(B^{3}\right)^{\gamma}, \qquad 
\max_{T}\,\Phi_{T}^q\left(\tuBT\right)\sim \left(B^{p}\right)^{\gamma} \quad \text{for} \ q > 3,
\label{eq:gamma}
\end{equation} 
we obtain the exponents $\gamma$ for different considered values of $q$ 
and in figure \ref{fig:compq}a we plot $(\gamma - 1)$ versus $q$ to 
highlight the part of the growth of $||\u(t)||_{L^q}$ resulting from 
the nonlinear amplification rather than merely from the increase of the 
constraint parameter $B$. We see that, interestingly, in solutions of 
Problems \ref{pb:PhiLq} and \ref{pb:PsiLq}, $(\gamma-1)$ is an 
increasing function of $q$. In solutions of Problems \ref{pb:PhiHs} and 
\ref{pb:PsiHs} the general trend is similar, but the increase with $q$ 
is not monotonic. The change of the trend at $q = 5$ in this case can 
possibly be attributed to the solutions of Problem \ref{pb:PhiHs} at 
this value of $q$ being only suboptimal local maximizers.

We now return to the behavior of the enstrophy in the extreme flows. 
Since in some of the flows obtained by solving Problems 
\ref{pb:PhiHs}--\ref{pb:PsiLq} the enstrophy tends to initially 
decrease, we define the quantity 
\begin{equation}
\Delta \E:=\max_{t\in\left[\argmin_{s\in[0,T]}\E(s),T\right]}\E(\u(t;\tuBT))-{\E_{\text{min}}}, \quad \text{where} \quad \E_{\text{min}}:=\min_{t\in[0,T]}\E(\u(t;\tuBT))
\label{eq:deltaE}
\end{equation}
measuring the total increase of the enstrophy relative to its minimum 
$\E_{\text{min}}$ which may be attained at an intermediate time (rather 
than at the initial time $t = 0$). It is plotted as a function of 
$\E_{\text{min}}$ for all extreme flows found by solving Problems 
\ref{pb:PhiHs}--\ref{pb:PsiLq} with $q = 3,4,5,9$ and three distinct 
values of $B$ in each case in figure \ref{fig:compq}b. The scatter 
evident in this plot is due to the fact that enstrophy is not directly  
controlled in these optimization problems, hence may in principle take 
arbitrary values. Despite this, we see that the upper envelope of all the 
data points exhibits a well-defined power-law relation $\Delta \E = 
\mathcal{O}\left( \E_{\text{min}} ^{3/2} \right)$ consistent with what 
was already observed by \citet{KangYunProtas2020} when solving Problem 
\ref{pb:maxET}, cf.~\eqref{eq:maxTET}, and by \citet{KangProtas2021} when 
solving Problem  \ref{pb:PhiHs} for $q = 4$.

\begin{figure}
\begin{center}
\mbox{
\subfigure[]{\includegraphics[width=0.5\textwidth]{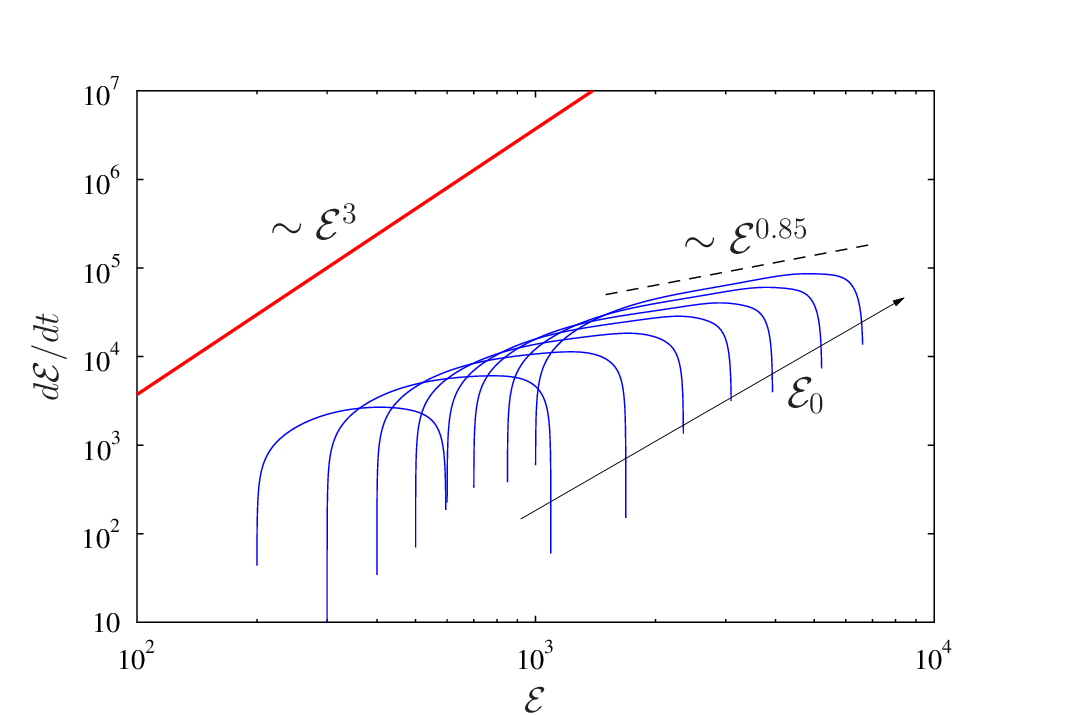}}
\Bmp{0.5\textwidth}
\vspace*{-2.75cm}
\subfigure[]{\includegraphics[width=1.0\textwidth]{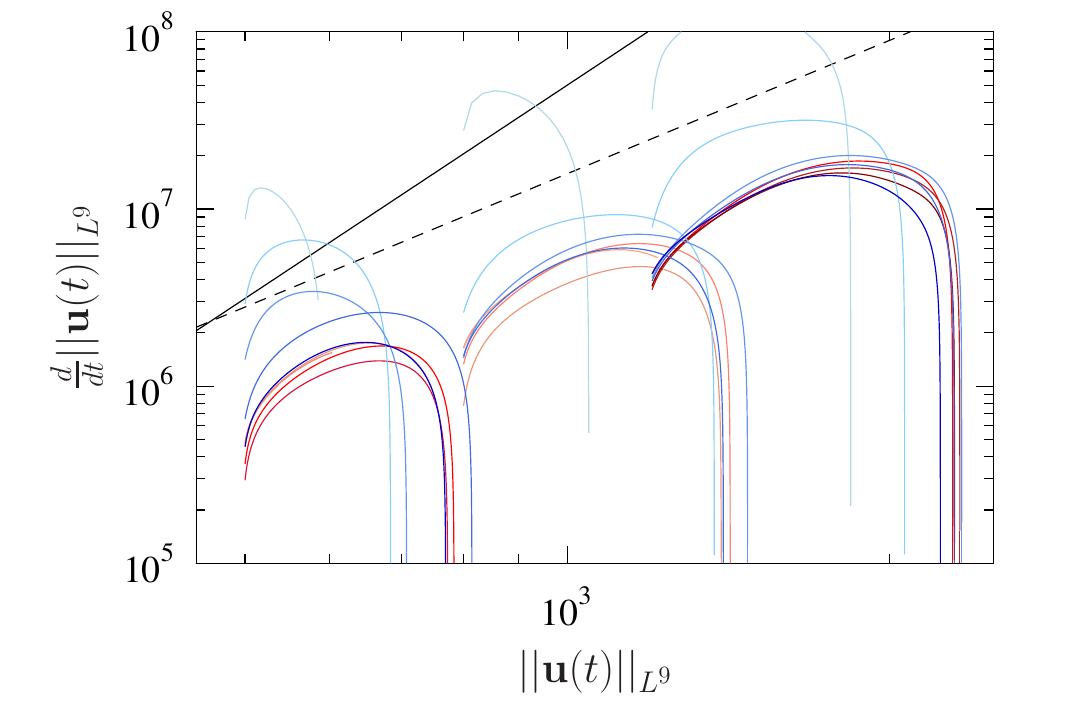}}
\Emp
} 
\caption{(a) Flow trajectories corresponding to the optimal initial 
data $\tuEtT$ obtained by solving Problem \ref{pb:maxET} with different $\E_0 \in 
  [100,1000]$ shown using the coordinates $\{ \E, d\E/dt\}$ (blue solid 
  lines with the arrow indicating the trend with the increase of 
  $\E_0$); the thick red line represents the relation $d\E/dt = 
  3.72\cdot 10^{-3} \, \E^3$ found by \citet{ap16} by solving Problem 
  \ref{pb:maxdEdt3D}, whereas the dashed black line represents the relation 
  $d\E/dt = 10^{2} \, \E^{0.85}$ \citep{KangYunProtas2020}. (b) 
  Flow trajectories corresponding to the optimal initial conditions 
  found by solving Problems \ref{pb:PhiHs} and \ref{pb:PhiLq} for 
  different $B$ and $T$ shown using the coordinates  
  $\left\{||\u(t)||_{L^{9}},(d/dt)||\u(t)||_{L^{9}}\right\}$;  the 
  black  solid line represents the upper bound 
  $(d/dt)||\u(t)||_{L^{9}}\sim||\u(t)||_{L^{9}}^4$ from 
  \eqref{eq:dLqdt}, whereas the dashed line shows the relation 
  $(d/dt)||\u(t)||_{L^{9}}\sim||\u(t)||_{L^{9}}^{5/2}$ from 
  \eqref{eq:noblowup_ub}; trajectories marked in blue and red 
  correspond to solutions of Problems \ref{pb:PhiHs} and 
  \ref{pb:PhiLq}, respectively \citep{RamirezProtas2026}.}
\label{fig:dEdtE}
\end{center}
\end{figure}

In order to assess how close the extreme flows found by solving 
Problems \ref{pb:maxET} and \ref{pb:PhiHs}--\ref{pb:PhiLq}, the latter 
with $q = 9$, come to forming a singularity, they are characterized 
using the coordinates $\{\E(t),d\E(t)/dt\}$ and 
$\left\{||\u(t)||_{L^{9}},(d/dt)||\u(t)||_{L^{9}}\right\}$  in figures 
\ref{fig:dEdtE}a and \ref{fig:dEdtE}b, respectively (these trajectories 
are parameterized with time $t$). Since the slope of the tangent to 
each of the trajectories represents the exponent $\alpha$ describing 
the instantaneous rate at which the quantity is amplified, $d\E(t)/dt 
\sim \E(t)^{\alpha}$ and $(d/dt) ||\u(t)||_{L^{9}} \sim 
||\u(t)||_{L^{9}}^{\alpha}$, this makes it possible to compare the 
observed behavior with the a priori bounds on the rate of growth of 
$\E(t)$ and $||\u(t)||_{L^{9}}$ discussed in \S\,\ref{sec:boundsNS}. As 
regards the data in figure \ref{fig:dEdtE}a, we observe that the rate 
of growth of enstrophy along the trajectories originating from 
the optimal initial conditions $\tuEtT$ is at all times and for all 
values of $\E_0$ several orders of magnitude smaller than the maximum 
rate of growth achieved by the instantaneous maximizers $\tuE$, 
cf.~figure \ref{fig:RvsE0}b. In fact, the sustained rate at which the 
enstrophy is amplified in solutions of Problem \ref{pb:maxET} is 
$d\E(t)/dt \sim \E(t)^{0.85}$ with an exponent significantly below 
the threshold value of 2 needed for singularity formation, cf.~the 
discussion after \eqref{eq:maxEtG}. As regards the data in figure 
\ref{fig:dEdtE}b, solutions of Problems \ref{pb:PhiHs}--\ref{pb:PhiLq} 
do not saturate the upper bound \eqref{eq:dLqdt} by a wide margin 
as well. However, along certain trajectories, the rate of growth $(d/dt) 
||\u(t)||_{L^{9}}$ does for some time exceed the level guaranteeing the 
regularity of the solution given in \eqref{eq:noblowup_ub}, although 
this time is not long enough for a singularity to form.

\begin{figure}
\begin{center}
\mbox{\subfigure[$\omega_1$]{\includegraphics[width=0.3\textwidth]{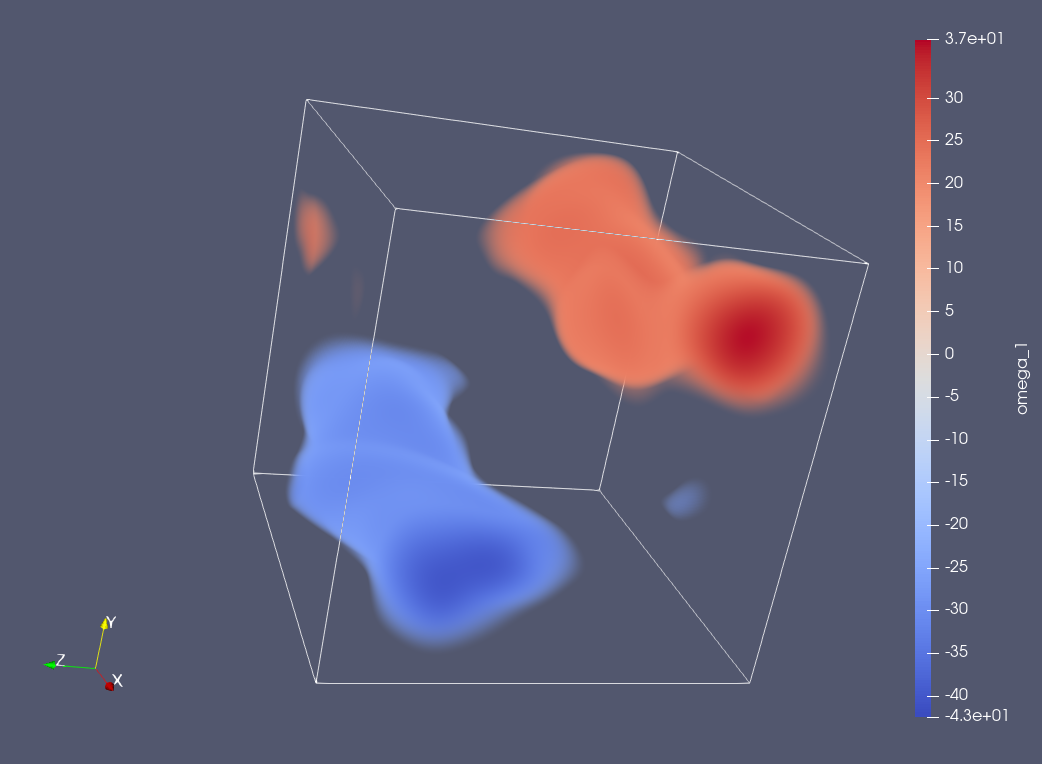}}\qquad
\subfigure[$\omega_2$]{\includegraphics[width=0.3\textwidth]{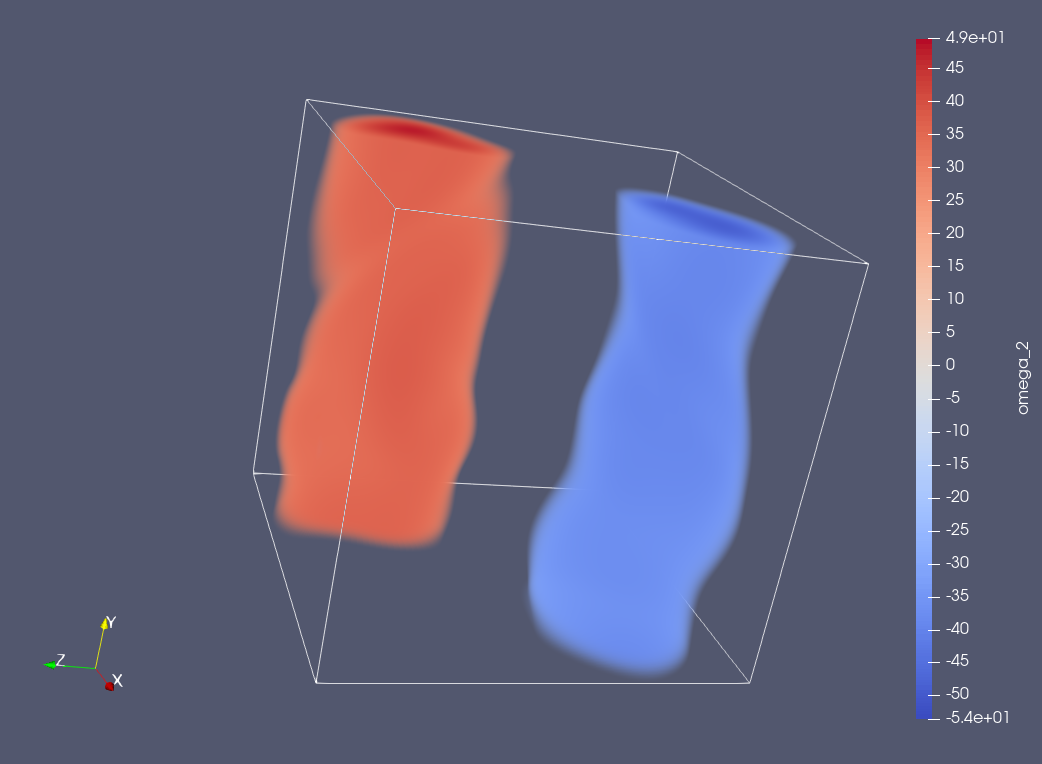}}\qquad
\subfigure[$\omega_3$]{\includegraphics[width=0.3\textwidth]{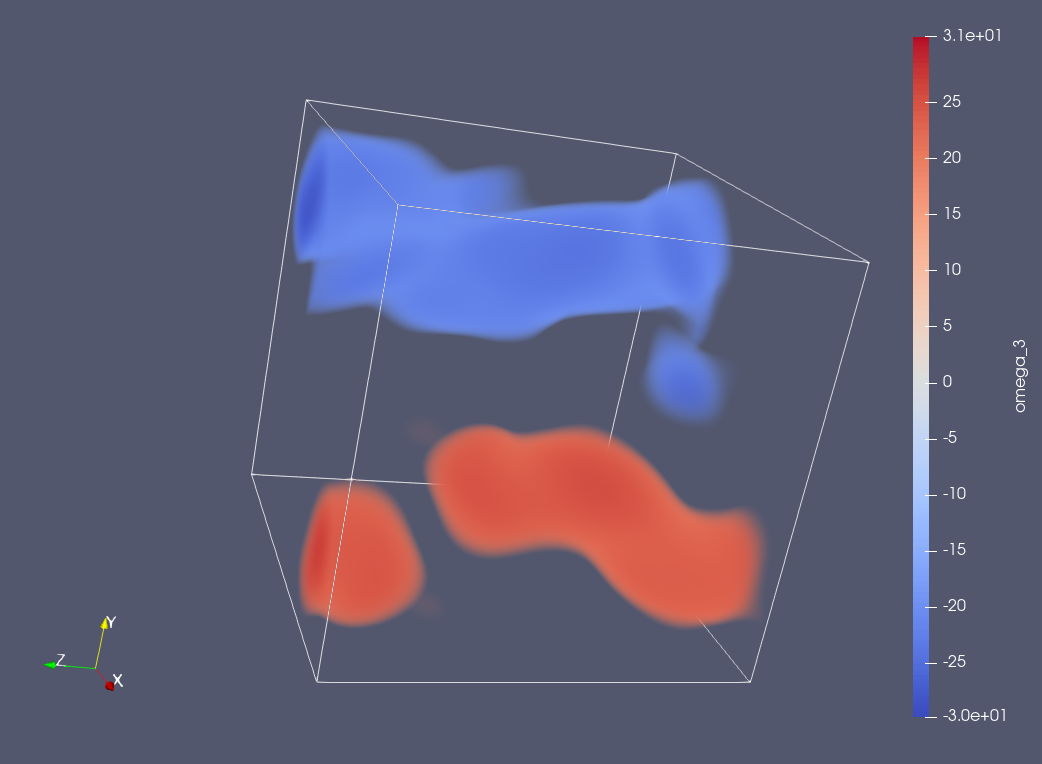}}}
\mbox{\subfigure[$t = 0.17 = \tTE$]{\includegraphics[width=0.3\textwidth]{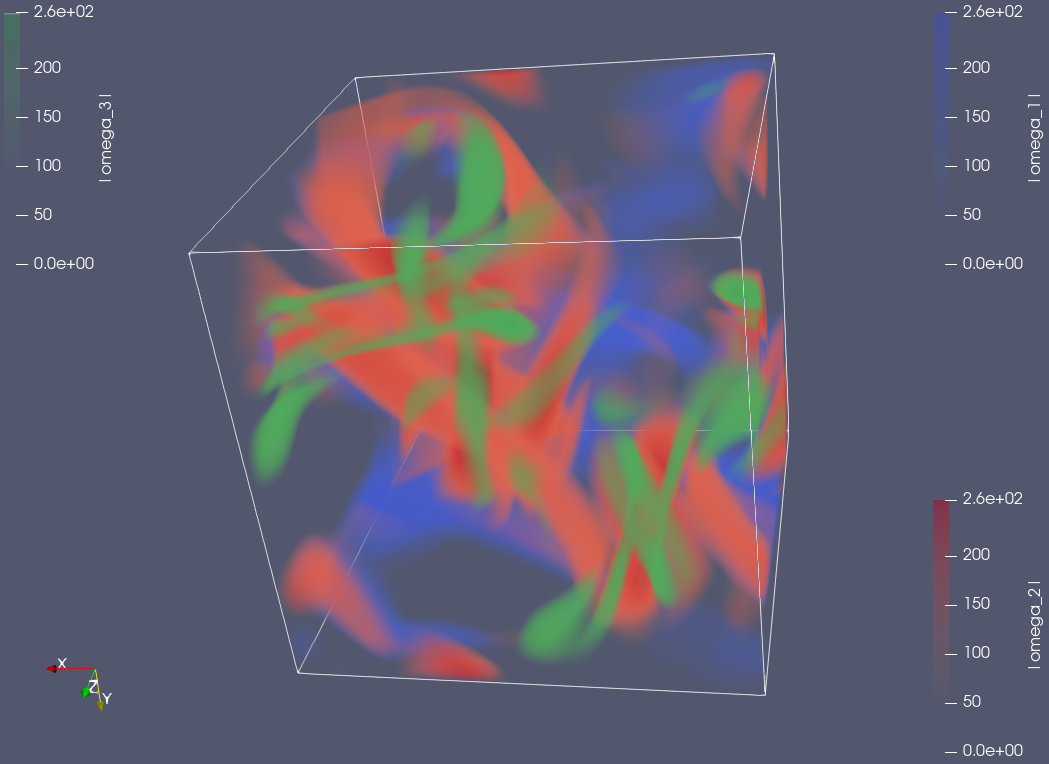}}\qquad
\subfigure[$t = 0$]{\includegraphics[width=0.28\textwidth]{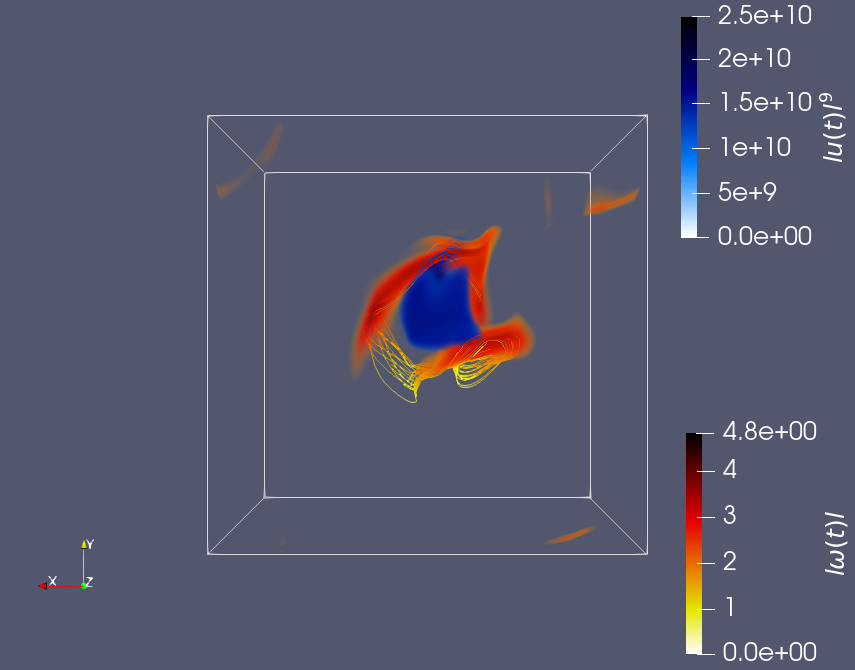}}\qquad
\subfigure[$t = T$]{\includegraphics[width=0.28\textwidth]{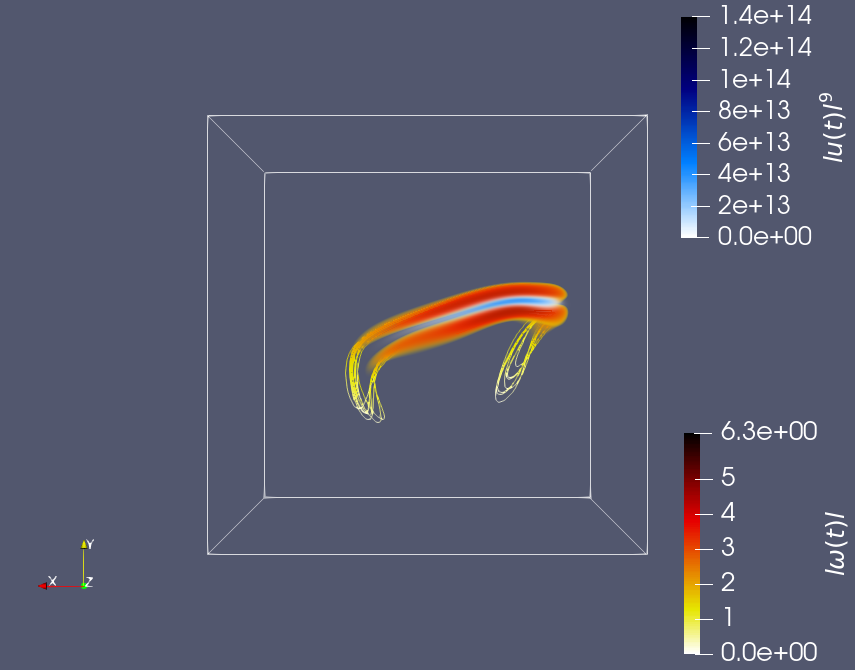}}}
\caption{(a--c) Vorticity components of the optimal initial condition 
$\tuEtT$ obtained by solving Problem \ref{pb:maxET} for the initial 
enstrophy $\E_0 = 500$ and the corresponding optimal length $\tTE = 
0.17$ of the time interval; the time evolution of the flow 
corresponding to this initial condition is visualized in 
\href{https://www.youtube.com/watch?v=tHU6gRNrVdo}{movie 2}. (d) The 
corresponding final flow state $\u(\tTE;\tuEtT)$ with the vorticity 
components (blue) $\omega_1$, (red) $\omega_2$ and (green) $\omega_3$ 
\citep{KangYunProtas2020}. (e,f) The magnitude of the vorticity 
{$|\bomega(t,\x;\tuBtB)|$} (red color scale) along with vortex lines 
(red) and the quantity {$|\u(t,\x;\tuBtB)|^9$} (blue color scale) in 
the Navier-Stokes flow obtained by solving Problem \ref{pb:PhiLq} with 
$B=800$ and $T=2\times 10^{-4}$ shown at $t = 0$  and $t = T$ 
\citep{RamirezProtas2026}.}
\label{fig:tubes}
\end{center}
\end{figure}

Finally, we analyze the structure of the extreme flows found by solving 
Problems \ref{pb:maxET} and \ref{pb:PhiHs}--\ref{pb:PhiLq}, the latter 
with $q = 9$. The optimal initial condition $\tuEtT$ obtained with 
$\E_{0} = 500$ and $\tTE = 0.17$ is visualized in figures 
\ref{fig:tubes}a--c, where three Cartesian vorticity components are 
shown. We see that this optimal initial condition has the form of three 
perpendicular pairs of antiparallel vortex tubes. As documented by 
\citet{KangYunProtas2020}, in the course of the flow evolution the 
vortex tubes collide leading to vortex reconnection events such that 
the flow at the final time $t = \tTE$ is turbulent, cf.~figure 
\ref{fig:tubes}d and 
\href{https://www.youtube.com/watch?v=tHU6gRNrVdo}{movie 2}. The 
initial condition $\tuBT$ and the corresponding final state 
$\u(T;\tuBT)$ of the flow obtained by solving Problem \ref{pb:PhiLq} 
with $B=800$ and $T=2\times 10^{-4}$ are visualized in figures 
\ref{fig:tubes}e,f and reveal an entirely different structure. In 
contrast to the flow obtained by solving Problem \ref{pb:maxET} in 
which the vorticity is effectively present in the entire domain 
$\Omega$, cf.~figure \ref{fig:tubes}d, here the main flow structure 
remains highly localized during the entire flow evolution. It has the 
form of a bent vortex ring that becomes more deformed at later stages. 
As documented by \citet{RamirezProtas2026}, the visualizations in 
figures \ref{fig:tubes}e,f  are representative of flows obtained by 
solving Problems \ref{pb:PhiHs}--\ref{pb:PsiLq} for different values of 
$q$, $B$ and $T$. Different flow structures embody distinct physical 
mechanisms for the amplification of the quantities of interest and it 
is therefore intriguing that very different extreme flows discussed 
here produce the same scaling of the maximum growth of enstrophy, 
cf.~figures \ref{fig:maxE3D}e and \ref{fig:compq}b.

\subsection{Possible Blow-up in Euler Flows}
\label{sec:EulerBlowup}

We now turn our attention to Question \ref{Q1} in the context of 3D 
Euler flows governed by system \eqref{eq:Eu}. The local existence of 
classical solutions in Sobolev spaces was established by \citet{Kato1972} 
and is summarized in the following theorem 
\begin{thm} \label{thm:Hm} If $\u_0 \in H^m (\mbb T^3)$ for some $m
  > 5/2$ and satisfies $\bnabla \cdot \u_0 = 0$, then there exists a
  time $T = T\left(\| \u_0 \|_{H^m}\right)>0$ such that
  \eqref{eq:Eu} has a unique solution $\bds u(\cdot;\u_0) \in
  C([0, T]; H^m)$ $\bigcap C^1([0,T]; H^{m-1})$.
\end{thm}
\noindent
Another well-known conditional regularity result is the
Beale-Kato-Majda (BKM) criterion \citep{bkm84} which states that a
smooth solution $\u(t)$ of the Euler system develops a singularity at
$t = t_{0}$ if and only if
\begin{equation}\label{eq:BKM:vort}
\lim_{t \rightarrow t_{0}} \int_{0}^{t} ||\bomega(\tau)||_{L^\infty} \;d\tau = \infty.
\end{equation}
The sufficiency of this condition can be deduced from
Theorem \ref{thm:Hm} using a Sobolev inequality \citep{af05}
\begin{equation}
\int_{0}^{t} ||\bomega(\tau)||_{L^\infty} \;d\tau \leq t \sup_{0\leq \tau\leq t} ||\bnabla \u(\tau)||_{L^\infty}
\leq C t \sup_{0\leq \tau \leq t} ||\bds u(\tau)||_{H^m}, \quad m > 5/2,
\end{equation}
whereas its necessity is a result of the inequality \citep{mb02}

\begin{equation}\label{eq:BKM:necessary}
||\u(t)||_{H^m} \leq ||\u_0||_{H^m}\exp\left\{C_1 \exp\left(C_2\int_{0}^{t} ||\bomega(\tau)||_{L^\infty} \;d\tau \right)\right\}.
\end{equation}
The double exponential on the RHS of \eqref{eq:BKM:necessary} suggests 
that, should blow-up indeed occur at some $t = t_{0}$, we can expect a 
much more rapid growth of $\| \u(t) \|_{H^m}$, $m > 5/2$, than of $\| 
\bomega(t)\|_{L^\infty}$, as $t \rightarrow t_{0}$. The regularity of 
weak solutions of the Euler system is related to Onsager's conjecture 
concerning energy dissipation in such flows and a significant progress 
has been made recently as regards this problem \citep{Eyink2024}.

Our search for singularities in Euler flows is guided by the local 
well-posedness result in Theorem \ref{thm:Hm} and we aim to find an 
initial condition $\u_{0} \in H^m$, $m > 5/2$, subject to certain 
constraints, such that the $H^m$ norm of the corresponding solution of 
the Euler system \eqref{eq:Eu} is maximized at a prescribed time $T$. 
The desired initial conditions are thus found as local maximizers of a 
constrained PDE optimization problem with the square of the $H^m$ 
(semi)norm used as the objective functional where for concreteness we 
set $m = 3$. We want to investigate whether the $H^{3}$ norm can grow 
without bound if the time window $[0,T]$ is sufficiently long. To this 
end, we adopt an indirect approach to distinguish between regular and 
singular evolution based on resolution refinement.  When solving the 
optimization problem on a ``short'' time interval $[0, T]$, if the 
objective functional approximated using increasing numerical 
resolutions converges to a finite value as the resolution is refined, 
then we can conclude the Euler system \eqref{eq:Eu} is well-posed on 
this short interval. On the contrary, when the interval $[0, T]$ is 
``long'', presumably longer than the minimum time of existence 
guaranteed by Theorem~\ref{thm:Hm}, the objective functional evaluated 
on the optimal solutions will diverge upon resolution refinement if a 
singularity occurs within the interval $[0,T]$.  Here ``short'' and 
``long'' times are defined in relation to the interval of local 
existence established by Theorem~\ref{thm:Hm}.  More specifically, 
"short" and "long" times are assumed to be, respectively, within and 
outside that interval.

\subsubsection{Optimization Formulation}
\label{sec:optEu}

Since the Euler system \eqref{eq:Eu} is locally well posed in 
$H^m(\Omega)$, $m>5/2$, cf.~Theorem \ref{thm:Hm}, it may appear natural 
to look for optimal initial data $\u_0$ in that space. However, we are 
interested in finite-time singularities potentially arising in smooth 
classical solutions, whereas initial conditions constructed in Sobolev 
spaces will in general not be smooth (real-analytic).  Unlike 
Navier-Stokes flows, Euler flows do not instantly become smooth at $t = 
0^{+}$. Therefore, solving the Euler system \eqref{eq:Eu} with such 
initial data would not allow us to benefit from the exponential 
convergence of the pseudospectral methods used in these studies, 
cf.~Appendix \ref{sec:numer}. \citet{zp23} thus considered an extended 
Gevrey space $G^\sigma$ with $\sigma > 0$ of real-analytic functions 
defined on $\TT^3$ and  endowed with the inner product
\begin{equation}\label{eq:Gevrey:def}
\begin{aligned}
\forall \v, \u\in G^{\sigma}, \quad 
\left\la\v, \u \right\ra_{G^{\sigma}} 
:= &\sum_{\k \in \mathbb Z^3} 
(1+|2\pi\k|^2)^me^{4\pi\sigma|\k|} \hat{\v}_{\k} \cdot \overline{\hat{\u}}_{\k}\\
=& \int_{\mbb T^3}(1+|D|^2)^me^{2\sigma|D|} \v \cdot \u \, d\x,
\end{aligned}
\end{equation}
where overbar denotes complex conjugation, whereas the operators $|D|$
and $e^{\sigma|D|}$ are defined via
\begin{equation}
\left[\widehat{|D| \v}\right]_{\k} := 2\pi|\k| \hat{\v}_{\k}, \qquad \qquad 
\left[\widehat{e^{\sigma |D|}\v}\right]_{\k} := e^{2\pi\sigma|\k|}\hat{\v}_{\k}.
\label{eq:|D|}
\end{equation}
In the setting of this problem, the Gevrey space can be regarded a
linear subspace of the Sobolev space $H^m(\Omega)$, i.e.,
\begin{equation}
G^\sigma := \left\{ \v \in H^m(\Omega) \ : \ \| \v \|_{G^\sigma} = \left\langle \v,\v \right\rangle^{1/2}_{G^\sigma} < \infty \right\}, \quad m > \frac{5}{2}, \quad \sigma > 0.
\label{eq:G}
\end{equation}

\begin{remark}
  It follows from definitions \eqref{eq:Gevrey:def} and \eqref{eq:G}
  that the Gevrey spaces have the property
\begin{equation}
G^{\sigma_2} \subset G^{\sigma_1} \subset G^{0} = H^m, 
\qquad 0 < \sigma_1 < \sigma_2.
\end{equation}
\end{remark}

Since the initial condition $\u_0$ in system \eqref{eq:Eu} needs to
be divergence-free and the quantity $\int_{\Omega} \u_0 \, d\x$ is an
invariant of motion, \citet{zp23} introduced the following subspace in 
which optimal initial conditions were sought
\begin{equation}\label{eq:space:initial:condition}
V := \{ \v  \in G^{\sigma}: \,
\bnabla \cdot \v = 0, \quad \int_{\Omega} \v \, d \x= \0\}.
\end{equation}
They then defined the objective functional $\Phi_T: V \to \mbb R^+$ as
\begin{equation}\label{eq:obj}
\Phi_{T}(\u_0) := \| \u(T; \u_0) \|^2_{\dot{H}^3}.
\end{equation}
Unlike the Navier-Stokes system \eqref{eq:NS}, the Euler system \eqref{eq:Eu} 
possesses a scaling symmetry such that if $\left\{ \u(t, \x), p(t, \x) 
\right\}$ is a solution, then for any $\lambda > 0$,
\begin{equation}\label{eq:scale}
\left\{ \u^\lambda(t, \x) := \lambda \u(\lambda t, \x), \qquad
p^\lambda(t, \x) := \lambda^2 p(\lambda t, \x) \right\}
\end{equation}
is also a solution of \eqref{eq:Eu}. This means that the time scale and 
the magnitude of an Euler flow are intrinsically linked. Hence, for any 
nonzero initial condition $\u_0$,  we have the identity
\begin{equation}
\Phi_{T}(\u_0) =  ||\u_0||^2_{\dot H^3}\Phi_{||\u_0||_{\dot H^3}T}\left(\frac{\u_0}{||\u_0||_{\dot H^3}}\right).
\label{eq:scalePhi}
\end{equation}
Therefore, one can restrict the discussion to initial conditions with
unit $\dot H^3$ seminorm which belong to a closed manifold
$\M_1 \subset V$ defined as
\begin{equation}\label{eq:M1}
\M_1 := \{\v \in V:\,  ||\v||_{\dot H^3} = 1\}.
\end{equation}
Thus, we  arrive at the following optimization problem \citep{zp23}
\begin{problem}\label{pb:Eu}
Given $T \in \mbb R_+$, find 
\begin{equation}
\tuT = \argmax\limits_{\u_0 \in \M_{1}} \Phi_T(\u_0).
\end{equation}
\end{problem}
Due to the scaling property \eqref{eq:scale} and the resulting identity 
\eqref{eq:scalePhi}, Problem \ref{pb:Eu} depends on one parameter only, 
the length $T$ of the optimization window. This is in contrast to 
Problems \ref{pb:maxET}--\ref{pb:PsiLq} formulated for the 
Navier-Stokes system which depend on two parameters, $T$ and the size 
of the initial data, $\E_{0}$ or $B$. This fact simplifies the search 
for singularities in Euler flows. If the time window $[0,T]$ falls 
within the interval of local existence guaranteed by Theorem 
\ref{thm:Hm}, then for any $\u_0 \in \M_1$, we expect $\Phi_T(\u_0)$ to 
be finite such that it will remain bounded upon resolution refinement.  
On the other hand, if there exists an $\u_0\in \M_1$, which will lead 
to a finite-time blow-up inside a sufficiently long time interval $[0, 
T]$, we anticipate $\Phi_T(\u_0)$ to diverge as the resolution is 
refined. These scenarios are investigated in the next subsection.

\subsubsection{Results}
\label{sec:resultsEu}

\begin{figure}
  \centering
\mbox{\subfigure[]{\includegraphics[width=0.48\textwidth]{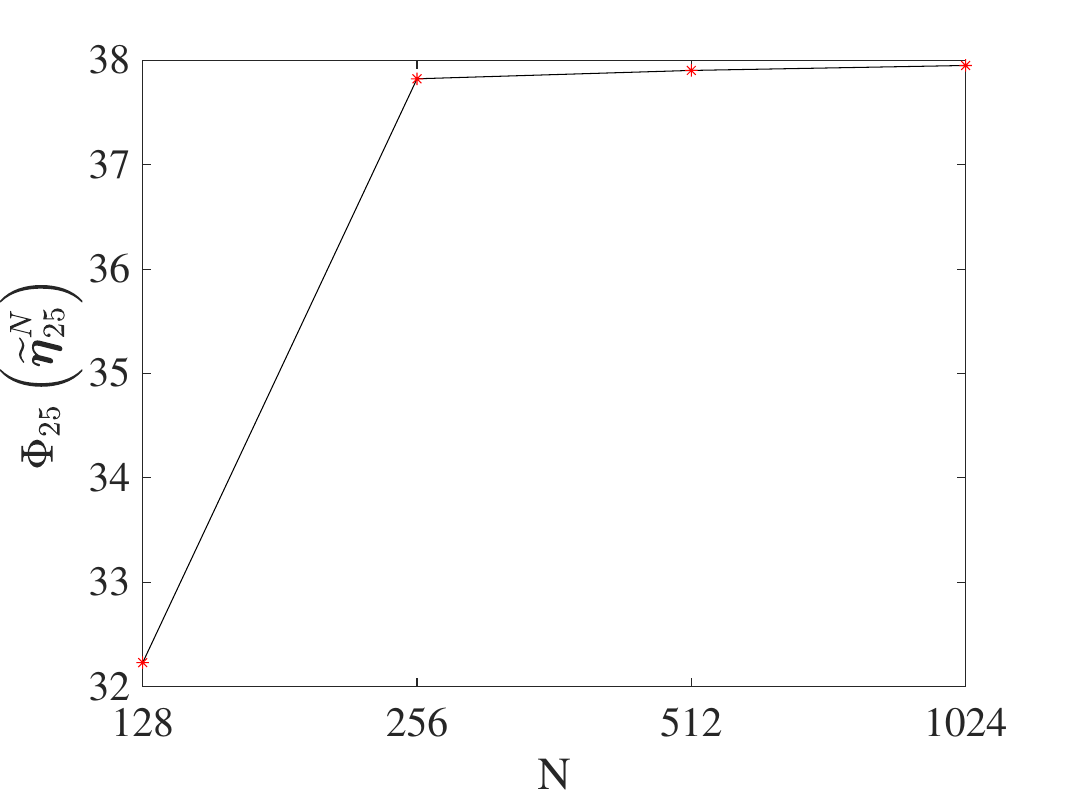}}
\subfigure[]{\includegraphics[width=0.48\textwidth]{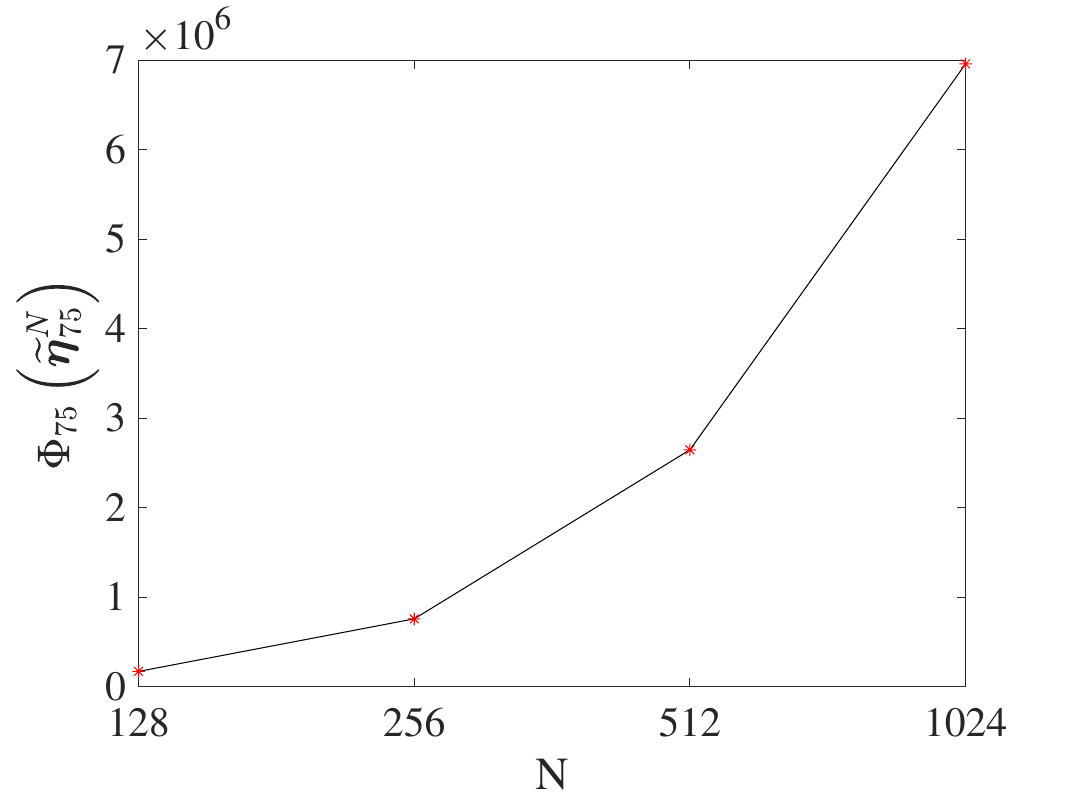}}}
\caption{Dependence of the maximum value of the objective functional 
$\Phi_{T}(\widetilde{\u}_{0;T}^{N}))$ obtained in the solution of 
Problem \ref{pb:Eu} with (a) $T = 25$ and (b) $T = 75$ on the numerical 
resolution $N$ \citep{zp23}.}
  \label{fig:H3N}
\end{figure}

We now move on to discuss the results obtained by solving Problem 
\ref{pb:Eu}. We focus on the main findings here and refer the reader to 
\citet{zp23} for all additional details. Problem \ref{pb:Eu} was solved 
on a "short" time window with $T = 25$ and on a "long" one with $T = 
75$, where the values of $T$ were determined empirically. In both cases 
a refinement of the numerical resolution was performed where the 
approximation $\tuT^{N}$ of the optimal initial condition obtained with 
the resolution $N$ was used as an initial guess in algorithm 
\eqref{eq:descHs} to solve Problem \ref{pb:Eu} with resolution $2N$ (the 
resolutions used were $N^3 = 128^3, 256^3, 512^3, 1024^3$) and the 
results are shown in figures \ref{fig:H3N}a,b. Extrapolating from the 
four resolutions used, we see that for $T = 25$ 
\begin{equation}
\lim_{N\to\infty} \Phi_{T}\left(\teta_T^N\right) < \infty,
\label{eq:H3T25} 
\end{equation}
i.e., the objective functional \eqref{eq:obj} remains finite upon
resolution refinement. On the other hand, for $T = 75$, we have 
\begin{equation}
\lim_{N\to\infty} \Phi_{T}  \left(\teta_T^N\right)= \infty
\label{eq:H3T75} 
\end{equation}
signalling the possibility of a singularity formation at some $t_{0} \in
[0,T]$. 

\begin{figure}
  \centering
\mbox{\subfigure[]{\includegraphics[width=0.48\textwidth]{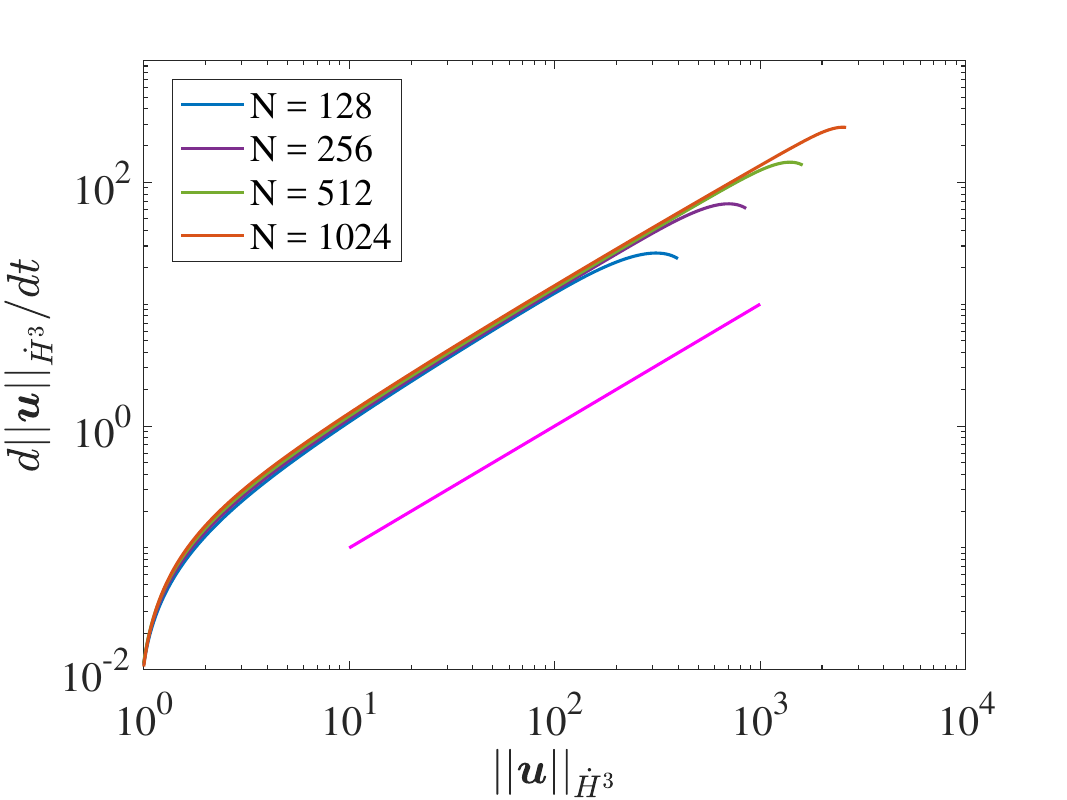}}
\hfill
\subfigure[]{\includegraphics[width=0.48\textwidth]{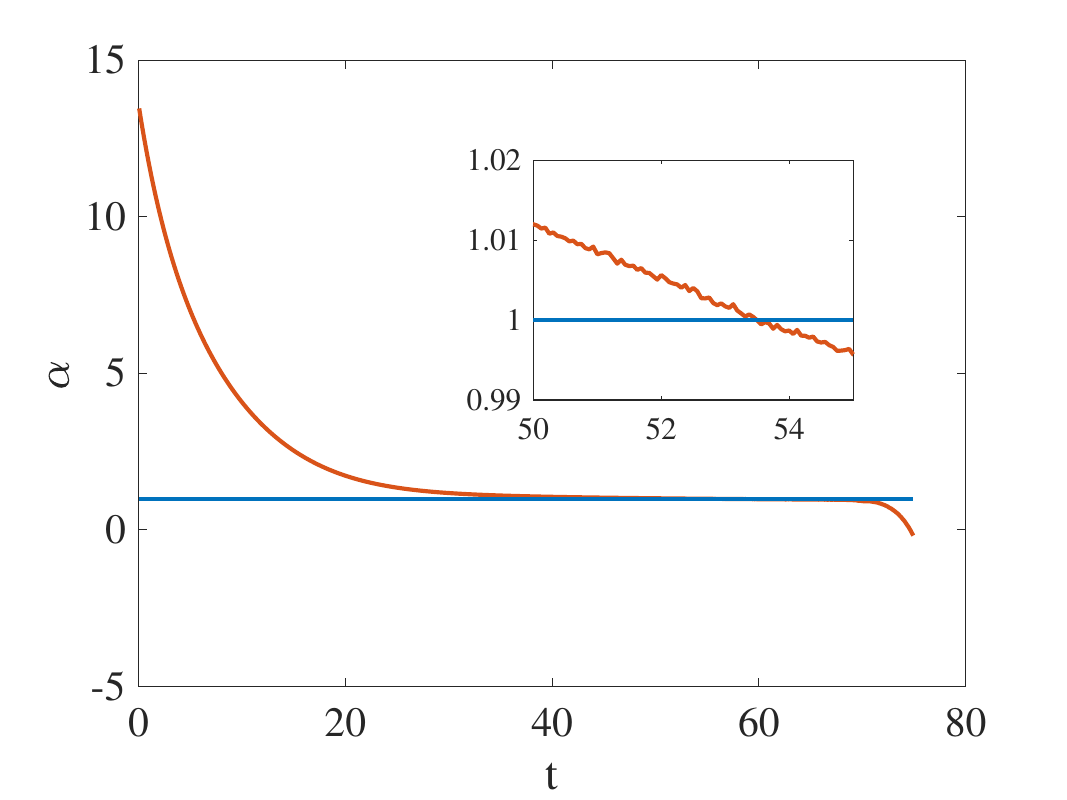}}}
\caption{(a) Dependence of $(d/dt)\| \u(t) \|_{\dot{H}^3}$ on $\| \u(t) 
\|_{\dot{H}^3}$ in the Euler flows with the optimal initial conditions 
$\tu_{0;75}^N$ for $t \in [0,75]$ and different resolutions $N^3$ and 
(b) the corresponding exponent $\alpha(t)$ in ansatz \eqref{eq:duH3dt} 
obtained for $N = 1024$ with a local fit as a function of time $t \in 
[0,75]$. The straight lines in both panels represent $\alpha = 1$ 
\citep{zp23}.}
  \label{fig:u:H3:rate}
\end{figure}

In order to better understand this behavior, we analyze the growth rate of $\|
\u(t) \|_{\dot{H}^3}$ and assume its evolution is described by the relation
 \begin{equation}
   \label{eq:duH3dt}
   \frac{d \|\u(t)\|_{\dot H^3}}{dt} = C(t) \|\u(t)\|_{\dot H^3}^{\alpha(t)}.
 \end{equation}
We note that if the exponent $\alpha(t)$ in \eqref{eq:duH3dt} remains 
larger than 1 over a sufficiently long time with the prefactor $C(t)$ 
bounded away from zero, then this will imply a finite-time blow-up of 
the norm $\| \u(t) \|_{\dot{H}^3}$, and thus formation of a singularity 
in the corresponding Euler flow, cf.~Theorem \ref{thm:Hm}. To 
investigate this possibility, in figure \ref{fig:u:H3:rate}a we plot 
$(d/dt)\| \u(t) \|_{\dot{H}^3}$ versus $\| \u(t) \|_{\dot{H}^3}$ for 
the flows found by solving Problem \ref{pb:Eu} using different 
resolutions. The plot uses log-log scaling, such that the exponent 
$\alpha(t)$ can be inferred from the slope of the tangent to the curves 
at $\| \u(t) \|_{\dot{H}^3}$. The evolution of the exponent $\alpha(t)$ 
determined by a local fitting procedure applied to ansatz 
\eqref{eq:duH3dt} with time $t \in [0,75]$ is shown for $N = 1024$ in 
figure \ref{fig:u:H3:rate}b. While a decreasing trend is evident, we 
nevertheless have $\alpha(t) > 1$ for $t \in [0, 53.495]$ with the 
computation becoming under-resolved at $t \approx 51.25$.  Thus, the 
exponent remains larger than unity as long as the flow is well 
resolved. As regards the prefactor $C(t)$ in \eqref{eq:duH3dt}, it 
reveals a slow growth with time $t$ (and with $\| \u(t) 
\|_{\dot{H}^3}$) which is well approximated by the expression $C(t) = 
0.0568 \left( \ln \| \u(t) \|_{\dot{H}^3} \right)^{0.5742}$ with 
parameters determined via a least-squares fit. We thus conclude that 
the time evolution of the norm $\| \u(t) \|_{\dot{H}^3}$ in the flow 
with the optimal initial condition $\tu_{0;75}^{1024}$ remains 
consistent with formation of a singularity as long as the computation 
remains well-resolved.

\begin{figure}
  \centering
\mbox{\subfigure[]{\includegraphics[width=0.48\textwidth]{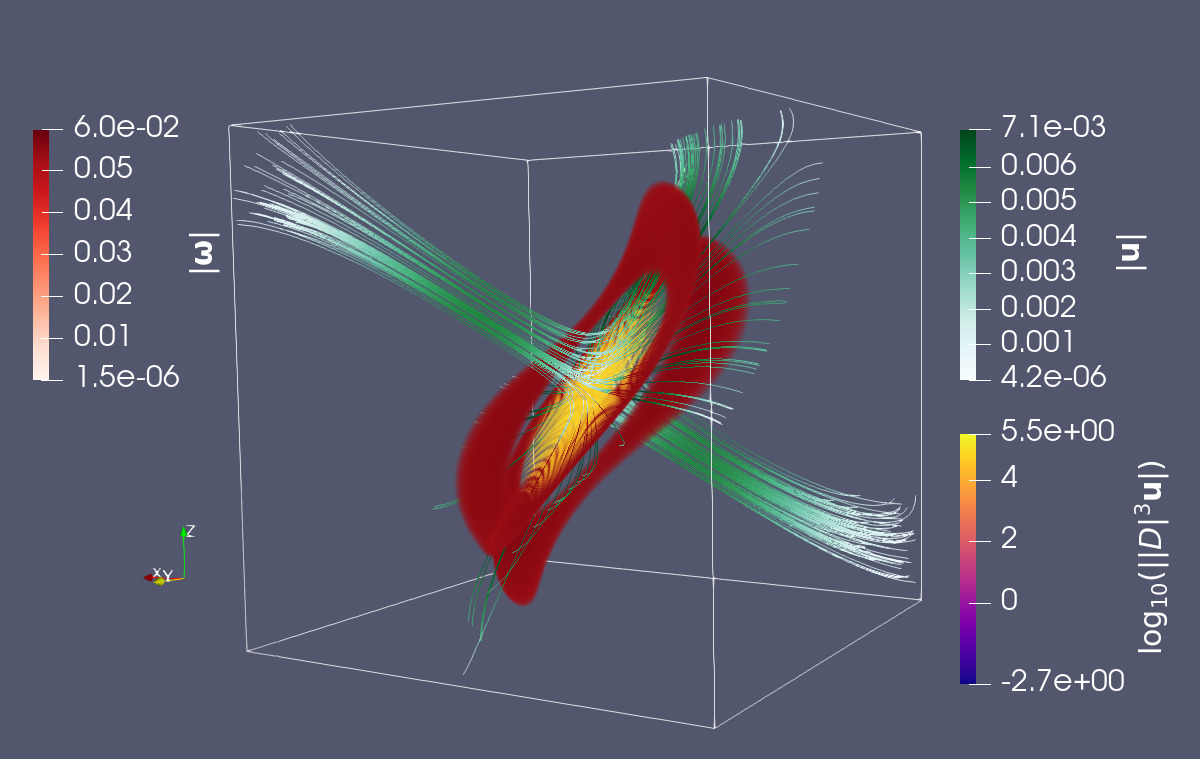}}
\subfigure[]{\includegraphics[width=0.48\textwidth]{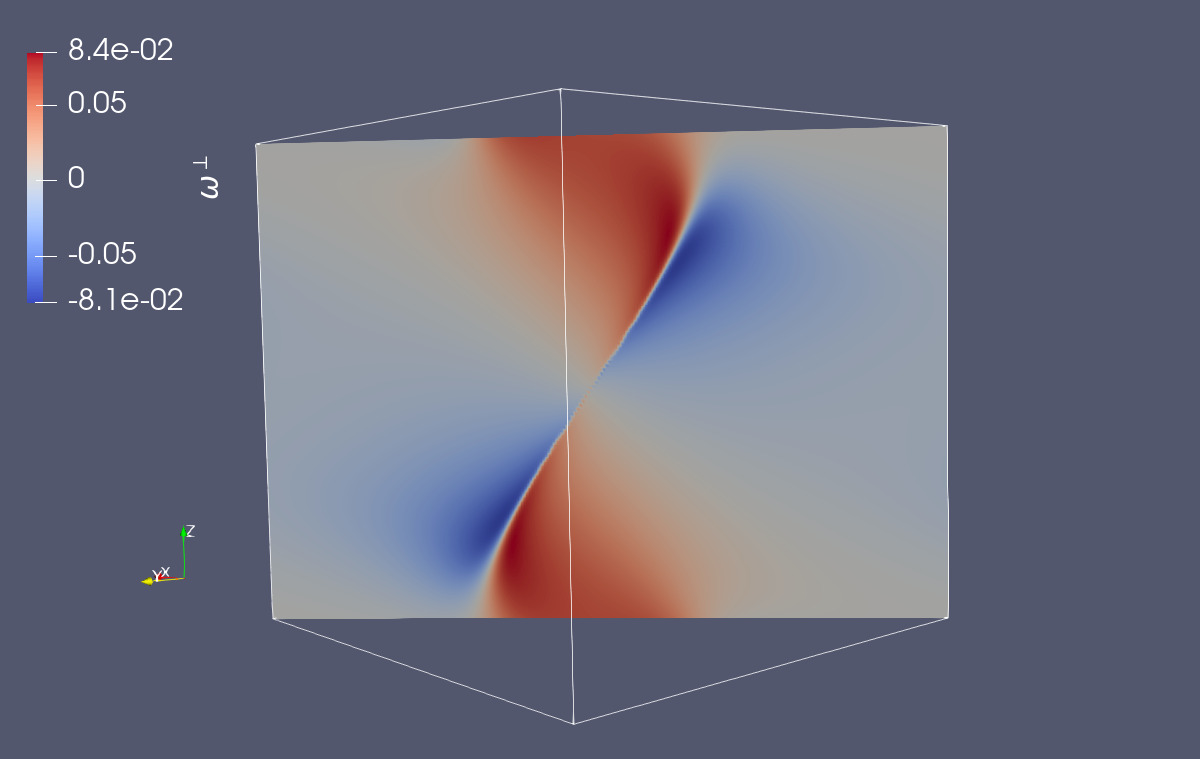}}}
\caption{(a) Isosurfaces of $|\bomega|$ and
  $\log_{10}(| |D|^3 \u|)$ in the terminal state $\u^{1024}\left(75; 
  \tu_{0;75}^{1024}\right)$ together with selected streamlines. (b) The 
  vorticity component $\omega^\perp$ normal to the symmetry plane $x_1 
  = x_2$ in  the terminal state $\u^{1024}\left(75; 
  \tu_{0;75}^{1024}\right)$. Animated versions of these figures showing 
  the time evolution for $t \in [0,75]$ are available as 
  \href{https://youtu.be/G_nfJNq6W1A}{movie 3} and 
  \href{https://youtu.be/6BcQ4LvcAE4}{movie 4} \citep{zp23}.}
  \label{fig:tuEu}
\end{figure}

Finally, we analyze the physical-space structure of the extreme Euler 
flow found by solving Problem \ref{pb:Eu} with $T = 75$. The optimal 
initial condition $\tu_{0;75}$ has a similar form to the optimal 
initial condition $\tuET$ found by solving Problem \ref{pb:maxET} for 
large $\E_{0}$ and features three perpendicular pairs of antiparallel 
vortex tubes, cf.~figures \ref{fig:tubes}a--c; it is therefore omitted 
here for brevity. However, the ensuing flow evolution for $t \in 
[0,75]$ is quite different. It is shown in 
\href{https://youtu.be/G_nfJNq6W1A}{movie 3} and the final state 
$\u^{1024}\left(75; \tu_{0;75}^{1024}\right)$ is visualized in figure 
\ref{fig:tuEu}a. We observe that at this final stage the optimal flow 
has the form of two jets colliding head-on with the vorticity 
concentrated into two strongly flattened vortex rings. The region with 
large values of $\log_{10}\left(\left| |D|^3 \u\right|\right)$, which 
is the quantity measured  by the objective functional \eqref{eq:obj}, 
is a flat disc located between the two rings.  The vorticity field at 
the final time $t = 75$ has three symmetry planes: $x_1 = x_2$, $x_1 = 
x_3$ and $x_2 = x_3$, in addition to discrete rotation symmetries with 
respect to the body diagonal passing through the center of the two 
rings.  Without loss of generality, we focus further discussion on the 
symmetry plane $x_1 = x_2$, and in figure \ref{fig:tuEu}b we visualize 
the vorticity component $\omega^\perp := \bomega \cdot \n$ normal to 
that plane ($\n = \left[-1/\sqrt{2}, 1/\sqrt{2}, 0\right]^T$ is the 
unit vector normal to the symmetry plane) at $t=75$ (see also 
\href{https://youtu.be/6BcQ4LvcAE4}{movie 4}). The streamline pattern 
in figure \ref{fig:tuEu}a indicates that the flow in the two jets 
colliding near the origin sharply transitions towards a radial outflow 
through the gap between the two vortex rings. The gap is quite  narrow 
resulting in a sharp transition between the regions of the symmetry 
plane characterized by opposite signs of the normal vorticity 
$\omega^\perp$, cf.~figure \ref{fig:tuEu}b. This is  the mechanism 
responsible for the possible singularity formation in the extreme flow 
analyzed here. It appears similar to one of the scenarios considered by 
\citet{DrivasElgindi2023}.
  
\subsection{Summary}  
\label{sec:summary}

Here we offer a brief summary of the findings presented in 
\S\,\ref{sec:extremeNS} and \S\,\ref{sec:EulerBlowup} in the context of 
steps S1--S3 listed in \S\,\ref{sec:bounds}. We conclude that the local 
maximizers of  Problems  \ref{pb:maxdEdt3D} and \ref{pb:maxdLqdt} 
obtained for increasing values of $\E_{0}$ and $B$ saturate estimates 
\eqref{eq:dEdt_estimate_E} and \eqref{eq:dLqdt}, in the sense that 
$\R_{\E_{0}}(\tuE)$ and $\R_{B}^{q}(\tuB)$ exhibit a power-law 
dependence on, respectively, $\E_{0}$ and $B$, with essentially the 
same exponents as the upper bounds in these estimates 
\citep{ld08,ap16,BleitnerProtas2026}. This suggest these estimates are 
sharp and therefore cannot be fundamentally improved, except for 
perhaps refining the prefactors. This last caveat notwithstanding, the 
goals of steps S1--S3 are achieved and this is a satisfactory situation 
from our point of view. However, we add that there seems to be no 
single family of extreme states that would saturate {\em both} 
estimates at the same time. The maximizers of the instantaneous 
problems produce a marginal only growth of the quantities of interest, 
$\E(\u(t))$ and $\| \u(t) \|_{L^{q}}$, $q > 3$, in finite time. These 
observations suggest that, should a singularity form in a Navier-Stokes 
flow in a finite time, it will likely occur along a trajectory which 
{\em does not} saturate bounds \eqref{eq:dEdt_estimate_E} and 
\eqref{eq:dLqdt}.

Solutions of the finite-time optimization problems, Problems 
\ref{pb:maxET}--\ref{pb:PsiLq}, obtained for different parameters did 
not reveal any evidence for unbounded growth of the quantities in 
question that would indicate singularity formation. This, however, does 
not preclude the possibility that a singularity may still form in a 
finite time. One reasons is that the maximizers discussed in 
\S\,\ref{sec:resultsNS} are "only" local. Problems 
\ref{pb:maxET}--\ref{pb:PsiLq} are all nonconvex and in general the 
task of finding global maximizers, or even determining if a given local 
maximizer is global, is intractable. Another reason is that the values 
of the constraint parameters $\E_{0}$ and $B$  for which Problems 
\ref{pb:maxET}--\ref{pb:PsiLq} could be solved may not be large enough, 
such that the resulting flows may belong to the small-data regime where 
global existence of classical solutions can be asserted 
\citep{RobinsonRodrigoSadowski2016}. In other words, it is possible 
that other, fundamentally different,  amplifications mechanisms may 
become unleashed in flows corresponding to larger values of parameters 
$\E_{0}$ and $B$ (and therefore also larger Reynolds numbers) leaving 
the door open to different conclusions.

Even though they do not involve any singularities, the extreme flows 
found by solving Problems \ref{pb:maxET}--\ref{pb:PsiLq} exhibit 
significant transient growth of different quantities of interest which 
is quantified by how their maximum values scale with $\E_{0}$ or $B$. 
In this regard, the maximum growth of enstrophy in the flows obtained 
by maximizing $\E(\u(T))$, $\| \u(T) \|_{L^3}$ and $\frac{1}{T} \int_0^T 
\| \u(\tau) \|_{L^q}^{p} \, d\tau$, $q > 3$, for different $T$ has been 
found to scale as $\mathcal{O}\left( \E_{\text{min}} ^{3/2} \right)$, 
cf. figures \ref{fig:maxE3D}e and \ref{fig:compq}b. It is interesting 
to note that this apparently universal behavior is realized by flows 
with a vastly different structure in the physical space, cf.~figures 
\ref{fig:tubes}a--f. Perhaps, if the Navier-Stokes system is one day 
shown to be globally well posed in the classical sense, this could be 
the form of the a priori estimate on the growth of enstrophy refining 
\eqref{eq:Et_estimate_E0}. Furthermore, it is intriguing that the 
observed scaling is the same as was established for 1D Burgers flows in 
\S\,\ref{sec:Burgers}.

In regard to Euler flows, our search did produce a solution with a 
behavior consistent with singularity formation in finite time, on a 
time window that is sufficiently long (longer than the interval of the 
local existence of classical solutions established in Theorem 
\ref{thm:Hm}). An interesting aspect of this flow is that the structure 
responsible for the singularity is in fact nearly axisymmetric, 
cf.~figure \ref{fig:tuEu}a, indicating that such may be the most 
singular geometry in 3D Euler flows. This observation is consistent 
with the studies by \citet{h09,lh14b,Hou:22:Euler}. However, we 
emphasize that here this structure was not imposed a priori and emerged 
as a solution of Problem \ref{pb:Eu}.

\FloatBarrier

\section{A Miscellany of Open Problems}
\label{sec:misc}

We present here a collection of various open problems loosely motivated 
by Questions \ref{Q1} and \ref{Q2}, respectively, in 
\S\,\ref{sec:extremeopen} and \S\,\ref{sec:anomalyopen} below. Then, in 
\S\,\ref{sec:bounded}, we mention some problems involving solid 
boundaries, an important topic which has not been addressed in this 
essay yet. The goal of this discussion is to point to some promising 
directions in which the research program outlined in 
\S\,\ref{sec:bounds} may evolve in the future.

\subsection{Extreme Growth and Singularity Formation}
\label{sec:extremeopen}

As is evident from the discussion in \S\,\ref{sec:search}, Question 
\ref{Q1} remains open. However, there is a growing body of evidence 
coming both from the mathematical PDE analysis 
\citep{DrivasElgindi2023,Elgindi2025} and numerical 
computations \citep{lh14b,Hou:22:Euler,zp23} indicating that a 
singularity may be possible in Euler flows corresponding to smooth 
initial data. A common theme in many of these investigations is that 
the Euler system is considered in the axisymmetric geometry which 
simplifies both analysis and computations by reducing  the number of 
independent variables to two (i.e., the radial and axial coordinates). 
Thus, a natural next step to follow the study discussed in 
\S\,\ref{sec:EulerBlowup} is to consider Problem \ref{pb:Eu} in such an 
axisymmetric setting.

Given that understanding the global well-posedness of the Navier-Stokes 
and Euler systems \eqref{eq:NS} and \eqref{eq:Eu} has so far proven 
intractable, a lot of attention has been focused on similar questions 
posed in the context of simpler models, especially in 1D and 2D, which 
nevertheless share some properties with systems \eqref{eq:NS} and 
\eqref{eq:Eu}. One such problem (although it is globally well posed) is 
the 1D viscous Burgers equation that was already discussed in 
\S\,\ref{sec:Burgers}. Another model problem in this category is the 
generalized Constantin-Lax-Majda (gCLM) equation
\begin{subequations}\label{eq:gCLM}
\begin{alignat}{2}
\partial_t \omega  + a u \partial_{x} \omega - \omega \partial_{x} u  & = 0 & &\qquad\mbox{in} \,\,(0,T]\times\Omega, \label{eq:gCLMa} \\
\partial_{x} u  & = \H \omega &   & \qquad\mbox{in} \,\,(0,T]\times\Omega,  \label{eq:gCLMb}
\end{alignat}
\end{subequations}
where $a \in \RR$ is a parameter and $\H$ denotes the Hilbert transform 
defined on a periodic domain $\Omega = [0,1]$ as $(\H\omega)(t,x) 
:= \int_{0}^{1} \omega(t,y) \cot(\pi(x-y)) \, dy$  with 
"PV" indicating that the integral is to be understood in Cauchy's 
principal-value sense. The nonlocal and singular form of 
\eqref{eq:gCLMb} mimics the relation between the velocity and vorticity 
in 3D, cf.~the discussion after \eqref{eq:EuVort}. Thus, equation 
\eqref{eq:gCLMa} models an interplay between vortex stretching 
represented by the last term and advection proportional to the 
parameter $a$. In the absence of advection ($a = 0$), system 
\eqref{eq:gCLM} is integrable and admits finite-time singularities. On 
the other hand, as $a$ increases from 0 to 1, advection has a growing 
regularizing effect and it is conjectured that blow-up is suppressed 
for $a = 1$ \citep{LushnikovSilantyevSiegel2021}, a case when 
\eqref{eq:gCLM} is referred to as the De Gregorio equation. While a 
number of partial results have recently been obtained, the question 
about the global regularity in the intermediate case with $0 < a \le 1$ 
remains open. Whether or not singularities can form in this regime 
under general initial conditions is a question that can be probed by 
formulating and solving variational optimization problems akin to 
Problem \ref{pb:maxET1D}.

Arguably the best known model problem in 2D for which the question about 
the global existence of classical solutions remains open is the 
generalized surface quasi-geostrophic (gSQG) system
\begin{subequations}\label{eq:gSQG}
\begin{alignat}{2}
\partial_t \theta  + \u\cdot\bnabla \theta  & = 0 & &\qquad\mbox{in} \,\,(0,T]\times\Omega, \label{eq:gSQGa} \\
\u = \bnabla^{\perp} \psi,  \quad  (-\Delta)^{\alpha/2}\psi & = \theta &   & \qquad\mbox{in} \,\,(0,T]\times\Omega  \label{eq:gSQGb}
\end{alignat}
\end{subequations}
which is a family of inviscid active-scalar equations parameterized by 
$\alpha \in (0,2]$. When $\alpha = 2$, we recover the 2D Euler 
equation, i.e., system \eqref{eq:EuVort} without the vortex-stretching 
term on the RHS in \eqref{eq:EuVorta}, which is known to be globally 
well posed for initial data in the Yudovich class $\omega_{0} \in 
L^{1}(\Omega) \cap L^{\infty}(\Omega)$ \citep{mb02}. However, as 
$\alpha$ decreases from 2 to 0, the velocity field $\u$ recovered from 
the scalar $\theta$ via \eqref{eq:gSQGb} becomes less and less regular 
and the question about the existence of classical solutions to 
\eqref{eq:gSQG} globally in time versus finite-time blow-up remains 
unresolved. While the 2D Euler system is globally well posed, it raises  
several open questions concerning the largest possible growth of various 
quantities, such as $\bnabla \omega(t)$, on domains without solid 
boundaries \citep{DrivasElgindi2023}. These problems as well can be 
studied using variational optimization formulations analogous to Problem 
\ref{pb:Eu}.

All the problems discussed up to this point concern extreme growth of 
various quantities. There are, however, important problems where one is 
interested in the converse, namely, the fastest possible decay of 
certain quantities. One such problem concerns the mixing of a passive 
scalar $\eta = \eta(t,\x)$ by an incompressible flow with the velocity 
$\u = \u(t,\x)$ which in most formulations need not satisfy the 
Navier-Stokes system \eqref{eq:NS}, but is subject to some norm 
constraints. This problem is governed by the advection-diffusion 
equation
\begin{equation}
\partial_t \eta  + \u\cdot\bnabla \eta - \kappa\Delta\eta  = 0 \qquad\mbox{in} \,\,(0,T]\times\Omega, 
\label{eq:mixing} 
\end{equation}
where $\kappa \ge 0$  is the diffusivity of the scalar and the spatial 
dimension is $d = 2,3$. The process of scalar mixing is quantified by 
the "mix norm" $\| \eta(t) \|_{H^{-1}}$ with lower values representing 
better mixing. The goal is to find suitably-constrained velocity fields 
$\u(t,\x)$ such that for a given initial distribution $\eta_{0}(\x)$ of 
the scalar, the mix norm vanishes as rapidly as possible. Mixing 
involves a complex interplay of advection and diffusion effects with 
the decay of the mix and other norms described by various a priori 
bounds. While there has been a lot of progress on this topic 
\citep{MilesDoering2018}, finding the fastest admissible mixing 
protocols remains an open problem where progress can be guided by 
variational optimization formulations.


\subsection{Dissipation Anomaly}
\label{sec:anomalyopen}

As discussed in \S\,\ref{sec:anomaly}, Question \ref{Q2} about the 
possibility of dissipation anomaly in forced 3D Navier-Stokes flows 
remains wide open from the mathematical point of view. On the other 
hand, anomalous enstrophy dissipation is ruled out in {\em unforced} 2D 
Navier-Stokes flows, cf.~\S\,\ref{sec:noanomaly}. However, the bounds 
recently obtained by \citet{MukherjeeGibbonVincenzi2025} leave the door 
open to an enstrophy dissipation anomaly in {\em forced} 2D 
Navier-Stokes flows. Thus, while addressing Question \ref{Q2} remains a 
long-term goal, investigating possible enstrophy dissipation anomaly in 
the simplified setting of forced 2D flows is a natural intermediate 
step. It can be studied based on a variational optimization formulation 
similar to Problem \ref{pb:maxchi} where the objective functional is a 
function of the forcing $\f$ in \eqref{eq:NSa}, which is also subject 
to suitable constraints, rather than of the initial condition 
$\varphi$.


\subsection{Extreme Behavior in the Presence of Solid Boundaries}
\label{sec:bounded}

In principle, Questions \ref{Q1} and \ref{Q2} can also be studied on 
bounded domains where systems \eqref{eq:NS} and \eqref{eq:Eu} are 
subject to suitable boundary conditions, usually no-slip and 
no-through-flow ($\u = \0$ on $\partial \Omega$) in the former case and 
no-through-flow ($\u\cdot\n = 0$ on $\partial \Omega$, where $\n$ is 
the unit vector normal to $\partial \Omega$) in the latter. This aspect 
adds an extra layer of complexity to these problems, not only because 
the presence of solid boundaries ushers new mechanisms of vorticity 
generation complementary to vortex stretching. In particular, they can 
lead to new forms of energy dissipation as discussed, e.g., by 
\citet{Nguyenetal2018}.

There are fundamental problems in fluid mechanics which are often 
especially relevant in the presence of solid boundaries. One such class 
of problems concerns the stability of fluid flows. For example, the 
nonlinear (in)stability of the Couette and Poiseuille flows at 
different Reynolds numbers can be studied using energy methods by 
searching for incompressible velocity fields $\v$ such that $d\K(\v)/dt 
> 0$ or $d\E(\v)/dt > 0$ (we emphasize that relations \eqref{eq:dKdt} 
and \eqref{eq:dEdtP} are not valid on domains with solid boundaries). 
Such problems can be investigated using variational optimization 
techniques similar to the approaches described in this essay, 
cf.~Problem \ref{pb:maxdEdt3D}. Another fundamental problem where solid 
boundaries play a key role is the Rayleigh–B{\'e}nard convection. 
Despite decades-long efforts, finding sharp a priori bounds on the 
Nusselt number in terms of the Rayleigh number (characterizing how the 
heat flux carried by the flow depends of the applied temperature 
gradient) in the so-called "ultimate regime" remains an elusive task. 
It is possible that some new insights about this problem may also be 
obtained by considering suitable variational formulations. Needless to 
say, numerical solution of such problems will require discretization 
techniques more complicated than those described in Appendix 
\ref{sec:numer}, although the general framework for solution of 
optimization problems, cf.~Appendix \ref{sec:solution}, would remain 
unchanged.

\section{Opportunities for Methodological Improvements}
\label{sec:improve}

In this section we outline a number of possible methodological 
improvements that have the potential to make both the task of finding 
sharp bounds (step S1 in \S\,\ref{sec:bounds}) and the search for 
solutions saturating these bounds (step S2 in \S\,\ref{sec:bounds}) 
more efficient. They concern the formulation of different problems as 
well as the numerical techniques used to solve them and reflect the 
lessons learned along the way while studying the questions discussed 
above. For brevity, this discussion will be  rather informal and we 
begin by outlining a connection with an entirely different bounding 
approach.

\subsection{Upper Bounds via Polynomial Sums-of-Squares Optimization}
\label{sec:SoS}

In order to introduce this framework we adopt an abstract perspective 
and consider solutions  $\u(t) \in \X$, where $\X$ is a suitable 
Hilbert space (finite or infinite dimensional), satisfying an 
autonomous system $d{\u}(t)/dt = \bF(\u(t))$ with some $\bF \; : \; \X 
\rightarrow \X$ which can represent system \eqref{eq:NS}, \eqref{eq:Eu} 
or \eqref{eq:B} and the initial condition $\u(0) = \u_0 \in X \subset 
\X$, where the set $X$ encodes the constraints imposed on $\u_0$.  
Denoting $\phi \; : \; \X \rightarrow \RR$ the quantity of interest in 
Problems \ref{pb:maxET1D}, \ref{pb:maxET}--\ref{pb:PsiLq} and 
\ref{pb:Eu},  these problems can be expressed as
\begin{equation}
\overline{\phi} := \sup_{\u_0 \in X} \phi(\u(\cdot;\u_0)).
\label{eq:supphi}
\end{equation}
Since Problems \ref{pb:maxET1D}, \ref{pb:maxET}--\ref{pb:PsiLq} and 
\ref{pb:Eu} are nonconvex, their solutions discussed in 
\S\S\,\ref{sec:Burgers}, \ref{sec:extremeNS} and \ref{sec:EulerBlowup} 
were obtained by {\em locally} maximizing $\phi$ over flow trajectories 
parameterized by the initial data $\u_0$ and hence may not attain the 
global maxima $\overline{\phi}$.

On the other hand, it is possible to obtain upper bounds on the
supremum $\overline{\phi}$ by exploiting the structure of the
governing equation, yet without reference to individual trajectories.
It has been shown by \citet{FantuzziGoluskin2020} that defining an
auxiliary function $V \; : \; \X \rightarrow \RR$ with the Lie
derivative $\L V(\u(t)) := \big\langle \bnabla V(\u(t)),
  \bF(\u(t)) \big\rangle_{\X} = dV(\u(t))/dt$, such upper bounds can
be deduced by solving the following optimization problem
\begin{subequations} 
\label{eq:infV}
\begin{align}
\overline{\phi} \ \le \ \inf_{V} \sup_{\u \in X} V(\u) &,  \label{eq:infVa}  \\
\L V(\u) & \le 0, \quad \u \in \X,  \label{eq:infVb} \\
\phi(\u) - V(\u) & \le 0, \quad \u \in \X,  \label{eq:infVc} 
\end{align}
\end{subequations}
which is independent of any particular solution trajectories. 
Importantly, in contrast to problem \eqref{eq:supphi}, the outer 
minimization problem in \eqref{eq:infVa} is convex. Numerical solution 
of problem \eqref{eq:supphi} is in principle possible provided the 
inner maximization subproblem in \eqref{eq:infVa} can be suitably 
relaxed and the set of auxiliary functions $V$ is made 
finite-dimensional. For example, when the auxiliary function $V(\u)$ 
and the function $\bF(\u)$ in the governing system are polynomial, 
inequality constraint \eqref{eq:infVb} can be interpreted as imposing 
the non-negativity of a polynomial which can then be expressed in terms 
of a sum of squares (SoS) of some polynomial basis functions.  For PDE 
problems a polynomial representation of $\bF(\u)$ can be constructed 
using a truncated Galerkin projection. These steps make it possible to 
approximate problem \eqref{eq:infV} in terms of a semi-definite 
optimization program for which many robust solution algorithms and 
software packages are available.

In relation to the results reviewed in \S\,\ref{sec:Burgers}, 
\citet{FantuzziGoluskin2020} solved problem \eqref{eq:infV} for a 
Galerkin truncation of the Burgers system \eqref{eq:B} and a range of 
values of $\E_0$. They obtained upper bounds on $\E_T(\u_0)$ consistent 
with relation \eqref{eq:maxETap}, which is remarkable since the lower 
and upper bounds found by solving problems \eqref{eq:supphi} and 
\eqref{eq:infV} coincide. In principle, Problems 
\ref{pb:maxET}--\ref{pb:PsiLq} and \ref{pb:Eu} can also be formulated 
in terms of the SoS framework \eqref{eq:infV} and it is interesting to 
see whether in practice it may be possible to develop suitable 
truncations and relaxations for the inner maximization problem what 
will yield computationally tractable semi-definite optimization 
programs.

\subsection{Self-Similar Blow-Up}
\label{sec:selfsimilar}

As regards the search for singularities in Euler flows, 
cf.~\S\,\ref{sec:EulerBlowup}, a promising approach is to exploit their 
possible self-similar structure. More specifically, assuming here 
$\Omega = \RR^{3}$ and introducing the functions $\bU,\bW \; : \; \Omega 
\rightarrow \RR^{3}$, one can use the ansatz
\begin{equation}
	\bomega(t,\x) =: \frac{1}{(t_{0} - t)^{\lambda}} \bW(\y), \qquad \y := \frac{\x}{(t_{0} - t)^{-\lambda}}
\label{eq:W}
\end{equation}
valid for $t < t_{0}$ and for some $\lambda \in \RR^{+}$ in 
\eqref{eq:EuVort} which leads to
\begin{subequations}
\label{eq:EuEval}
\begin{align}
\bW + \bU \cdot\bnabla_{\y}\bW - \bW \cdot\bnabla_{\y}\bU & = - \lambda \y\cdot\bnabla_{\y}\bW \qquad \text{in} \ \Omega 
\label{eq:EuEvala} \\
\bW & = \bnabla \times \bU \label{eq:EuEvalb}
\end{align}
\end{subequations}
where $\bW$ is subject to suitable decay boundary conditions at 
infinity. We note that, due to the presence of the viscous term in 
\eqref{eq:NS}, such a reduction is not possible for Navier-Stokes 
flows. With the time variable eliminated, this system can be 
interpreted as a nonlinear eigenvalue problem for the pair $\left\{ 
\lambda, \bW \right\}$. \citet{Wangetal2023} solved a related problem 
in 2D using machine-learning techniques and it is an interesting question 
whether system \eqref{eq:EuEval} can be tackled using standard methods 
of numerical analysis.

\subsection{Time-reversibility of Euler Flows}
\label{sec:trev}

In contrast to Navier-Stokes flows, solutions of the Euler system 
\eqref{eq:Eu} have the remarkable property of being time-reversible. 
This means that if $\left\{ \u(t,\x), p(t,\x) \right\}$ satisfies 
\eqref{eq:Eu}, then so does $\left\{ - \u(-t,\x), p(-t,\x) \right\}$, 
$t \in \RR$, $\x \in \Omega$. Among the many ramification of this fact, 
this means in particular that the terminal-value problem for the Euler 
system is well-posed for the same class of data as the initial-value 
problem, which is not the case for the Navier-Stokes system 
\eqref{eq:NS}. We add that the linearized Euler system is also 
time-reversible provided the base flow around which the linearization 
is performed is time-reversible as well. These properties make it 
possible to reframe the search for singularities in Euler flows. More 
specifically, in Problem \ref{pb:Eu} we maximize the $\dot{H}^{3}$ 
seminorm of the solution $\u(t)$ at time $t = T$ with respect to the 
initial data $\u_{0}$ constrained such that $\| \u_{0} \|_{\dot{H}^{3}} 
= 1$. Time-reversibility allows us to "flip" this formulation such that 
one can instead maximize $\left( \| \u(T) \|_{\dot{H}^{3}} - \| \u_{0} 
\|_{\dot{H}^{3}} \right)$ with respect to $\u(T)$, i.e., the state 
"close" to the hypothetical blow-up state $\u(t_{0})$, rather than with 
respect to the initial condition $\u_{0}$. Such a "flipped" formulation 
will likely offer advantages as regards the numerical resolution of a 
nearly singular behavior and will require backward-in-time integration 
of the governing Euler system and forward-in-time integration of the 
adjoint system, cf.~Algorithm \ref{alg:optimAlg}, both of which are 
possible due time time-reversibility.

\subsection{Log-lattices}
\label{sec:log}

Solution of optimization problems of the type discussed in this essay 
is quite costly and for each set of parameters requires roughly 
$\O(10-10^{2})$ iterations in \eqref{eq:descHs}, each consisting of an 
adjoint solve \eqref{eq:aNSE3D} and $\O(10)$ integrations of the 
governing system to solve the arc-search problem \eqref{eq:tau_nHs}. As 
a result, solution of optimization problems in 3D is usually limited to 
small values of the constraint parameters or, equivalently, small 
Reynolds numbers. A recently introduced approach to increase the 
effective Reynolds number in a numerical solution of hydrodynamic 
models relies on a Fourier-Galerkin representation of the solution, but 
with wavevectors spaced logarithmically (on the so-called 
"log-lattice") rather than diadically \citep{CampolinaMailybaev2018}. 
This thus makes it possible to resolve fluid motions with much smaller 
length scale. However, since the nature of the triadic interactions in 
the nonlinear terms in \eqref{eq:NS} and \eqref{eq:Eu} changes, this 
approach must be viewed as a form of a turbulence model. In the context 
of the research program discussed in this essay, it is particularly 
interesting to adopt such an approach to investigate different aspects 
of the dissipation anomaly problems motivated by Question \ref{Q2}.

\subsection{Adaptive Discretizations}
\label{sec:AMR}

The results reported in \S\,\ref{sec:search} highlight a fundamental 
limitation of the computational approach based on pseudospectral 
methods, cf.~Appendix \ref{sec:numer}. Hypothetical singularities in 
Navier-Stokes and Euler flows may only arise through spontaneous 
emergence of motions with very small, infinitesimal in fact, length 
scales which are localized in space and time. In order to capture these 
flows structures in numerical computations, one must refine the 
numerical resolution and the main shortcoming of the pseudospectral 
methods is that this can only be done globally by increasing the number 
of grid points everywhere in the domain $\Omega$, which is inefficient. 
Thus, in order to improve the ability to resolve small-scale features, 
one needs to develop discretization techniques allowing for adaptive 
mesh refinement (AMR) to be used in lieu of the Fourier-Galerkin 
pseudospectral methods in the solution of optimization problems. While 
AMR is an active research area in computational mathematics, there are 
currently no techniques expressly designed to handle extreme flows. The 
challenges that will need to be overcome are related to development of 
rigorous mesh refinement criteria suitable for such problems, balancing 
accuracy with conservation and other mimetic properties in addition to 
ensuring good parallel execution efficiency. For the Euler system, 
additional opportunities for AMR arise in the context of Lagrangian 
formulations such as the method based on characteristic mappings 
\citep{YinSchneiderNave2023}.

\subsection{Computer-Assisted Proofs}
\label{sec:computerproofs}

Finally, we mention what is arguably an emerging frontier in applied 
mathematics research, namely, computer-assisted proofs. This is a 
framework where numerical computations can be elevated to the level of 
mathematically rigorous results. When dealing with problems involving 
PDEs, this is achieved by deriving and verifying bounds on the 
truncation errors introduced in the discretization of different 
operators and combining them with bounds on round-off errors due to 
finite-precision arithmetics, typically obtained using interval 
arithmetics \citep{Gomez-Serrano2019}. Such techniques have been 
successfully used to study steady-state problems and solutions of 
evolutionary problems. It is interesting whether techniques for 
constructing computer-assisted proofs can also be developed to study 
PDE optimization problems such as the ones discussed in this essay.


\section{Final Comments}
\label{sec:final}

Leveraging recent developments in scientific computing, specifically, 
in solution of large-scale PDE optimization problems, the research 
program described in this essay aims to bridge the flow physics with 
mathematically rigorous bounds. This synergy has paved the way to 
results of a new type where sharpness of such a priori estimates is 
revealed by the existence of flows saturating these bounds found by 
solving suitable optimization problem, effectively closing the gap 
between "abstract" mathematical analysis and physical reality. We 
discussed two model problems where this goal has been achieved, namely the 
maximum growth of enstrophy in 1D Burgers flows in 
\S\,\ref{sec:Burgers} and the vanishing of the enstrophy dissipation 
rate in the inviscid limit in unforced 2D Navier-Stokes flows in 
\S\,\ref{sec:noanomaly}. There are, in fact, other problems where this 
goal has also been realized and they are discussed by \citet[see Table 
1]{p21a}. In addition, the proposed framework offers a systematic way 
to search for singularities in hydrodynamic models which, despite some 
recent developments, remains one of the most important goals of this 
research program. 

We should also mention limitations of the proposed framework. The most 
fundamental one is that all of the optimization problems considered in 
\S\,\ref{sec:twoproblems} and \S\,\ref{sec:search} are nonconvex, such 
that the maximizers found with the approach described in Appendix 
\ref{sec:solution} are generally only local. Thus, the search for 
extreme behavior based on solutions of these problems is usually 
nonexhaustive. Another major limitation is that the computational cost 
of solving these optimization problems is high, typically by two orders 
of magnitude larger than the cost of solving the governing equation 
alone. This limits the search to relatively small values of the 
constraint parameters or, equivalently, the Reynolds number. Some 
avenues for addressing these limitations were discussed in 
\S\,\ref{sec:improve}.

\section*{Acknowledgements}

The perspective presented in this essay was largely inspired by my 
interactions with the late Charlie R.~Doering who left us all too 
early. Progress with the research program surveyed here would not have 
been possible without the dedicated efforts of my graduate students and 
post-doctoral fellows including Diego Ayala, Fabian Bleitner, Di Kang, 
Pritpal ``Pip'' Matharu, Elkin Ram{\'i}rez, Dongfang Yun and Xinyu 
Zhao. I also wish to thank Miguel Bustamante, Sergei Chernyshenko, John 
Gibbon, David Goluskin, Thomas Y.~Hou, Anna Mazzucato, Evan Miller, 
Koji Ohkitani, Dmitry Pelinovsky, Takashi Sakajo, Roman Shvydkoy and 
Tsuyoshi Yoneda for many enlightening and enjoyable discussions. 
Funding for this research was provided by the Natural Sciences and 
Engineering Research Council of Canada (NSERC) under its Discovery 
Grants program whereas computational resources were made available by 
the Digital Research Alliance of Canada (DRAC).

\appendix

\section{Inequalities}
\label{sec:ineq}

To make this essay self-contained, we collect here a number of key 
inequalities used in the derivations of the different results presented 
above. Proofs can be found in standard textbooks on PDE analysis.
\begin{itemize}
\item 
Young's generalized inequality
\begin{equation}
\forall a,b > 0 \qquad ab \le \frac{\beta^{m} a^{m}}{m} + \frac{ b^{n}}{n\beta^{n}}, 
\qquad \text{where} \ \frac{1}{m} + \frac{1}{n} = 1, \ \beta > 0.
\label{eq:Y}
\end{equation}

\item The Cauchy-Schwarz inequality
\begin{equation}
	\left|\langle \u, \v \rangle_{X}\right| = \| \u \|_{X} \| \v \|_{X}, \qquad \u,\v \in X,
\label{eq:CS}
\end{equation}
where $X$ is an inner-product (Hilbert) space, finite or infinite dimensional.

\item{}
The Poincar{\'e} inequality
\begin{equation}
	\| u \|_{L^{2}} < C_{P} \| \bnabla u \|_{L^{2}}, \qquad C_{P} = \frac{1}{2\pi}.
	\label{eq:Poincare}
\end{equation}

\item The 1D Gagliardo-Nirenberg interpolation inequality
\begin{equation}
	\| u \|_{L^{\infty}} < C \| u \|_{L^{2}} \| \partial_{x} u \|_{L^{2}}, \qquad C = 2 \sqrt{\frac{2}{\pi}}.
	\label{eq:GN}
\end{equation}
	
\item 
\begin{lemma}[Gr{\"o}nwall]
Suppose $f \in C^{1}(I)$, $g \in C(I)$, where $I := [a,b]$ with 
$-\infty < a < b \le \infty$, such that $\int_{a}^{b} g(\tau)\, d\tau < 
\infty$ and $df(t)/dt \le g(t) f(t)$ for all $t \in I$. Then
\begin{equation}
f(t) \le |f(a)| \exp\left( \int_{a}^{t} g(\tau) \right) \qquad \forall t \in I.
\label{eq:Gro}
\end{equation}

\end{lemma}	
\end{itemize}

\section{Solution of PDE Optimization Problems}
\label{sec:solution}

In this appendix we describe the computational approach employed to 
solve Problems \ref{pb:maxET1D}, \ref{pb:maxchi} and 
\ref{pb:maxdEdt3D}--\ref{pb:Eu}. Since  these optimization problems 
have the same general mathematical structure, the methods employed to 
solve them are similar and for concreteness  we focus here on the 
specific approach to solve Problem \ref{pb:PhiHs} which is arguably one 
of the most complex problems considered here. Comments  are provided 
about modifications needed to solve some of the other problems and 
additional details can be found in the original studies by 
\citet{ap11a,ap16,KangYunProtas2020,KangProtas2021,MatharuProtasYoneda2022,zp23,RamirezProtas2026,BleitnerProtas2026}. 
An important aspect of Problems \ref{pb:maxET1D}, \ref{pb:maxchi} and 
\ref{pb:maxdEdt3D}--\ref{pb:Eu} is  that they are Riemannian which is 
because the maximizers are sought over constraint manifolds with a 
Riemannian structure represented in terms of a suitable inner product 
\citep{ams08}. To account for this property, we have developed a 
Riemannian gradient method described below in \S\,\ref{sec:RG}. A key 
element of this approach is evaluation of the gradient of the objective 
functional \eqref{eq:Phi}  with respect to the control variable which 
is the initial condition $\u_{0}$ in \eqref{eq:NS}, 
cf.~\S\,\ref{sec:grad}. Additional tools needed to implement the 
Riemannian gradient method rely on various concepts from differential 
geometry and are introduced in \S\,\ref{sec:proj}. Finally, in 
\S\,\ref{sec:cont}, we describe a continuation approach allowing one to 
find branches of local maximizers by sequentially solving optimization 
problems with varying parameters. In the research program surveyed here 
we follow the ``optimize-then-discretize" strategy where the 
optimization approach is first formulated in the infinite-dimensional 
(continuous) setting and only then the resulting equations and 
expressions are discretized for the purpose of numerical solution. As 
regards the latter aspect, some details of the numerical implementation 
are presented in Appendix \ref{sec:numer}.

\subsection{Riemannian Gradient Method}
\label{sec:RG}

To fix attention, we we focus here on solution of Problem \ref{pb:PhiHs} 
\citep{KangProtas2021}. To locally characterize the constraint manifold 
$\M_{B}$, we define the tangent space $\T_{\z}\mathcal{M}_B$ at a point 
$\z \in \mathcal{M}_B$. To do so, the fixed-norm constraint can be 
expressed in terms of the function ${G}_{q} \; :\; H^s(\Omega) 
\rightarrow \RR_+$, where ${G}_{q}(\z) := \|\z\|^q_{L^q}$ and the 
exponents $s$ and $q$ are related via Sobolev embedding 
\eqref{eq:SobEmbb}. Computing the G\^{a}teaux differential of 
${G}_{q}(\z) = B$ and using the Riesz representation theorem 
\citep{b77}, we obtain
\begin{equation}
\forall \z' \in H^{s}(\Omega) \qquad {G'}_{q}(\z;\z') := \frac{d}{d\epsilon} {G}_{q}(\z + \epsilon \z')\big|_{\epsilon = 0} = \left\langle \bnabla {G}_{q}(\z), \z' \right\rangle_{H^s} = 0,
\label{eq:dGHs}
\end{equation}
where $\bnabla {G}_{q}(\z) \in H^{s}(\Omega)$ is the gradient of the 
function ${G}_{q}$ at $\z \in \M_{B}$ and can be interpreted as an 
element orthogonal to the subspace $\T_{\z}\M_B$ in $H^{s}(\Omega)$. 
Thus, the tangent subspace is given by
\begin{equation}
\T_{\z}\M_B  := \left\{ \v \in H^s(\Omega) \, : \, \bnabla\cdot\v = 0, \; \int_{\Omega} \v \, d\x = \0,  \; \left\langle \bnabla {G}_{q}(\z), \v \right\rangle_{H^s} = 0 \right\}.
\label{eq:TzHS} 
\end{equation}
A local maximizer $\tuBT$ will then be found by constructing a sequence
of divergence-free zero-mean vector fields with a fixed $L^q(\Omega)$ norm, 
$\left\{\uBT^{(n)}\right\}_{n\in\mathbb{N}}$, such that
$\tuBT = \lim_{n\rightarrow\infty} \uBT^{(n)}$.
This sequence is defined using the following iterative procedure 
representing a discretization of the gradient 
flow for $\Phi_T^q(\u_{0})$ projected onto the manifold $\M_B$
\begin{equation}
\uBT^{(n+1)} =  \R_{\M_B}\left(\;\uBT^{(n)} + \uptau_n \, \P_{\T_n\mathcal{M}_B}\bnabla^{H^s}\Phi_T^q\left(\uBT^{(n)}\right)\;\right), \quad n = 1,2,\dots, \qquad 
\uBT^{(1)}  =  \u^0,
\label{eq:descHs}
\end{equation}
where $\u^0$ is an initial guess, $\P_{\T_n\mathcal{M}_B} \; : \; 
H^{s}(\Omega) \rightarrow \T_n \mathcal{M}_B:= 
\T_{\uBT^{(n)}}\mathcal{M}_B$ is an operator representing the orthogonal 
projection onto the tangent subspace \eqref{eq:TzHS} at the $n$th iteration, 
$\uptau_n$ is the step size, $\bnabla^{H^s}\Phi_T^q$ is the gradient of 
the functional $\Phi_T^q(\u_{0})$ computed in the Sobolev space 
$H^s(\Omega)$, whereas $\R_{\mathcal{M}_B} \: : \; \T_n\mathcal{M}_B 
\rightarrow \mathcal{M}_B$ is a retraction from the tangent subspace to 
the constraint manifold \citep{ams08}.  Precise definitions of 
$\P_{\T_n \mathcal{M}_B},$ $\uptau_{n}$ and $\R_{\mathcal{M}_B}$ are 
given in \S\,\ref{sec:proj} while computation of the gradient 
$\bnabla^{H^s}\Phi_T^q$ is described below.

\subsection{Evaluation of the Gradient}
\label{sec:grad}

A key element of the iterative procedure \eqref{eq:descHs} is 
evaluation of the Sobolev gradient $\bnabla^{H^s}\Phi_T^q \in 
H^{s}(\Omega)$ of the objective functional $\Phi_T^q(\u_{0})$, 
cf.~\eqref{eq:Phi}. The first step in determining it is to find the 
gradient of \eqref{eq:Phi} with respect to the $L^2$ topology. We begin 
by considering the G\^{a}teaux (directional) differential of the 
objective functional \eqref{eq:Phi} 
\begin{equation}
(\Phi_T^q)'(\u_0;\u_0') := \lim_{\epsilon \rightarrow 0}
\epsilon^{-1} \left[ \Phi_T^q(\u_0+\epsilon \u_0') - \Phi_T^q \right]
\label{eq:dPhi}
\end{equation}
which, when viewed as a function of its second argument with $\u_{0}$ 
fixed, is a bounded linear functional on $L^{2}(\Omega)$. Then, 
invoking the Riesz representation theorem \citep{b77}, we have 
\begin{equation}
(\Phi_T^q)'(\u_0;\u_0')
= \Big\langle \bnabla^{L^2}\Phi_T^q(\u_0), \u_0' \Big\rangle_{L^2}
\label{eq:rieszL2}
\end{equation}
which means that the evaluation of the G\^{a}teaux differential in the 
direction $\u'_{0}$ can be expressed as an inner product in 
$L^{2}(\Omega)$  with the Riesz representer interpreted as the $L^{2}$ 
gradient $\bnabla^{L^2}\Phi_T^q(\u_0)$. This tells us the $L^{2}$ 
gradient exists, but does not yet give us a recipe to compute it.

We can now use its definition \eqref{eq:dPhi} to directly
evaluate the G\^{a}teaux differential of functional \eqref{eq:Phi} as
\begin{equation}
(\Phi^q_T)'(\u_0;\u_0') = \frac{2q}{(q-3)T} \int_0^T \left( \|\u(t)\|_{L^q}^{\frac{q(5-q)}{q-3}} 
\int_{\Omega} |\u(t,\x)|^{q-2} \u(t,\x) \cdot \u'(t,\x) \, d\x \right) \, dt,
\label{eq:dPhiT}
\end{equation}
where the perturbation field $\u' = \u'(t,\x)$ is a solution of the
Navier-Stokes system linearized around the trajectory $\u(t;\u_{0})$, 
i.e.,
\begin{subequations}
\label{eq:lNSE3D}
\begin{alignat}{2}
 \mathcal{L}\begin{bmatrix} \u' \\ p' \end{bmatrix} := 
& \begin{bmatrix}
\partial_{t}\u'+\u'\cdot\bnabla\,\u+\u\cdot\bnabla\,\u'+\bnabla\, p'-\nu\Delta\u' \\
\bnabla\cdot\u'
\end{bmatrix} = \begin{bmatrix} \mathbf{0} \\ 0\end{bmatrix} & \qquad & \text{in} \ [0,T] \times\Omega, \label{eq:lNSE3Da} \\
 \u'(0)= &\u_0'& & \text{in} \ \Omega \label{eq:lNSE3Db}
\end{alignat}
\end{subequations}
which is subject to the periodic boundary conditions and where $p'$ is 
a perturbation to the pressure. Expression \eqref{eq:dPhiT} for the 
G\^{a}teaux differential is not yet in the Riesz form 
\eqref{eq:rieszL2}, because the perturbation $\u_0'$ of the initial 
data does not enter in it explicitly as a factor, but instead appears 
in the initial condition \eqref{eq:lNSE3Db} of the perturbation system. 
In order to transform the G\^{a}teaux differential to the required 
Riesz form \eqref{eq:rieszL2}, we introduce the {\em adjoint states} 
$\u^* \, : \, [0,T]\times\Omega  \rightarrow \RR^3$, $p^* \, : \, 
[0,T]\times\Omega  \rightarrow \RR$, and the following duality-pairing 
relation
\begin{equation}
\begin{aligned}
\left\langle \mathcal{L}\begin{bmatrix} \u' \\ p' \end{bmatrix}, \begin{bmatrix} \u^* \\ p^* \end{bmatrix} \right\rangle
:= & \int_0^T \int_{\Omega} \mathcal{L}\begin{bmatrix} \u' \\ p' \end{bmatrix} \cdot \begin{bmatrix} \u^* \\ p^* \end{bmatrix} \, d\x \, dt 
= \overbrace{\left\langle \begin{bmatrix} \u' \\ p' \end{bmatrix}, \mathcal{L}^*\begin{bmatrix} \u^* \\ p^* \end{bmatrix}\right\rangle}^{(\Phi^q_T)'(\u_0;\u_0')} + \\
\phantom{=} & \int_\Omega \u'(t,\x)\cdot\u^*(T,\x)  \,d\x - 
\int_\Omega \u'(0,\x)\cdot\u^*(0,\x)  \,d\x = 0.
\end{aligned}
\label{eq:dual}
\end{equation}
Using \eqref{eq:lNSE3D}, performing integration by parts with respect 
to both time and space (where all boundary terms resulting from 
integration by parts vanish due to periodicity) and defining the {\em 
adjoint system} as 
\begin{subequations}
\label{eq:aNSE3D}
\begin{alignat}{2}
 \mathcal{L}^*\begin{bmatrix} \u^* \\ p^* \end{bmatrix} := 
& \begin{bmatrix}
-\partial_{t}\u^*-\left[\bnabla\,\u^*+\left(\bnabla\,\u^{*}\right)^T\right]\u-\bnabla\, p^*-\nu\Delta\u^* \\
-\bnabla\cdot\u^*
\end{bmatrix}  = \begin{bmatrix} \f  \\ 0\end{bmatrix} & \qquad & \text{in} \ [0,T] \times\Omega, \label{eq:aNSE3Da} \\
\quad \f(t,\x)  := & \frac{2q}{(q-3)T}  \|\u(t)\|_{L^q}^{\frac{q(5-q)}{q-3}} \,  |\u(t,\x)|^{q-2} \u(t,\x), &&  \x\in \Omega, \, t \in [0,T], \label{eq:aNSE3Df} \\
 \u^*(T)= & \0 & & \text{in} \ \Omega \label{eq:aNSE3Db}
\end{alignat}
\end{subequations}
with a judicious choice of the source term $\f$ and the terminal 
condition $\u^{*}(T)$ yields the identity
\begin{equation}
(\Phi^q_T)'(\u_0;\u_0') = \int_\Omega \u'_0(\x)\cdot\u^*(0,\x)  \,d\x=\Big\langle \u^*(0), \u'_{0} \Big\rangle_{L^2}.
\label{eq:dET2}
\end{equation}
It is a Riesz representation \eqref{eq:rieszL2} of the G\^{a}teaux 
differential and noting that the perturbation $\u_{0}'$ is 
arbitrary, the $L^2$ gradient is obtained as
\begin{equation}
\bnabla^{L^2}\Phi_T^q(\u_{0}) = \u^*(0).
\label{eq:gradL2}
\end{equation}
We note that the adjoint system \eqref{eq:aNSE3D} is a linear 
terminal-value problem and hence needs to be integrated backwards in 
time. The coefficients in \eqref{eq:aNSE3Da} and the source term 
\eqref{eq:aNSE3Df} are determined by the state $\u = \u(t,\x;\u_{0})$ 
around which linearization is performed at a given iteration.

On the other hand, when solving Problems 
\ref{pb:maxdEdt3D}--\ref{pb:maxdLqdt}, which do not involve time 
evolution, the $L^{2}$ gradient can be obtained directly from the 
G\^{a}teaux differential of the objective functional without the need 
to introduce an adjoint system. For Problem \ref{pb:maxdEdt3D} we thus 
have using the Riesz representation theorem \citep{ap16}
\begin{align}
\R'_{\E_{0}}(\u;\u') & = \int_{\Omega}\left[\u'\cdot\bnabla\u\cdot\Delta\u + 
\u\cdot\bnabla\u'\cdot\Delta\u + 
\u\cdot\bnabla\u\cdot\Delta\u' \right]\,d\x 
-2\nu\int_{\Omega}\Delta^2\u\cdot\u'\,d\x
\label{eq:dR} \\
& = \Big\langle \nabla^{L^2}\R_{\E_{0}}(\u), \u' \Big\rangle_{L^2}
\end{align}
such that after performing integration by parts we obtain
\begin{equation}
\nabla^{L^2}\R_{\E_{0}}(\u) = \Delta\left( \u\cdot\bnabla\u \right) + (\bnabla\u)^T\Delta\u - 
\u\cdot\bnabla(\Delta\u) - 2\nu\Delta^2\u \qquad \text{in} \ \Omega.
\label{eq:gradRL2}
\end{equation}
For Problem \ref{pb:maxdLqdt} we proceed in an analogous manner and 
since the technicalities are more involved in this case, we refer the 
reader to \citet{BleitnerProtas2026} for all details.

Once the $L^{2}$ gradient is found as in \eqref{eq:gradL2} or 
\eqref{eq:gradRL2}, it is possible to construct the corresponding 
Sobolev and Lebesgue gradients $\bnabla^{H^s}\Phi_T^q$ and 
$\bnabla^{L^q}\Phi_T^q$ as described in the following subsections.

\subsubsection{Gradient in the Sobolev Space $H^s(\Omega)$}

The Sobolev gradient is obtained by reinterpreting the G\^{a}teaux 
differential \eqref{eq:dPhiT}, viewed as a function of its second 
argument  with $\u_{0}$ fixed, as a bounded linear functional on the 
Sobolev space $H^{s}(\Omega)$ \citep{pbh04}. 
Therefore, the differential also admits the Riesz representation, 
cf.~\eqref{eq:dET2},
\begin{equation}
(\Phi_T^q)'(\u_0;\u_0')
= \Big\langle \bnabla^{H^s}\Phi_T^q(\u_0), \u_0' \Big\rangle_{H^s}.
\label{eq:rieszHs}
\end{equation}
Motivated by computational considerations, here we use an equivalent 
form of the inner product in $H^{s}(\Omega)$, cf.~\eqref{eq:Hs}, 
namely, $\big\langle \f,\g\big\rangle_{H^s}:=\int_\Omega \f\cdot\g 
\,d\x +\ell^{2s}\int_\Omega \Delta^{s/2}\f\cdot\Delta^{s/2}\g\, d\x$, 
where $\ell$ is an adjustable parameter. Clearly, the corresponding 
definition of the norm is equivalent to \eqref{eq:Hs} in the precise 
sense of norm equivalence \citep{b77} as long as $0 < \ell < 
\infty$. Identifying the two Riesz representations \eqref{eq:dET2} 
and \eqref{eq:rieszHs} of the G\^{a}teaux differential, using 
\eqref{eq:gradL2} together with this definition of the inner product, 
performing integration by parts and noting the arbitrariness of the 
perturbation  $\u_{0}'$, we obtain the elliptic boundary-value problem 
\begin{equation}
\left[\mbox{Id}-\ell^{2s}\Delta^{s}\right]\bnabla^{H^s}\Phi_T^q=\bnabla^{L^2}\Phi_T^q \quad \mbox{in}\;\Omega,
\label{eq:HsBVP}
\end{equation}
subject to periodic boundary conditions which allows us to determine 
the Sobolev gradient $\bnabla^{H^s}\Phi_T^q$ when the $L^{2}$ gradient 
is available from \eqref{eq:gradL2}. It is clear that 
\eqref{eq:HsBVP} preserves the divergence-free and zero-mean properties 
of the $L^{2}$ gradient. Transforming equation \eqref{eq:HsBVP} to the 
Fourier-space representation, we obtain 
\begin{subequations}
\label{eq:SobvgradientFourier}
\begin{align}
\left[\widehat{\bnabla^{H^s}\Phi_T^q}\right]_{\k}&= \overbrace{\frac{1}{1+\ell^{2s}|\k|^{s}}}^{\mathcal{F}(k)} \left[\bnabla^{L^2}\Phi_T^q\right]_{\k}, \qquad \k\in\mathbb{Z}^3\setminus \{\0\},\label{eq:SobvgradientFouriera}\\
\left[\widehat{\bnabla^{H^s}\Phi_T^q}\right]_{\0}&=\0. \label{eq:SobvgradientFourierb}
\end{align}
\end{subequations}
This demonstrates that the computation of the Sobolev gradient 
$\bnabla^{H^s}\Phi_T^q$ can be regarded as an application of the 
low-pass filter $\mathcal{F}(k)$, which is a smoothing operation, to 
the $L^{2}$ gradient \eqref{eq:gradL2} with the cut-off wavenumber 
given by $\ell^{-1}$ \citep{pbh04}. It can be therefore viewed as a 
form of preconditioning. While problems with different values of $\ell 
\in (0, \infty)$ are mathematically equivalent, adjusting the value of 
this parameter can in practice have a significant effect on the rate of 
convergence of iterations \eqref{eq:descHs}.

\subsubsection{Gradient in the Lebesgue Space $L^q(\Omega)$}
\label{sec:gradLq}

Determination of the gradient in the Lebesgue spaces $L^{q}(\Omega)$, 
$q \ge 3$, needed in Problems \ref{pb:PhiLq} and \ref{pb:PsiLq} is more 
involved as we do not have an inner-product structure in these spaces 
and hence there is no Riesz identity. One thus needs to invoke the 
concept of a {\em metric} gradient, which is a generalization of the 
notion of the gradient to normed spaces. It relies on the observation 
that the gradient is the element maximizing the directional derivative 
of a function under certain constraints. This then leads to a 
constrained variational optimization subproblem defining the Lebesgue 
gradient $\bnabla^{L^{q}} \Phi^q_{T}$ in terms of the $L^{2}$ gradient 
$\bnabla^{L^{2}} \Phi^q_{T}$. The corresponding Euler-Lagrange 
equations have the form of a nonlinear boundary-value problem that can 
be solved with a variant of Newton's method. These calculations are 
rather technical and are therefore omitted here; the reader is referred 
to \citet{RamirezProtas2026} for all details. In the Hilbert case with 
$q = 2$, this formulation trivially reduces to the approach described 
above with \eqref{eq:gradL2}. A key distinction with respect to the 
formulation in a Hilbert space is that now the map from the $L^{2}$ 
gradient to the Lebesgue gradient is in general nonlinear. In contrast, 
the Sobolev gradient is obtained from the $L^{2}$ gradient by inverting 
a linear operator, 
cf.~\eqref{eq:HsBVP}--\eqref{eq:SobvgradientFourier}.

\subsection{Projection, Retraction and Arc-Maximization}
\label{sec:proj}

In Problem \ref{pb:PhiHs}, the condition characterizing the subspace 
tangent to the manifold $\M_{B}$ at some $\z \in \M_{B}$ has the form 
$\left\langle\bnabla G_{q}(\z),\z'\right\rangle_{H^s} =\left\langle 
q\,|\z|^{q-2}\z,\z'\right\rangle_{\dot{H}^s} = 0$ for all $\z'\in 
H^s(\Omega)$. cf.~\eqref{eq:dGHs}. We note that given the nonlinearity 
of the expression $|\z|^{q-2}\z$, the element $\bnabla G_{q}(\z)$ does not, 
in general, satisfy the divergence-free and zero-mean conditions, even 
if they are satisfied by $\z$. Therefore, the projection operator 
$\mathcal{P}_{\T_n\mathcal{M}_{B}}:\,H^{s}(\Omega)\rightarrow 
\T_{n}\mathcal{M}_{B}$ in iterations \eqref{eq:descHs} is defined as 
\citep{ams08,KangProtas2021}
\begin{align}
\mathcal{P}_{\T_n\mathcal{M}_{B}} \z & := 
\overline{\z} - \frac{\left\langle \z,\bnabla {G}_{q}\left(\uBT^{(n)}\right)\right\rangle_{H^s}}{
\left\langle\overline{\bnabla F_{q}\left(\uBT^{(n)}\right)},\bnabla {G}_{q}\left(\uBT^{(n)}\right)\right\rangle_{H^s}}\, \overline{\bnabla {G}_{q}\left( \uBT^{(n)}\right)}, \label{eq:PHs} \\
& \text{where} \quad  \overline{\v} := \v - \bnabla \Delta^{-1} (\bnabla \cdot \v) - \int_{\Omega} \v \, d\x \nonumber
\end{align}
which ensures the divergence-free and zero-mean 
properties are satisfied by construction. 
\begin{figure}[t]
\centering
\includegraphics[width=0.5\textwidth]{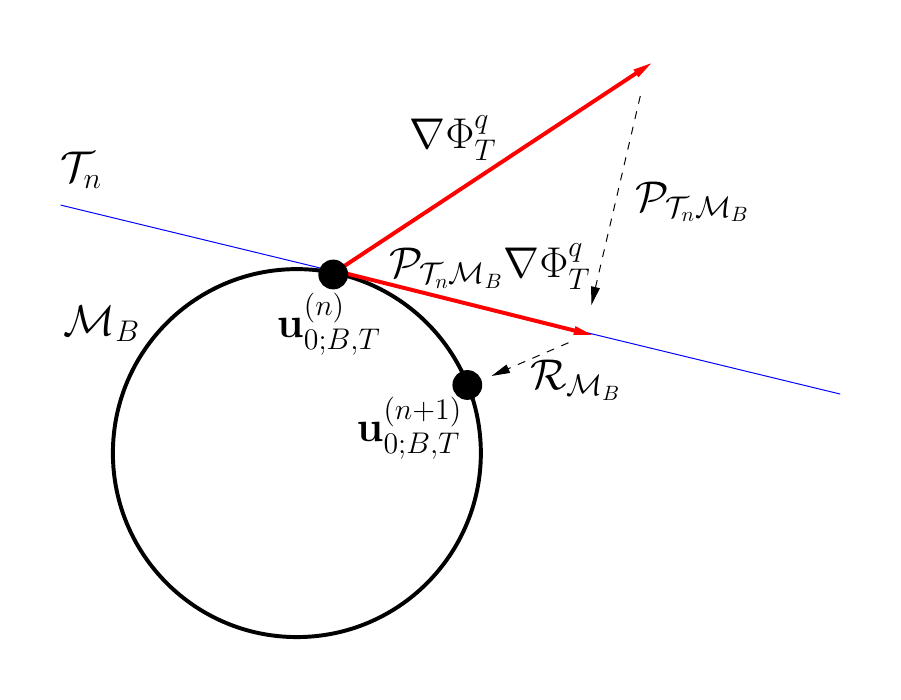}
\caption{Schematic representation of the operations performed at each 
iteration in \eqref{eq:descHs} with the projection 
$\mathcal{P}_{\T_n\mathcal{M}_{B}}$  and retraction 
$\mathcal{R}_{\mathcal{M}_B}$ defined, respectively, in \eqref{eq:PHs} 
and \eqref{eq:R_MB} \citep{KangProtas2021}.}
\label{fig:ProjHs}
\end{figure}

The retraction operator 
$\mathcal{R}_{\mathcal{M}_B}:\,\T_{n}\mathcal{M}_{B}\rightarrow\mathcal{M}_{B}$ 
in \eqref{eq:descHs} is defined as the normalization 
\citep{ams08,KangProtas2021}
\begin{equation}
\mathcal{R}_{\mathcal{M}_B}(\z) := \frac{B}{\|\z\|_{L^q}}\,\z \qquad \mbox{for all}\; \z \in \T_n\mathcal{M}_{B}.
\label{eq:R_MB}
\end{equation}
This allows us to find the step size $\uptau_{n}$ in iterative 
algorithm  \eqref{eq:descHs} by solving the arc-search problem
\begin{equation}\label{eq:tau_nHs}
\uptau_n = \mathop{\arg\max}_{\uptau>0} \Phi^q_T\left[ \mathcal{R}_{\M_B}\left(\;\uST^{(n)} + \uptau \, \mathcal{P}_{\T_n{\mathcal{M}_{B}}}\bnabla^{H^s}\Phi^q_T\left(\uBT^{(n)}\right)\;\right) \right].
\end{equation}
It can be regarded as a generalization of the standard line-search 
problem with maximization performed following an arc (a geodesic in the 
limit of infinitesimal step sizes) lying on the constraint manifold 
$\M_B$, rather than along a straight line. The computations performed 
at each iteration of the Riemannian gradient method
\eqref{eq:descHs} applied to solve Problem \ref{pb:PhiHs} are 
illustrated schematically in Figure \ref{fig:ProjHs}.

\begin{algorithm}[t]
\small
\begin{algorithmic}
\STATE
\STATE set $B = 0$, $T = 0$
\REPEAT 
\STATE \COMMENT{-------------- loop over increasing values of the constraint parameter $B$ --------------}
\STATE $B = B + \Delta B$
\STATE compute $\tuB$ {by solving Problem \ref{pb:maxdLqdt}}, as described by \citet{BleitnerProtas2026}
\STATE $\uBT^{(0)} = \tuB$
\REPEAT 
\STATE \COMMENT{------------------------ loop over {expanding} time intervals $T$ --------------------------}
\STATE $T = T + \Delta T$
\STATE $n = 0$
\REPEAT 

\STATE \COMMENT{-------------------------- optimization iterations \eqref{eq:descHs}  -------------------------------}

\STATE Solve the Navier-Stokes system with initial condition $\uBT^{(n)}$, see equation \eqref{eq:NS}

\STATE Solve the adjoint system to obtain $\u^*$ and $p^*$, see equation \eqref{eq:aNSE3D}

\STATE Compute the $L^2$ gradient $\bnabla^{L^2}\Phi^q_T\left(\uBT^{(n)}\right)$, see equation \eqref{eq:gradL2}

\STATE Compute the Sobolev gradient $\bnabla^{H^s}\Phi^q_T\left(\uBT^{(n)}\right)$,
            see system \eqref{eq:HsBVP}--\eqref{eq:SobvgradientFourier}

\STATE Compute the optimal step size $\uptau_n$, see equation \eqref{eq:tau_nHs}

\STATE Set $\uBT^{(n+1)} = \mathcal{R}_{\mathcal{M}_{B}}\left(\;\uBT^{(n)} + \uptau_n \mathcal{P}_{\mathcal{T}_{n}{\mathcal{M}_{B}}}\bnabla^{H^s}\Phi_T^q\left(\uBT^{(n)}\right)\;\right)$

\STATE Evaluate the termination condition \texttt{relative\_change $ =\frac{\Phi_T^q\left(\uBT^{(n+1)}\right)-\Phi_T^q\left(\uBT^{(n)}\right)}{\Phi_T^q\left(\uBT^{(n)}\right)}$}

\STATE Set $n=n+1$

\UNTIL{ \ \texttt{relative\_change} $<$ $\epsilon$}
\STATE $\tuBT = \uBT^{(n+1)}$, \quad $\uBT^{(0)} = \tuBT$
\UNTIL {\ $T > T_{\text{max}}$}
\UNTIL {\ $B > B_{\text{max}}$}

\end{algorithmic}
\caption{
\small
 Computation of branches of local maximizers in Problem \ref{pb:PhiHs} parameterized by $T$ for different $B$ via the continuation approach.  \newline
     \textbf{Input:} \newline
 \hspace*{0.22cm} $B_{\text{max}}$ --- maximum value of the constraint parameter \newline
 \hspace*{0.22cm} $T_{\text{max}}$ --- maximum time interval \newline
 \hspace*{0.22cm}    $\Delta B$ --- (adjustable) increment of the constraint parameter  \newline
 \hspace*{0.22cm}    $\Delta T$ --- (adjustable) increment of the length of the time interval \newline
 \hspace*{0.22cm}    $\epsilon$ --- tolerance in the solution of optimization problem via iterations \eqref{eq:descHs} \newline
 \hspace*{0.22cm}    $\ell_{2s}$ --- adjustable length scale defining the inner product in $H^{s}(\Omega)$, see also \eqref{eq:HsBVP}--\eqref{eq:SobvgradientFourier}\newline
 \textbf{Output:} \newline
 \hspace*{0.22cm}    branches of optimal initial data $\tuBT$, \ $0 \le B \le B_{\text{max}}$, $0 \le T \le T_{\text{max}}$ 
}
\label{alg:optimAlg}
\end{algorithm}

\subsection{Continuation}
\label{sec:cont}

Maximizing branches are computed using a continuation approach where we 
fix one parameter, e.g., $B$, and then solve Problem \ref{pb:PhiHs} 
with procedure \eqref{eq:descHs} repeatedly for increasing values of 
$T$. In this process the maximizer $\tuBT$ obtained for some $B$ and 
$T$ is employed as the initial guess $\u^0$ in \eqref{eq:descHs} to 
compute the maximizer $\widetilde{\u}_{0;B,T+\Delta T}$ on a larger 
time interval $[0,T+\Delta T]$, or $\widetilde{\u}_{0;B+\Delta B,T}$ 
for a larger value of the constraint parameter $B+\Delta B$, for some 
sufficiently small $\Delta T$ or $\Delta B$. Since in the limit $T 
\rightarrow 0$ solutions of Problem \ref{pb:PhiHs} coincide with the 
solutions of Problem \ref{pb:maxdLqdt} \citep{KangProtas2021}, for 
small values of the constraint parameter $B$ the instantaneous 
maximizers $\tuB$, cf.~figure \ref{fig:tuEB}b, can be used as ``seeds'' 
to initiate the computation of the maximizing branches, i.e., as the 
initial guess for $\widetilde{\u}_{0;B,\Delta T}$\footnote{Since 
Problem \ref{pb:maxdLqdt} was solved by \citet{BleitnerProtas2026} only 
afterwards, this approach was not in fact used by 
\citet{RamirezProtas2026} to solve Problem \ref{pb:PhiHs}. However, it 
was successfully employed by \citet{KangYunProtas2020} to solve Problem  
\ref{pb:maxET} based on solutions of Problem \ref{pb:maxdEdt3D}.}. The 
procedure outlined here is summarized as Algorithm \ref{alg:optimAlg}. 
While there exist alternatives to the continuation approach, provided 
$\Delta B$ and $\Delta T$ are sufficiently small, this technique in 
fact results in the fastest convergence of iterations \eqref{eq:descHs} 
and also ensures that the computed optimal initial data lie on a single 
maximizing branch.

\section{Numerical Approximations}
\label{sec:numer}

In this essay we have surveyed results obtained for problems formulated 
in 1D (in \S\,\ref{sec:Burgers}), 2D (in \S\,\ref{sec:noanomaly}) and 
in 3D (in \S\,\ref{sec:extremeNS} and \S\,\ref{sec:EulerBlowup}). All 
the PDE systems involved in the solution of the corresponding 
optimization problems, namely, the governing systems and the 
corresponding adjoint systems, were solved using standard 
pseudo-spectral methods \citep{canuto:SpecMthd}. They involve 
Fourier-Galerkin discretization in space with nonlinear products 
evaluated using collocation in the physical space and dealiasing 
performed based on the 3/2 rule. FFT routines were used to perform 
transforms between the real- and Fourier-space representations of the 
solution. The systems of ordinary differential equations resulting 
from the discretization in space were discretized in time using a hybrid 
approach typically combining an explicit Runge-Kutta RK3 method applied 
to the nonlinear terms and terms with non-constant coefficients with an 
implicit Crank–Nicolson method applied to the linear terms, which 
offers a good balance between accuracy and favourable stability 
properties. For Problems \ref{pb:maxchi} and 
\ref{pb:maxdEdt3D}--\ref{pb:Eu} formulated in 2D and 3D,  massively 
parallel implementations based on the Message-Passing Interface (MPI) 
were used with solutions of the largest problems typically requiring 
$\mathcal{O}(10^{2})$ hours on $\mathcal{O}(10^{2})$ CPUs. The Fourier 
spectra of all solutions were carefully monitored to ensure the 
calculations were well resolved. In addition, in the case of Problem 
\ref{pb:Eu} involving the Euler system \eqref{eq:Eu}, it was also 
ensured that the inviscid invariants (kinetic energy \eqref{eq:K} and 
helicity) were conserved with a sufficient accuracy.

\bigskip
\noindent
{\bf Declaration of Interests.} The author reports no conflict of interest.

\FloatBarrier


\end{document}